\documentclass[twocolumn]{aastex631}
\usepackage[utf8]{inputenc}
\usepackage{amsmath,enumitem}
\usepackage{amssymb}
\usepackage{pifont}
\usepackage{textcomp}
\usepackage{gensymb}
\usepackage{verbatim}
\usepackage{graphicx}
\usepackage{lipsum}
\usepackage{hyperref}

\newcommand\W{$\lambda$}
\newcommand\niioii{N$^+$/O$^+$}
\newcommand\niiioiii{N$^{+2}$/O$^{+2}$}
\newcommand\nivoiii{N$^{+3}$/O$^{+2}$}
\newcommand\niiinivoiii{(N$^{+2}$ + N$^{+3}$)/O$^{+2}$}

\newcommand\noUVopt{(N$^+$ + N$^{+2}$)/(O$^+$ + O$^{+2}$)}
\newcommand\noUVoptfrc{$\frac{\rm N^+ + N^{+2}}{\rm O^+ + O^{+2}}$}
\newcommand\Te{$T_e$}
\newcommand\den{$n_e$}
\newcommand{\CH}[1]{\colhead{#1}}
\newcommand\sfru{$(\frac{M_\odot}{\rm{yr}})$}
\newcommand\sigsfru{$(\frac{M_\odot}{\rm{yr}\ \rm{kpc^2}})$}

\newcommand\lz{low-$z$}
\newcommand\hz{high-$z$}
\newcommand\p{$\pm$}
\newcommand\dg{$^{\dagger}$}

\newcommand\rarr{$\rightarrow$}

\newcommand\tran{$l_u\;\rightarrow\;l_l$}

\newcommand\OIII{\ion{O}{3}]}

\newcommand\NIII{\ion{N}{3}]}
\newcommand\NIV{\ion{N}{4}]}
\newcommand\NII{[\ion{N}{2}]}
\newcommand\ArIV{[\ion{Ar}{4}]}
\newcommand\ArIII{[\ion{Ar}{3}]}
\newcommand\CIII{\ion{C}{3}]}
\newcommand\SII{[\ion{S}{2}]}
\newcommand\OII{[\ion{O}{2}]}

\newcommand{\cmark}{\ding{51}}
\newcommand{\xmark}{\ding{55}}

\definecolor{dab}{RGB}{134, 109, 180}
\definecolor{zm}{HTML}{C71585}

\begin{document}

\title{Under Pressure: Decoding the Effect of High Densities on Derived Nebular Properties}

\author[0009-0000-2997-7630]{Zorayda Martinez}
\affiliation{Department of Astronomy, The University of Texas at Austin, 2515 Speedway, Stop C1400, Austin, TX 78712, USA}
\affiliation{Cosmic Frontier Center, The University of Texas at Austin, Austin, TX 78712, USA} 

\author[0000-0002-4153-053X]{Danielle A. Berg}
\affiliation{Department of Astronomy, The University of Texas at Austin, 2515 Speedway, Stop C1400, Austin, TX 78712, USA}
\affiliation{Cosmic Frontier Center, The University of Texas at Austin, Austin, TX 78712, USA} 

\author[0000-0003-4372-2006]{Bethan L. James}
\affiliation{AURA for ESA, Space Telescope Science Institute, 3700 San Martin Drive, Baltimore, MD 21218, USA}

\author[0000-0002-2644-3518]{Karla Z. Arellano-C\'{o}rdova}
\affiliation{Institute for Astronomy, University of Edinburgh, Royal Observatory, Edinburgh, EH9 3HJ, UK}

\author[0000-0001-6106-5172]{Daniel P. Stark}
\affiliation{Department of Astronomy, University of California, Berkeley, Berkeley, CA 94720, USA}

\author[0000-0002-9132-6561]{Peter Senchyna}
\affiliation{Observatories of the Carnegie Institution for Science, 813 Santa Barbara Street, Pasadena, CA 91101, USA}

\author[0000-0003-0605-8732]{Evan D.\ Skillman}
\affiliation{Minnesota Institute for Astrophysics, University of Minnesota, 116 Church St. SE, Minneapolis, MN 55455}

\author[0000-0002-0361-8223]{Noah S. J. Rogers}
\affiliation{Center for Interdisciplinary Exploration and Research in Astrophysics (CIERA), Northwestern University, 1800 Sherman Avenue, Evanston, IL 60201, USA}

\author[0000-0002-0302-2577]{John Chisholm}
\affiliation{Department of Astronomy, The University of Texas at Austin, 2515 Speedway, Stop C1400, Austin, TX 78712, USA}
\affiliation{Cosmic Frontier Center, The University of Texas at Austin, Austin, TX 78712, USA} 

\shorttitle{Nitrogen and Density Evolution}
\shortauthors{Martinez et al.}
\correspondingauthor{Zorayda Martinez}
\email{zorayda@utexas.edu}

\begin{abstract}
    Recent JWST observations have uncovered a population of compact, high-redshift ($z>6$) galaxies exhibiting extreme nebular conditions and enhanced nitrogen abundances that challenge standard chemical evolution paradigms.
    We present a joint UV and optical abundance analysis using a new suite of \texttt{Cloudy} photoionization models covering a wide density range ($n_e=10^2-10^9$ cm$^{-3}$), combined with HST and JWST spectroscopy for a sample of star-forming galaxies across $0.0\lesssim~z~\lesssim10.6$. 
    We find that assuming uniform, low-density conditions ($n_e\sim10^2$ cm$^{-3}$) in high-density environments ($n_e\sim10^5$ cm$^{-3}$) can bias nebular diagnostics by overestimating \Te\ (up to 1800 K), overpredicting $\log~U$ (by $>1$~dex), and underestimating O/H (up to 0.67~dex), while only modestly inflating N/O.
    Therefore, robust abundance determinations at \hz\ require a multi-phase density model. Using this model, we recalculate O/H and N/O abundances for our sample and present the first $\log~U$ diagnostics and ICFs for high-ionization UV N lines. We find that the UV tracers systematically overestimate N/O by $\sim0.3-0.4$ dex relative to the optical benchmark.
    We find that N/O increases with redshift, correlating with both \den\ and star formation rate surface density ($\rm\Sigma_{SFR}$), suggesting that N/O is temporarily enhanced in compact, high-pressure environments. 
    However, the \den\ evolution with $z$ is more gradual than the $(1+z)^3$ scaling of virial halo densities, suggesting that \den\ evolution is shaped by both cosmological structure growth and baryonic processes. 
    These trends point to prompt N/O enrichment potentially driven by very massive stars, with key implications for interpreting UV emission and determining reliable chemical abundances from JWST observations of the early universe.
\end{abstract}

\section{Introduction} \label{sec:intro}
Observed nebular properties of galaxies are a powerful tool for probing galaxy evolution, and accurate interpretation of these properties requires robust methods of characterizing the physical conditions in galaxies across cosmic time.
Notably, the chemical composition of a galaxy encodes vital details about star formation and the dynamic processes that reshape galaxies over time.
Furthermore, total O/H and relative N/O abundances are useful measures of cumulative and recent star formation history, respectively, since each one is sensitive to enrichment from stars of different masses.
However, accurate abundance determinations require observations of faint emission lines, which are both highly sensitive to gas-phase physical conditions (e.g., temperature, density, ionization) and difficult to observe at higher redshifts.
While many assumptions about ionized gas physical conditions, notably electron densities (\den), are made based on galaxy properties from detailed $z \sim 0$ measurements, recent JWST studies suggest that ionized gas conditions and chemical compositions evolve significantly across redshift \citep[e.g.,][]{isobe23a, abdurrouf24, topping25b}.
As such, environments with high \den~in the ionized gas may significantly bias UV and optical abundance determinations.

JWST abundance studies of high-redshift ($z > 6$) galaxies have revealed a {\it nitrogen overabundance} where some very young galaxies exhibit enhanced relative N/O abundances, as probed by their high-ionization rest-frame ultraviolet (UV) \NIII~\W1750 and \NIV~\W\W1483,1487 emission lines.
While these high-ionization N lines are rarely seen at low-redshift \citep[e.g.,][]{mingozzi22}, which is in part due to the significant instrument challenges involved in obtaining UV and optical spectra for low-$z$ galaxies, a surprising number of high-$z$ galaxies, whose UV and optical spectra can both be traced by JWST, have been observed with very strong UV N emission line intensities \citep[e.g.,][]{pascale23, bunker23, ji24, marqueschaves24, topping24, schaerer24}.
Furthermore, the relative N/O abundances derived from the UV lines are significantly higher than the N/O abundances measured from low-ionization, optical N$^{+}$ and O$^{+}$ emission lines.

While persistent UV-optical discrepancies may suggest fundamentally different N production mechanisms in early galaxies, this discrepancy could also be driven by the observed high values of \den\ at high-$z$ \citep{isobe23a, abdurrouf24, topping25b}.
Noting that most abundance studies adopt a low-density assumption (\den~$\sim 10^2$ cm$^{-3}$), \cite{hayes25} recently used stacks of high-$z$ JWST spectra to show that oxygen abundances changed significantly when electron temperatures were derived from a high-density assumption rather than a low-density one. 
To date, the effects of electron density evolution on abundances have yet to be explored in individual galaxies, and continued use of low-$z$ assumptions will potentially bias inferred galaxy properties.

Density, in particular, is extremely important for careful determination of other nebular properties.
Substantially increasing the electron density of ionized gas causes a rise in gas pressure according to the ideal gas law, where pressure scales with both density and temperature.
At high-densities, recombination rates accelerate, which alters the balance between ionized and neutral species, shifting the ionization equilibrium and reducing the ionization parameter.
In these extreme environments, key observables such as emission line strengths, temperature diagnostics, and inferred metallicities will experience notable changes.
Additionally, collisional de-excitation begins to occur for different emission lines at different densities meaning that carefully deriving gas conditions will be more crucial than ever in the high-density environments that are more common at \hz.
Finally, while past works have yet to systematically explore these effects, many traditional diagnostics may overestimate the ionization state and underestimate the metallicity in high-densities, which in turn create potential biases in our understanding of star-forming environments, active galactic nuclei, and galaxy evolution.

Frameworks for interpreting ionized gas consist of ionization zones, where the gas closer to an ionizing source has higher ionization.
In the common three-zone framework, the inner high-ionization zone is defined by the O$^{+2}$ gas ($> 35.1$ eV), the intermediate-ionization zone is defined by the S$^{+2}$ gas ($23.3 - 34.8$ eV), and the outer low-ionization zone is defined by the N$^+$ gas ($14.5  - 29.6$ eV).
Observationally, it has been suggested that the ionized gas in galaxies exhibit a gradient in their electron temperatures and densities across the ionization structure of the gas \citep[e.g.,][]{berg21, mingozzi22}.
Essentially, measured densities from low-ionization emission lines are found to be lower than the densities derived from intermediate- and high-ionization emission lines.
If these high-ionization densities evolve with redshift, similar to low-ionization densities, at \hz~our derived nebular properties may be significantly biased by these extreme densities.

Notably, within a given ionization zone density inhomogeneities can also have a significant impact on the inferred low-ionization nebular properties of local objects \citep[e.g.,][]{seaton57, peimbert71, rubin89, mendezdelgado23, rickardsvaught24}.
These density fluctuations can be negligible \citep[$<100$ cm$^{-3}$; e.g.,][]{garciabenito10} or more significant \citep[$\simeq600$ cm$^{-3}$; e.g.,][]{hamelbravo24} such that density diagnostics with low critical densities (e.g., $n_{e,crit}$([\ion{O}{2}]) $\sim 10^3$ cm$^{-3}$) will be significantly affected.
While these \den-fluctuations may evolve with $z$, only a handful of works have spatially mapped \den~at $z \gtrsim$ 2 \citep[e.g.,][]{cresci23, marconcini24} and find variations that are similar to local results suggesting minimal redshift evolution in \den-fluctuations.
However, integrated densities from a given ion have been observed to increase at \hz~(i.e., $n_e\gtrsim10^5$ cm$^{3}$) such that diagnostics with low critical densities are in the high-density limit,
where most of their flux will be collisionally suppressed.
As a result, even when such diagnostics are detected, they trace only the diffuse component and systematically underestimate the true densities of the ionized gas.
This limitation is especially important for interpreting high-redshift measurements that rely exclusively on optical or NIR rest-frame lines.
Fortunately, the low-critical density diagnostics tend to characterize the lower ionization ions used in this work, but UV N emitters are inherently high-ionization objects and so the low-ionization emission lines only make a small contribution to the total abundance.

Building on the \lz~foundation of density studies, we explore the more extreme nebular conditions of \hz~galaxies.
We examine the effects of high-densities on determinations of nebular properties including electron temperature, ionization parameter, total O/H abundances, relative N/O abundances, and the overall impact on chemical evolution trends.
Additionally, we only explore stellar populations as an ionizing source, but future follow-up work will examine other ionizing sources.
We detail our sample of UV N-emitting galaxies from $0 < z < 11$ in Section \ref{sec:samp} and our robust emission line measurements, where possible, in Section \ref{sec:emlinemeasure}.
We present a new custom suite of \texttt{Cloudy} photoionization models that accounts for the observed ionized gas conditions up to \den~$\sim10^9$ cm$^{-3}$ and significant super-solar N/O enhancements in Section \ref{sec:models}.
We examine the effects of increasing \den~in a uniform density model in Section \ref{sec:const_den}, exploring emission line diagnostics in \S~\ref{sec:emline_varyden}, electron temperatures in \S~\ref{sec:tem_varyden}, and ionization parameters, ionization correction factors, and abundances in \S~\ref{sec:abund_varyden}. 
A summary of these \den~effects along with a more observationally-motivated high-density core model is discussed in Section \ref{sec:den_effects}.
In Section \ref{sec:implications_sample}, we recalculate abundances for our sample using a multi-phase density model.
We discuss the implications of our results on the evolution of N/O abundances in Section \ref{sec:Nevolution}, with special consideration for the evolution of density ( \S~\ref{sec:evolve_conditions}) and star formation rate surface density (\S~\ref{sec:galprop}).  
Finally, we provide a summary of our work and key conclusions in Section \ref{sec:conclude}.
Note that we use vacuum wavelengths throughout this work. 

\section{Sample \& Data} \label{sec:samp}
In order to explore the conditions producing extreme UV N emission at high redshifts and the role that density plays, we select a sample of galaxies spanning $0 \lesssim z \lesssim 10.6$ with UV N line detections, either \NIV~\W\W1483,1487 or \NIII~\W1750\footnote{\NIII~is a quintuplet composed of emission lines at \W1746.82, \W1748.65, \W1749.67, \W1752.16, and \W1754.00.
Note that the \W1749.67 and \W1752.16 lines are typically the strongest, but their ratio is density sensitive. 
In this work, we use \NIII~\W1750 to represent the sum of the quintuplet.}, and intermediate- or high-ionization electron density measurements from \CIII~or \NIV, respectively.
The resulting sample consists of 17 objects that we divide into two subsamples: the Low$-z$ Sample ($z < 1$) and the High$-z$ Sample ($z > 1$).
The Low$-z$ Sample consists of 8 nearby objects that we describe below in Section \ref{sec:lzsamp}.
The remaining 9 objects of the High$-z$ Sample are intermediate- to \hz~ objects  selected from the literature as described in Section \ref{sec:hzsamp}.
Relevant information about our sample and corresponding references can be found in Table \ref{tab:samp_prop}.
We require that every object in our sample has at least an intermediate- or high-ionization density diagnostic.
The available high-ionization temperature and density diagnostics for each object can found in Table \ref{tab:avail_diag}.

\subsection{Low-\texorpdfstring{$z$}{z} Sample} \label{sec:lzsamp}
The Low$-z$ Sample was difficult to construct because very few rest-UV N emission line detections exist in the literature.
We searched the literature for \ion{N}{4}] or \ion{N}{3}] detections, finding nine galaxies with $z < 0.1$.
We note that no galaxies with high-ionization UV N detections have yet been found for $0.1 < z < 1.0$\footnote{From our experience with COS gratings, the sensitivity extends only to about 2000\AA, which skews detections of towards \ion{N}{4}] or \ion{N}{3}] to $z < 0.1$}.
Seven of these galaxies are taken from the COS Legacy Archive Spectroscopy SurveY \cite[CLASSY;][]{berg22,james22}.
CLASSY consists of 45 nearby star-forming galaxies with high-resolution rest-frame far-UV (FUV) {\it HST}/COS spectra, which allows detailed studies of nebular emission lines that are only possible with local galaxies.

One additional object is the extremely metal-poor ($Z<0.05Z_\odot$) galaxy Leo P \citep{skillman13}, whose rest-UV spectra are taken from \citet{telford23} and the HST-GO-17102 program (PI D. Berg).
Following the CLASSY coadding processes described in \citet{berg22}, we combined the high-resolution G130M and G160M gratings \citep[from][]{telford23} with the low-resolution G140L grating (from HST-GO-17102) to cover both the \ion{N}{4}] and \ion{N}{3}] features, as well as \ion{C}{3}] \W\W1907,1909. The multi-step process includes extracting, reducing, aligning, and coadding the spectra from the different gratings. The final spectrum is shown in Figure \ref{fig:LeoP_spec} in the Appendix. 

Properties of the Low-$z$ Sample are listed in Table~\ref{tab:samp_prop}.
The entire Low-$z$ Sample has high-quality rest-UV and optical spectra that cover density diagnostics in both low-ionization (e.g., [\ion{O}{2}] and [\ion{S}{2}]) and high-ionization gas (e.g., \ion{N}{4}], \ion{C}{3}], and [\ion{Ar}{4}]), as well as low- and high-ionization N emission from [\ion{N}{2}] \W6585 and \ion{N}{3}] \W1750 and/or \ion{N}{4}] \W\W1483,1487, respectively, \citep{mingozzi22}.
Thus, the power of the Low-$z$ Sample lies in its detailed dataset, which enables calculations of multi-phase gas properties and the first self-consistent comparison of N/O abundances derived from low- and high-ionization emission lines.

\begin{figure*}[htp]
    \begin{center}
    \includegraphics[width=0.95\textwidth]{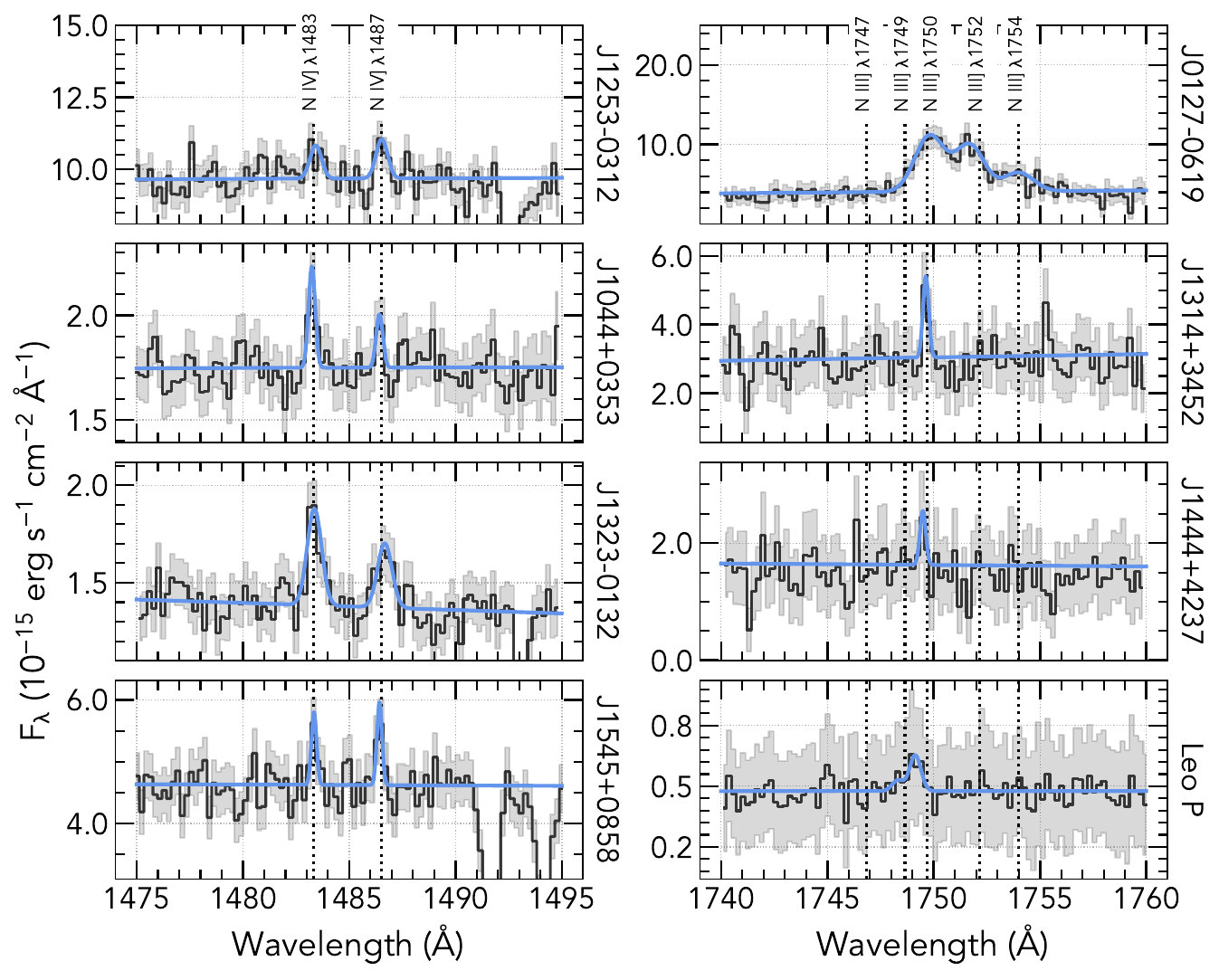}
    \caption{Rest-UV spectra for the Low-$z$ Sample (see Section~\ref{sec:lzsamp} and Table~\ref{tab:samp_prop}).
    The detected \NIV~\W\W1483,1487 (left) and \NIII~\W1750 (right) emission profiles are shown; four galaxies have \NIV~\W\W1483,1487 emission and five have \NIII~\W1750 emission, but none have both.
    We plot the emission line fits (blue line) for the eight of these galaxies without reported \NIV~and \NIII~line fluxes in the literature.}
    \label{fig:spectra}
    \end{center}
\end{figure*}

The \ion{N}{3}] and \ion{N}{4}] emission line portions of the rest-UV spectra of the Low-$z$ Sample are shown in Figure \ref{fig:spectra}.
Five of the nine galaxies have visible \ion{N}{3}] \W1750 emission and four have \ion{N}{4}] \W\W1483,1487.
Interestingly, no objects have significant emission from both \ion{N}{3}] and \ion{N}{4}].
Given that N$^{+3}$ emission requires higher ionization and excitation energies than N$^{+2}$, different physical conditions are needed to produce the observed dichotomy in \ion{N}{4}] and \ion{N}{3}] detections in our sample. 

\subsection{High\texorpdfstring{$-z$}{-z} Sample} \label{sec:hzsamp}
The primary goal of this work is to assess the evolution of gas density and its effect on observed nebular conditions and abundances, including the significantly N-enriched \hz~galaxies observed with JWST.
Therefore, we searched the literature for $z > 1$ galaxies\footnote{We utilize $z > 1$ as the qualifier for the High$-z$ Sample since this is the redshift at which [\OIII~\W5008 will redshift past 1 $\mu$m, which is usually beyond the range of most ground based optical spectrographs.} with reported \NIII~\W1750 or \NIV~\W\W1483,1487 detections to form our the High-$z$ Galaxy Sample.
In order to perform robust calculations of nebular properties, we also limited our sample to sources that have published attenuation values ($E(B-V)$) or multiple Balmer line measurements from which we could derive $E(B-V)$.
As a result, we found nine galaxies in the redshift range of $1.3 < z < 10.6$ with \ion{N}{3}] or \ion{N}{4}] detections.

The properties of the High-$z$ Sample are listed in Table~\ref{tab:samp_prop}.
In contrast to the Low-$z$ Sample, all nine \hz~galaxies have \ion{N}{3}] detections, and six of these also have \NIV~\W\W1483,1487 detections.
However, only four of these galaxies have sufficiently red optical spectra from which we can derive N/O abundances using the low-ionization \NII~\W6585 and \OII~\W\W3727,3730 emission lines.
Note that this sample includes the extreme GN-z11 galaxy, enabling us to examine its properties against a robust sample of high-ionization N emitters with density measurements across cosmic time.

\begin{deluxetable*}{llRRcCcCccccc}[ht]
\tabletypesize{\footnotesize}
\caption{UV N-Emitting Sample Properties \label{tab:samp_prop}}
\tablewidth{\textwidth}
\tablehead{
\CH{} & \CH{} & \CH{Coords.} &\CH{} &\CH{$\log M_\star$}& \CH{$\log$ SFR} & \CH{R$_e$} & \CH{$\log \Sigma_{\rm SFR}$} & \CH{12+} & \CH{} & \multicolumn{3}{c}{N Detections} \\ [-1.25ex] \cline{11-13} 
\CH{Label} & \CH{Name} & \CH{(J2000)} &\CH{$z$}& \CH{($M_\odot$)} & \CH{\sfru} & \CH{(pc)} & \CH{\sigsfru} & \CH{log(O/H)} & \CH{Ref.} & \CH{\NII} & \CH{\NIII} & \CH{\NIV}
}
\startdata
\multicolumn{13}{l}{\bf Low-$z$ Sample:} \\
Lz-1  & J0127$-$0619 &   21.89796,\ -6.32668 & 0.0054 & 8.74 & -0.75 & 20.0   & +1.85 & 7.68 & (1)   & \cmark & \cmark & \xmark \\
Lz-2  & J1314$+$3452 & 198.69736,+34.88328   & 0.0029 & 7.56 & -0.67 & 30.0   & +1.58 & 8.26 & (1)   & \cmark & \cmark & \xmark \\
Lz-3  & J1444$+$4237 & 221.04776,+42.62655   & 0.0023 & 6.48 & -1.94 & 45.0   & -0.04 & 7.64 & (1)   & \cmark & \cmark & \xmark \\
Lz-4  & Leo P        & 155.43792,+18.08811   & 0.0009 & 5.75 & -4.31 & 0.4    & +1.69 & 7.17 & (2)   & \cmark & \cmark & \xmark \\
Lz-5  & J1253$-$0312 &  193.27487,\ -3.21637 & 0.0227 & 7.65 & +0.56 & 390.0  & +0.58 & 8.06 & (1)   & \cmark & \xmark & \cmark \\
Lz-6  & J1044$+$0353 &  161.24080,\ +3.88699 & 0.0129 & 6.80 & -0.59 & 100.0  & +0.61 & 7.45 & (1)   & \cmark & \xmark & \cmark \\
Lz-7  & J1323$-$0132 &  200.94775,\ -1.54776 & 0.0225 & 6.31 & -0.72 & 100.0  & +0.48 & 7.71 & (1)   & \cmark & \xmark & \cmark \\
Lz-8  & J1545$+$0858 &  236.43147,\ +8.96704 & 0.0377 & 7.52 & +0.37 & 250.0  & +0.78 & 7.75 & (1)   & \cmark & \xmark & \cmark \\
[2ex]
\multicolumn{13}{l}{\bf High-$z$ Sample:} \\
Hz-1  & RXCJ2248-ID  & 342.17972,-44.53300   & 6.106  & 8.05 & +1.80 & 22.0   & +4.32  & 7.43 & (4)  & \cmark & \cmark & \cmark \\ 
Hz-2  & GN-z11       & 189.10605,+62.24205   &10.621  & 8.73 & +1.27 & 64.0   & +2.86  & 7.82 & (5)  & \xmark & \cmark & \cmark \\ 
Hz-3  & GDS 3073     &  53.07888,-27.88416   & 5.500  & 9.52 & +1.95 & 110.0  & +3.07  & 8.00 & (6)  & \cmark & \cmark & \cmark \\
Hz-4  & Sunburst     & 237.52013,-78.18314   & 2.371  & 9.04 & +1.00 & 7.8    & +4.42  & 7.96 & (7)  & \cmark & \cmark & \xmark \\ 
Hz-5  & J1723$+$3411 & 260.90067,+34.19946   & 1.329  & 8.77 & +0.90 & \ldots & \ldots & 8.30 & (8)  & \cmark & \cmark & \xmark \\
Hz-6  & CEERS 1019   & 215.03539,+52.89066   & 8.678  & 9.30 & +2.17 & 112.0  & +3.27  & 7.70 & (9)  & \xmark & \cmark & \cmark \\
Hz-7  & SL2S J0217   &   34.40515,\ -5.22494 & 1.844  & 8.26 & +1.35 & 300.0  & +1.60  & 7.50 & (10) & \xmark & \cmark & \xmark \\
Hz-8  & Lynx         & 132.20317,+44.93044   & 3.300  & 7.67 & +1.68 & \ldots & \ldots & 7.76 & (11) & \xmark & \cmark & \cmark \\
Hz-9  & A1703-zd6    & 198.75423,+51.83466   & 7.044  & 7.70 & +1.49 & 120.0  & +2.53  & 7.47 & (12) & \xmark & \cmark & \cmark 
\enddata
\tablecomments{
Properties for our rest-UV N-emitter Low-$z$ ($z \sim 0$) and High-$z$ ($1 < z < 11$) Samples.
Columns 1 and 2 list the label used in this work and the literature name of
each object, followed by the coordinate R.A. and Decl. of their spectral observations and redshifts in Columns 3 and 4.
Galaxy stellar mass, SFR, rest-UV effective radius, SFR surface density, and metallicity are listed in Columns 5, 6, 7, 8, and 9 respectively, with their corresponding references in Column 10.
Finally, we note which rest-UV and optical N lines are detected at a $2\sigma$ significance in each object in Columns 11, 12, and 13. \\
{\bf References:}
(1) \citet{berg22}, \citet{mingozzi22}, \citet{xu22};
(2) \citet{skillman13}, \citet{mcquinn15}, \citet{telford23}; 
(3) \citet{berg19}
(4) \citet{topping24};
(5) \citet{tacchella23}, \citet{bunker23}, \citet{maiolino24}, \citet{senchyna24}, \citet{cameron23}, \citet{alvarezmarquez25};
(6) \citet{vanzell10}, \citet{grazian20}, \citet{barchiesi23}, \citet{ubler23}, \citet{ji24};
(7) \citet{vanzella22}, \citet{mestric23}, \citet{pascale23}, \citet{riverathorsen24}, \citet{welch25};
(8) \citet{rigby21}, \citet{welch24}; 
(9) \citet{marqueschaves24};
(10) \citet{berg18};
(11) \citet{fosbury03}, \citet{villarmartin04}, \citet{sanders20};
(12) \citet{topping25a};
}
\end{deluxetable*}

\begin{deluxetable}{lcccc}
\tabletypesize{\footnotesize}
\caption{Electron Temperature and Density Diagnostics\label{tab:avail_diag}}
\tablewidth{\textwidth}
\tablehead{
\CH{} & \multicolumn{4}{c}{Available High-Ion. Diagnostics} \\
\cline{2-5}
\CH{}     & \CH{$T_{e,high}$} & \CH{$n_{e,int.}$} & \CH{$n_{e,high}$} & \CH{$n_{e,high}$} \\ [-2ex]
\CH{Name} & \CH{O$^{+2}$}       & \CH{C$^{+2}$}        & \CH{Ar$^{+3}$}        & \CH{N$^{+3}$}        
}
\startdata
\multicolumn{5}{l}{\bf Low-$z$ Sample:}    \\
J0127-0619    & \cmark & \cmark & \xmark & \xmark \\
J1314+3452    & \cmark & \cmark & \cmark & \xmark \\
J1444+4237    & \cmark & \cmark & \xmark & \xmark \\
Leo P         & \cmark & \cmark & \cmark & \xmark \\
J1253-0312    & \cmark & \cmark & \cmark & \cmark \\
J1044+0353    & \cmark & \cmark & \cmark & \cmark \\
J1323-0132    & \cmark & \cmark & \cmark & \cmark \\
J1545+0858    & \cmark & \cmark & \cmark & \cmark \\
[2ex]
\multicolumn{5}{l}{\bf High-$z$ Sample:} \\
RXCJ2248-ID   & \cmark & \cmark & \xmark & \cmark \\ 
GN-z11        & \cmark & \xmark & \xmark & \cmark \\ 
GDS 3073      & \cmark & \xmark & \xmark & \cmark \\
Sunburst      & \cmark & \cmark & \xmark & \xmark \\ 
J1723+3411    & \cmark & \cmark & \xmark & \xmark \\
CEERS 1019    & \cmark & \xmark & \xmark & \cmark \\
SL2S J0217    & \cmark & \cmark & \xmark & \xmark \\
Lynx          & \cmark & \cmark & \xmark & \cmark \\
A1703-zd6     & \cmark & \cmark & \xmark & \cmark 
\enddata
\tablecomments{
The high-ionization, high-excitation diagnostics available for each object in our sample are denoted by a checkmark if their emission line ratios are detected (SNR $> 2$). 
Each objects has both the \Te~diagnostic and one of the \den~diagnostics, enabling a calculation of $T_{e,high}$ with a \den~from the a similar ionization zone.}
\end{deluxetable}

\section{Emission Line Measurements} \label{sec:emlinemeasure}
We collated emission line measurements from the literature when available and measured emission line fluxes from archival spectra in all other cases.
While we were able to collect all of the necessary emission line measurements from the literature for the High-$z$ Sample, the Low-$z$ Sample emission line measurements are mixed.
For the Low-$z$ Sample, all necessary optical emission lines have been previously reported, but few rest-UV N fluxes exist.
In curating this sample, we find three CLASSY galaxies without prior N III] emission line measurements in the literature.
Therefore to maintain consistency in the UV emission line measurements of our Low-$z$ Sample, we fit all the UV emission lines for the eight objects in the Low-$z$ Sample.
We find that when compared to \cite{mingozzi22}, our N III], N IV], and O III] emission lines measurements agree within 3$\sigma$.
In the case of the High-$z$ Sample, for RXCJ2248-ID, GDS 3073, J1723+3411, CEERS 1019, SL2S J0217, Lynx, and A1703-zd6, we use the flux measurements presented in \cite{topping24}, \cite{ji24}, \cite{rigby21}, \cite{marqueschaves24}, \cite{berg18}, \cite{fosbury03}, and \cite{topping25a}, respectively.
For GN-z11, all fluxes are from \cite{bunker23} except [\OIII~\W5008, which we scaled from \cite{alvarezmarquez25} to match the other fluxes.
Additionally, we use the \NIV~\W1483/\W1487 ratio for GN-z11 from \cite{maiolino24} in order to decompose the combined flux \NIV~\W\W1483,1487 reported in \cite{bunker23}.
Finally, for the Sunburst Arc, we obtain the UV flux measurements from \cite{pascale23} and the optical fluxes from \cite{riverathorsen24}, which agrees within uncertainties with the measurements presented by \cite{welch25}.

Emission-lines were fit using the \texttt{LMfit} package in \texttt{python}.
Each line was fit simultaneously using a Gaussian plus linear continuum and constrained by a single full width at half maximum (FWHM) velocity width.
The resolution of our spectra enables us to fit the \NIV~emission as a doublet, whereas the \NIII~emission is often blended and weak in the individual lines and so best fit with a single Gaussian whose width is allowed to vary.
There are two exceptions: J0127-0619 and Leo P.
For J0127-0619, we fit three individual lines in the \NIII~\W1750 complex and for Leo P, we fit the two strongest lines.
The uncertainties on the line fluxes are the estimated standard error derived from the least-squares minimization in \texttt{LMfit}, which considers the uncertainty on the Gaussian profile and linear continuum.

We present the flux measurements for each of the 9 objects in our Low-$z$ Sample in Table~\ref{tab:Lz_ELs} in the Appendix.
To be reported, we require at least one \NIV~line detection with signal to noise (SN) greater than 2 and the integrated \NIII~line flux to have SN$>2$.
Figure \ref{fig:spectra} shows the UV N profiles for the Low-$z$ Sample and the emission line fits (red) for the eight galaxies without previous N lines reported.

Dereddened emission line intensities ($I_\lambda$/$I_{H\beta}$\footnote{We present UV emission line measurements for our Low$-z$ Sample relative to $I_{\text{\ion{C}{3}]}}$ to avoid introducing further uncertainties due to flux calibrations between the UV and optical spectra.}) are determined using the nebular $E(B-V)$ values in the literature for each object in our sample.
We note that all the $E(B-V)$ used in this work can be found in Tables \ref{tab:Lz_ELs} and \ref{tab:Hz_ELs} in the Appendix.
For objects that lack published $E(B-V)$ values, we determine $E(B-V)$ using the published H$\alpha$, H$\beta$, and H$\gamma$ fluxes with the theoretical Balmer ratio assuming case B, $T_e = 1.5\times10^4$ K, $n_e = 10^2$ cm$^{-3}$, and the parameterized extinction law of \cite{cardelli89}.
We note that for three of these objects the reddening derived from the Balmer series emission returns a negative $E(B-V)$.
In these cases, we assume that there is no reddening from the host galaxy, but recent works have explored scenarios where the Balmer ratios would deviate from case B recombination \citep[e.g.][]{scarlata24, yanagisawa24, mcclymont25, arellanocordova25a}.
The $E(B-V)$ values are then used to deredden the rest-optical emission lines assuming the \cite{cardelli89} extinction law and the rest-UV emission lines assuming the \cite{reddy16} extinction law.
All line intensities used for this work, including those adapted from literature sources are reported in Tables \ref{tab:Lz_ELs} and \ref{tab:Hz_ELs} for the \lz~and \hz~samples, respectively, in the Appendix.

\section{Photoionization Models} \label{sec:models}
In order to properly evaluate the effects of density structure on derived nebular properties, we computed new \texttt{Cloudy} 23.01 photoionization models \citep{chatzikos23, gunasekera23} over a large range of densities.
We assumed a spherical geometry and used ``Binary Population and Spectral Synthesis" \citep[\texttt{BPASS}v2.14]{eldridge17} burst models combined with an ionization parameter to define the input ionizing radiation field.
For these models we intentionally covered an extended parameter space appropriate to the conditions of our sample galaxies observed across cosmic time. 
This included an age range of $10^{6.0}-10^{7.6}$ yr to probe the young bursts likely associated with the very-high-ionization emission observed, a range of stellar metallicities ($Z_\star = Z_{\rm neb} = 0.001, 0.002, 0.003, 0.004, 0.006, 0.008 = 0.05, 0.10, 0.15, 0.20, 0.30, 0.40 Z_\odot$) that cover the observed gas-phase abundances of our sample and the expected values for early galaxies, an expanded range of N/O and C/O abundances (described below), and a large range of uniform gas densities ($10^2 < n_e < 10^9$) to explore the effects of high density (i.e., pressure). 
Note that great care must be taken to consider the sensitivity range of observed density diagnostics in comparison to photoionization models.

We also explored a large range in ionization parameter of $-3.5<\log U<-1.0$.
\texttt{Cloudy} defines the dimensionless ionization parameter as
\begin{equation}
    U\equiv \frac{Q({\rm H})}{4\pi r_o^2 n_{\rm H} c},
\end{equation}
where $Q({\rm H})$ is the number of H-ionizing photons at the illuminated face of the gas cloud, $r_o$ is the distance from the ionizing source to the cloud face, $n_{\rm H}$ is the total H density at the cloud face, and $c$ denotes the speed of light.
Note that in a fully ionized nebula, $n_{\rm H}$ will be approximately equivalent to the electron density, \den, but not exactly because He contributes a non-negligible fraction of electrons through photoionization.
The density dependence suggests that the ionization parameters inferred for high$-z$ galaxies will be significantly reduced when large gas densities are present \citep[e.g.,][]{reddy23}. 
Overall, $\log U$ is an important parameter that gives insight into the strength of the radiation field and is critical for understanding the physical conditions in the ISM.

The GASS10 solar abundance ratios within \texttt{Cloudy} were used to initialize the relative gas-phase abundances. 
These abundances were then scaled to cover the observed range in nebular metallicity and relative C/O and N/O abundances.
The expanded ranges in relative N/O and C/O were motivated to test the validity of the extremely large range of values reported at high$-z$ from JWST observations \citep[e.g.,][]{cameron23, pascale23, ji24, marqueschaves24}\footnote{We note that UV emission lines are extremely sensitive to \Te~and some current measurements lack a robust \Te~detection, which could severely bias these derived values.}, resulting in coverage of $0.05 \lesssim$ (C/O)/(C/O)$_\odot \lesssim 3$, corresponding to a range of $-1.75 <$ log(C/O) $<+0.25$, and $0.25 \lesssim$ (N/O)/(N/O)$_\odot \lesssim 75$, corresponding to a range of $-1.5 <$ log(N/O) $<+1.0$.

\begin{figure*}[ht]
    \begin{center}%
    \includegraphics[width=0.85\textwidth]{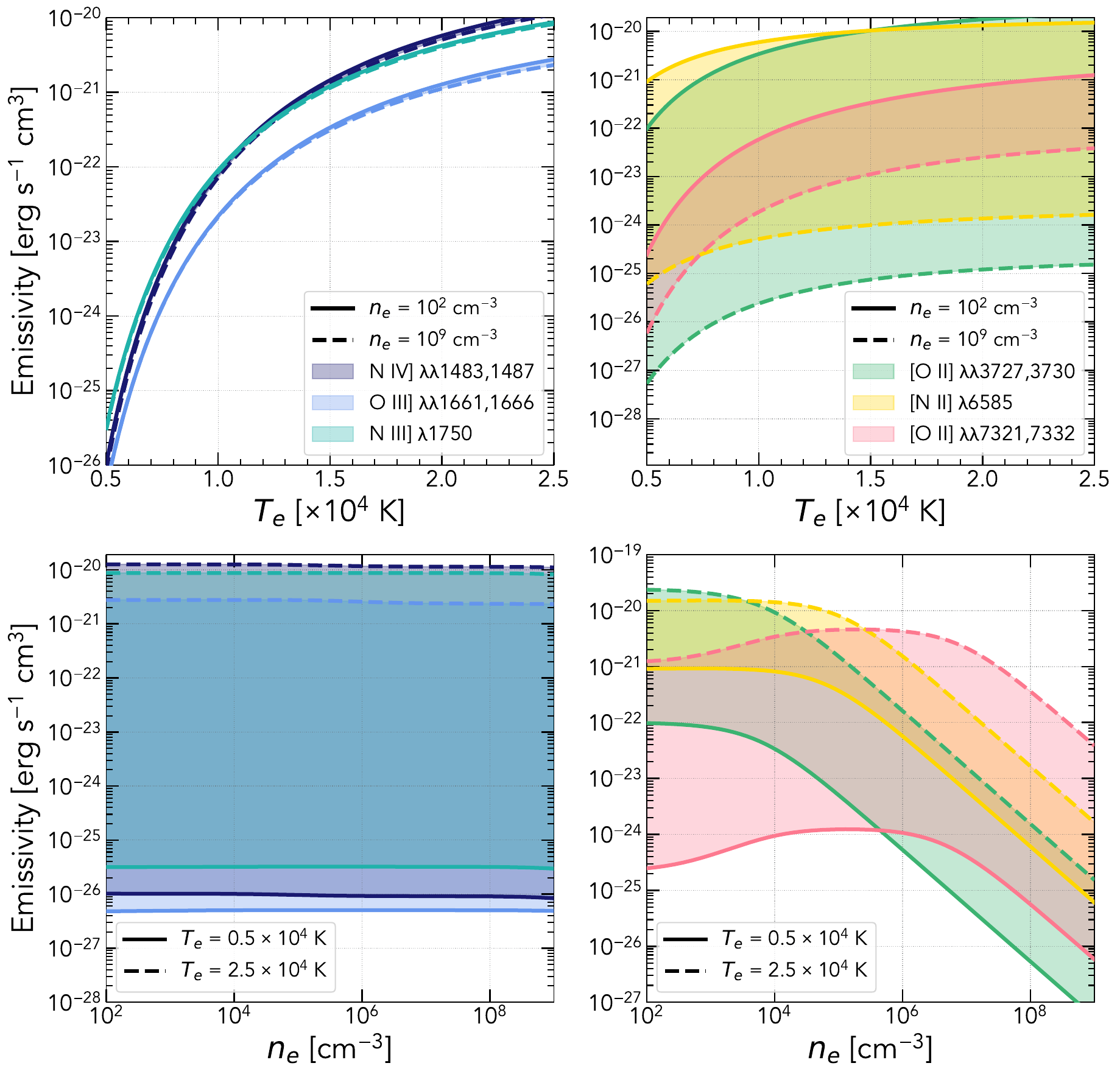}
    \caption{The emissivities of \NIV~\W\W1483,1487, \NIII~\W1750, and \OIII~\W\W1661,1666, determined with \texttt{PyNeb}, are shown in the left column, while the right column shows the same for the emissivities of \OII~\W\W3727,3730, \NII~\W6584, and \OII~\W\W7321,7332.
    The upper panels show emissivities as a function of \Te~(5,000 to 25,000 K), and the lower panels describe these emissivities as a function of \den~(10$^2$ - 10$^9$ cm$^{-3}$).
    Note that we show the full range of densities used in photoionization models to show the different ranges each line is sensitive over.
    The top panels illustrate that \den~has a larger impact on optical lines that have lower critical \den~, while the bottom panels show that UV lines are more impacted by \Te~than the optical lines given that they have higher excitation energies.}
    \label{fig:physcond_sensitivity}
    \end{center}
\end{figure*}

\section{Low- to High- Density Models} \label{sec:const_den}
N and O are two of the most important elements for understanding chemical evolution.
O is predominantly synthesized in short-lived, massive stars \citep[MSs; $M_\star \gtrsim 10\;M_{\odot}$, e.g.,][]{kobayashi20} and ejected into the ISM via core-collapse supernovae (CCSNe) on timescales of $\lesssim 10$ Myr (the peak of CCSNe).
It is an important element for cooling processes and dust formation, and is used as a proxy for the gas-phase metallicity of a galaxy ($Z=$ 12+log(O/H)).
Since O/H abundance increases with each new generation of stars \citep{wheeler89}, it serves as a robust tracer of the cumulative SFH and the current evolutionary state of a galaxy.
N, in contrast, is generated in both MSs and intermediate mass stars \citep[IMSs; $10\;M_{\odot} \gtrsim M_\star \gtrsim 2\;M_{\odot}$, e.g.,][]{kobayashi20}, and, therefore, produced on longer time scales ($> 200$ Myr).
As a result, N/O is sensitive to the relative contribution of high-mass versus intermediate-mass stars to a galaxy's chemical enrichment.

We use the new photoionization model grid described in Section \ref{sec:models} to explore the effects of increasing density on measured nebular conditions and abundances in the simple case of a constant, uniform density model. 
Specifically, we explore the effect on diagnostic emission line ratios (\S~\ref{sec:emline_varyden}), derived electron temperatures (\S~\ref{sec:tem_varyden}), direct-method abundances (\S~\ref{sec:abund_varyden} - \ref{sec:NO_varyden}), ionization parameters (\S~\ref{sec:ICF_varyden}), and ionization correction factors (\S~\ref{sec:ICF_varyden}).

\begin{deluxetable}{lccccc}[ht]
\tabletypesize{\footnotesize}
\caption{Emission Line Characteristics\label{tab:em_characteristics}}
\tablehead{
\CH{}          & \CH{$\lambda_{vac}$} & Transition & \CH{I.P.}   & \CH{$E_{exc}$} & \CH{$n_{e, crit}$} \\ [-1.0ex]
\CH{Line}      & \CH{(\AA)}   & (\tran)    & \CH{(eV)} & \CH{(eV)} & \CH{(cm$^{-3}$)}
}
\startdata
\multicolumn{6}{c}{\bf UV Emission Lines} \\
\hline
\ion{N}{4}]    & 1483.321 & 4 \rarr~1 & 47.445 & 8.359 & 1.83$\times 10^5$    \\
\ion{N}{4}]    & 1486.496 & 3 \rarr~1 & 47.445 & 8.341 & 5.62$\times 10^9$    \\
\ion{O}{3}]    & 1660.809 & 6 \rarr~2 & 35.121 & 7.479 & 4.87$\times 10^{10}$ \\
\ion{O}{3}]    & 1666.150 & 6 \rarr~3 & 35.121 & 7.479 & 4.87$\times 10^{10}$ \\
\ion{N}{3}]    & 1746.823 & 4 \rarr~1 & 29.601 & 7.098 & 1.16$\times 10^9$    \\
\ion{N}{3}]    & 1748.646 & 3 \rarr~1 & 29.601 & 7.090 & 1.12$\times 10^{10}$ \\
\ion{N}{3}]    & 1749.674 & 5 \rarr~2 & 29.601 & 7.108 & 6.84$\times 10^9$    \\
\ion{N}{3}]    & 1752.160 & 4 \rarr~2 & 29.601 & 7.098 & 1.16$\times 10^9$    \\
\ion{N}{3}]    & 1753.995 & 3 \rarr~2 & 29.601 & 7.090 & 1.12$\times 10^{10}$ \\
\ion{C}{3}]    & 1906.683 & 4 \rarr~1 & 24.384 & 6.503 & 8.59$\times 10^4$    \\
{[\ion{C}{3}]} & 1908.734 & 3 \rarr~1 & 24.384 & 6.496 & 1.14$\times 10^9$    \\
{[\ion{O}{3}]} & 2321.663 & 5 \rarr~2 & 35.121 & 5.354 & 3.12$\times 10^7$    \\
\hline
\multicolumn{6}{c}{\bf Optical Emission Lines} \\
\hline
{[\ion{O}{2}]} & 3727.092 & 3 \rarr~1 & 13.618 & 3.327 & 4.77$\times 10^3$    \\
{[\ion{O}{2}]} & 3729.875 & 2 \rarr~1 & 13.618 & 3.324 & 1.38$\times 10^3$    \\
{[\ion{O}{3}]} & 4364.435 & 5 \rarr~4 & 35.121 & 5.354 & 2.88$\times 10^7$    \\
{[\ion{Ar}{4}]}& 4712.579 & 3 \rarr~1 & 40.735 & 2.613 & 1.73$\times 10^4$    \\
{[\ion{Ar}{4}]}& 4741.448 & 2 \rarr~1 & 40.735 & 2.615 & 1.56$\times 10^5$    \\
{[\ion{O}{3}]} & 5008.239 & 4 \rarr~3 & 35.121 & 2.514 & 7.83$\times 10^5$    \\
{[\ion{N}{2}]} & 6585.270 & 4 \rarr~3 & 14.534 & 1.899 & 1.04$\times 10^5$    \\
{[\ion{S}{2}]} & 6718.291 & 3 \rarr~1 & 10.360 & 1.845 & 1.92$\times 10^3$    \\
{[\ion{S}{2}]} & 6732.670 & 2 \rarr~1 & 10.360 & 1.842 & 5.07$\times 10^3$    \\
{[\ion{Ar}{3}]}& 7137.755 & 4 \rarr~1 & 27.630 & 1.737 & 5.11$\times 10^6$    \\
{[\ion{O}{2}]} & 7320.935 & 5 \rarr~2 & 13.618 & 5.018 & 4.70$\times 10^6$    \\
{[\ion{O}{2}]} & 7322.002 & 4 \rarr~2 & 13.618 & 5.017 & 6.88$\times 10^6$    \\
{[\ion{O}{2}]} & 7331.680 & 5 \rarr~3 & 13.618 & 5.018 & 4.70$\times 10^6$    \\
{[\ion{O}{2}]} & 7332.751 & 4 \rarr~3 & 13.618 & 5.017 & 6.88$\times 10^6$
\enddata
\tablecomments{To aid in the discussion provided in Section \ref{sec:emline_varyden}, we provide the  vacuum wavelengths, transitions, ionization potential energies, excitation energies, and critical densities for the emission lines relevant to this work.}
\end{deluxetable}

\subsection{Emission Line Diagnostics} \label{sec:emline_varyden}
Emission line ratios that are essential to determining nebular abundances in galaxies are sensitive to both electron temperature (\Te) and density (\den).
In particular, the collisionally-excited metal lines used in these ratios experience an exponential dependence on \Te, which is dictated by the Maxwell-Boltzmann distribution of electron kinetic energies ($\propto e^{-\Delta E/kT_e}$).
The critical density, $n_{e,crit}$, marks the point where the spontaneous emission and collisional de-excitation rates are equal \citep[e.g.,][]{osterbrock06, draine11}.
At low-densities ($n_e \ll n_{e,crit}$), the dominant decay mechanism is spontaneous emission; however, at higher densities ($n_e \gtrsim n_{e,crit}$) the contribution from collisional deexcitation will become increasingly important.
Note that while $n_{e,crit}$ marks the equivalence between deexcitation mechanisms, the contribution from collisional deexcitation becomes significant at much lower densities.
To facilitate the discussion in this section, we present Table \ref{tab:em_characteristics} which contains useful emission line characteristics for many of the most relevant emission lines in this work. 

In Figure \ref{fig:physcond_sensitivity} we use the \texttt{getEmissivity} function in \texttt{PyNeb} \citep{luridiana12, luridiana15} to plot the emissivities of the UV and optical O and N emission lines as a function of electron \Te~(upper panels) and \den~(lower panels). 
The upper left panel of Figure \ref{fig:physcond_sensitivity} shows that strong emission ($\gtrsim 10^{-21}$ erg s$^{-1}$ cm$^{-3}$) from the high-ionization UV N and O emission lines require high \Te~($T_e \gtrsim 1.5\times10^4$ K) due to their high-excitation energies ($\gtrsim 7.1$ eV).
In general, UV emission lines have higher excitation energies than optical emission lines, which are plotted with respect to \Te~in the upper right panel of Figure \ref{fig:physcond_sensitivity} \citep[e.g.,][]{osterbrock89, draine11}.
For example, the excitation energies of the most common O$^{+2}$ emission lines in increasing energy are [\OIII~\W5008, [\OIII~\W4364, [\OIII~\W2322, and \OIII~\W1666 ($E_{exc} = $ 2.51, 5.35, 5.35, 7.48 eV; for more see Table \ref{tab:em_characteristics}). 
Thus, given their higher excitation energies, strong detections of the UV O and N emission lines require high electron temperatures and/or high densities near or less than their respective critical densities ($n_{e, crit} = $ 7.8$\times 10^5$, 2.9$\times 10^7$, 3.1$\times 10^7$, 4.9$\times 10^{10}$ cm$^{-3}$; for more see Table \ref{tab:em_characteristics}).

The lower right panel of Figure \ref{fig:physcond_sensitivity} shows emissivities of the low-ionization optical O and N emission lines as a function of density. 
Note that while our density diagnostics are only sensitive up to $\sim$10$^7$ cm$^{-3}$ but we show larger range of densities over which emission lines are affected.
Emission from a given line begins to decrease as the density approaches its $n_{e, crit}$ and collisional de-excitation becomes dominant over spontaneous emission.
The [\ion{O}{2}] \W\W3727,3730 transitions have the lowest $n_{e,crit}$ of the lines plotted (see Table \ref{tab:em_characteristics}  for $n_{e,crit}$). 
Thus, for $n_e \gtrsim 10^3$ cm$^{-3}$, the [\ion{O}{2}] \W\W3727,3730 emission significantly decreases \citep{juandedios21, mendezdelgado23}.
On the other hand, the $n_{e,crit}$ of \NII~\W6585 is higher (see Table \ref{tab:em_characteristics}) such that the observed \NII~\W6585/[\ion{O}{2}] \W\W3728,3730 ratio will increase for $10^3 \lesssim n_e \lesssim 10^5$ cm$^{-3}$.
At higher densities ($n_e \gtrsim 10^5$ cm$^{-3}$), the \NII~\W6585/\OII~ \W\W3728,3730 ratio levels out as both the N and O emission decrease at similar rates due to collisional de-excitation.
The \NII~\W6585/\OII~\W\W7321,7332 will decrease slightly at $n_e < 10^7$ cm$^{-3}$, but at even higher densities, $n_e \gtrsim 10^7$ cm$^{-3}$, the ratio will level out due to the \OII~\W\W7321,7332 emission lines reaching their $n_{e,crit}$ and becoming weaker (see Table \ref{tab:em_characteristics} for $n_{e,crit}$).

The lower left panel of Figure \ref{fig:physcond_sensitivity} shows similar density sensitivity trends for the emission of the UV N and O lines, but with the turnover to lower emissivities occurring at higher densities (\den~$< 10^6$).
This is because in addition to higher excitation energies, the UV emission lines (\NIV~\W\W1483,1487, \OIII~\W\W1661,1666, and \NIII~\W1750) also have higher $n_{e,crit}$.
The \NIV~\W\W1483,1487 lines have $n_{e,crit}$ of $1.8 \times 10^5$ and $5.6 \times 10^9$ cm$^{-3}$, respectively, while the lines in the \OIII~\W\W1661,1666 doublet both have an $n_{e,crit}$ of $4.9 \times 10^{10}$  cm$^{-3}$.
As a result, the \NIV~\W\W1483,1487/\OIII~\W\W1661,1666 ratio will be relatively robust to varied \den~\citep[e.g.,][]{zhang25}.
Each of the lines in the \NIII~\W1750 quintuplet has $n_{e, crit} \approx 10^9 - 10^{10}$  cm$^{-3}$, which implies little change in the \NIII~\W1750/\OIII~\W\W1661,1666 across a large range in \den.

In summary, the relatively low-critical densities of optical lines will lead them to collisionally de-excite in extreme environments like those observed at \hz.
This highlights the importance of robust physical condition measurements for interpreting emission line strengths, which in turn has significant implications for diagnostic emission line ratios.
However, for more details on the physics of nebular emission, see \cite{osterbrock06} and \cite{draine11}.

\begin{figure}[ht]
    \begin{center}
    \includegraphics[width=0.45\textwidth]{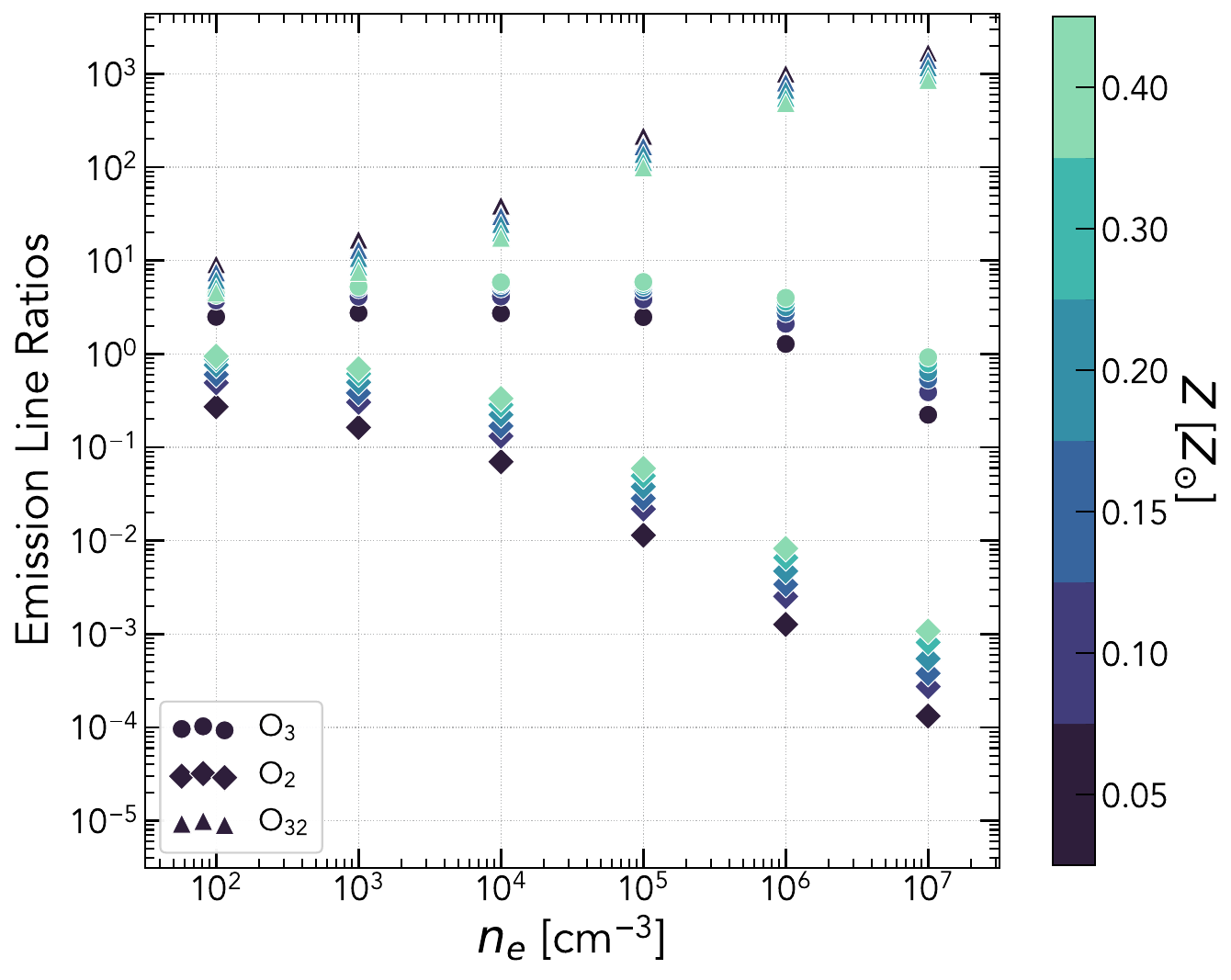}
    \caption{The modeled emission line ratios, O$_3$, O$_2$, and O$_{32}$, from \texttt{Cloudy} across \den~$= 10^2 - 10^7$ cm$^{-3}$, the range over which we expect these diagnostics to be sensitive.
    The circles show O$_{3}$, the diamonds are O$_{2}$, and the triangles describe O$_{32}$.
    The colors show different metallicities ranging from 0.05 $Z_{\odot}$ and 0.40 $Z_{\odot}$ given in the color bar on the right.
    O$_3$ and O$_2$ begin to change rapidly after the $n_{e,crit}$ of [\OIII~and \OII, respectively, and, as a result, the O$_{32}$ ratio changes drastically at \den~$> 10^3$ cm$^{-3}$.}
    \label{fig:ratios_varyden}
    \end{center}
\end{figure}

We explore several common diagnostic emission line ratios using the suite of \texttt{cloudy} photoionization models from Section \ref{sec:models}.
To match the properties of our sample ($<0.5\ Z_{\odot}$), we focus on the models that correspond to $0.05-0.40\ Z_{\odot}$. 
We explore the effects of the \den, \Te, and ionization structure on the observed emission lines and how they influence the derived nebular properties.
Hereafter, we adopt the following emission line ratio naming convention:
\begin{enumerate}[itemsep=-0.5ex,partopsep=0.5pt]
    \item O$_{3}$: [\OIII~\W5008/H$\beta$\W4863
    \item O$_{2}$: \OII~\W\W3727,3730/H$\beta$\W4863
    \item O$_{32}$: [\OIII~\W5008/\OII~\W\W3727,3730
    \item N$_{43}$: \NIV~\W\W1483,1487/\NIII~\W1750
    \item Ar$_{43}$: \ArIV~\W\W4713,4742/\ArIII~\W7138
    \item S$_{32}$: [\ion{S}{3}] \W\W9069,9532/[\ion{S}{2}] \W\W6717,6731
    \item N2O2: \NII~\W6585/\OII~\W\W3727,3730
    \item N3O3: \NIII~\W1750/\OIII~\W\W1661,1666
    \item N4O3: \NIV~\W\W1483,1487/\OIII~\W\W1661,1666
\end{enumerate}

In Figure \ref{fig:ratios_varyden}, we examine O$_{3}$, O$_{2}$, and O$_{32}$ as a function of density.
For this exercise, we use $10^2 \leq n_e \leq 10^7$ cm$^{-3}$, which is the range over which the diagnostics we examine are sensitive.
The O$_{3}$ and O$_{2}$ emission line ratios are relatively insensitive to \den~at low-\den, but begin to change rapidly as they approach their respective $n_{e,crit}$ values (see Table \ref{tab:em_characteristics} for $n_{e,crit}$).
More specifically, O$_{3}$ is robust against variations for \den~$\lesssim 10^5$  cm$^{-3}$, while O$_{2}$ and O$_{32}$ begin to be severely impacted by densities of \den~$\gtrsim 10^3$ cm$^{-3}$.
In the subsequent sections, we examine the effects of increasing density on the derivation of physical gas conditions from these emission lines.

O$_{3}$ and O$_{2}$ are commonly used tracers of the gas-phase metallicity.
Individually, these ratios trace the ionic O abundances in the high- and low-ionization zones, which are then added together to derive a total O abundance.
In the absence of temperature-sensitive line ratios, strong-line calibrations utilizing the sum of O$_{3}$ and O$_{2}$ are used to determine gas-phase metallicity.
Therefore, both direct-method and strong-line calibrations will be impacted by notable effects on these line ratios due to high-\den.\footnote{Note that strong-line calibrations are statistical with individual measurements varying as much as \p0.4 dex between different calibrations \citep{kewley08}; therefore, applying these calibrations to small samples may lead to severe errors.}
Significant efforts have been made to calibrate strong-line methods at higher-redshifts using samples with low-to-moderate \den~in the low-ionization gas \citep[e.g.,][]{bian18,sanders23,scholte25}.
However, such studies lack density measurements in the high-ionization gas and so cannot account for collisional de-excitation effects in high-ionization emission lines. 
O$_{32}$ is also often used as an observational tracer for $\log U$, which was introduced in Section \ref{sec:models}, and can be used to parameterize the ionization correction factors \citep[e.g.][]{berg16}.
Therefore, our ability to accurately account for unseen ionic species will also be impacted by the effects of high-\den\ ($n_e > 10^3 {\rm\ cm}^{-3}$) on O$_{32}$ (see Figure~\ref{fig:ratios_varyden}).

\subsection{Electron Temperatures and Densities} \label{sec:tem_varyden}
Accurate \Te~and \den~determinations across their relevant ionization zones are not only important to understand the gas conditions within galaxies, but are also critical for determining the metallicities and relative abundances that shape many galaxy evolution trends.
A three-zone ionization model is commonly used to characterize ionized nebular gas, where the low-ionization zone is typically defined by the ionization potential range associated with N$^+$ (14.5$-$29.6 eV), the intermediate-ionization zone defined by the range of S$^{+2}$ (23.3$-$34.8 eV), and the high-ionization zone traced by the range of O$^{+2}$ (35.1$-$54.9 eV). 
In this ionization zone model, each zone has a characteristic electron temperature and density such that $T_{e, low}$, $T_{e, int}$, and $T_{e, high}$ are used to describe the gas temperatures of nebula and $n_{e, low}$, $n_{e, int}$, and $n_{e, high}$ detail the densities of the gas.
When temperature is measured for only a single ionization zone, empirical or photoionization model based temperature-temperature relationships can then be used to estimate the temperatures of the other ionization zones \citep[e.g.,][]{garnett92}.
Note, however, that similar relationships do not exist for connecting the densities.

\begin{figure}[ht]
    \begin{center}
    \includegraphics[width=0.45\textwidth]{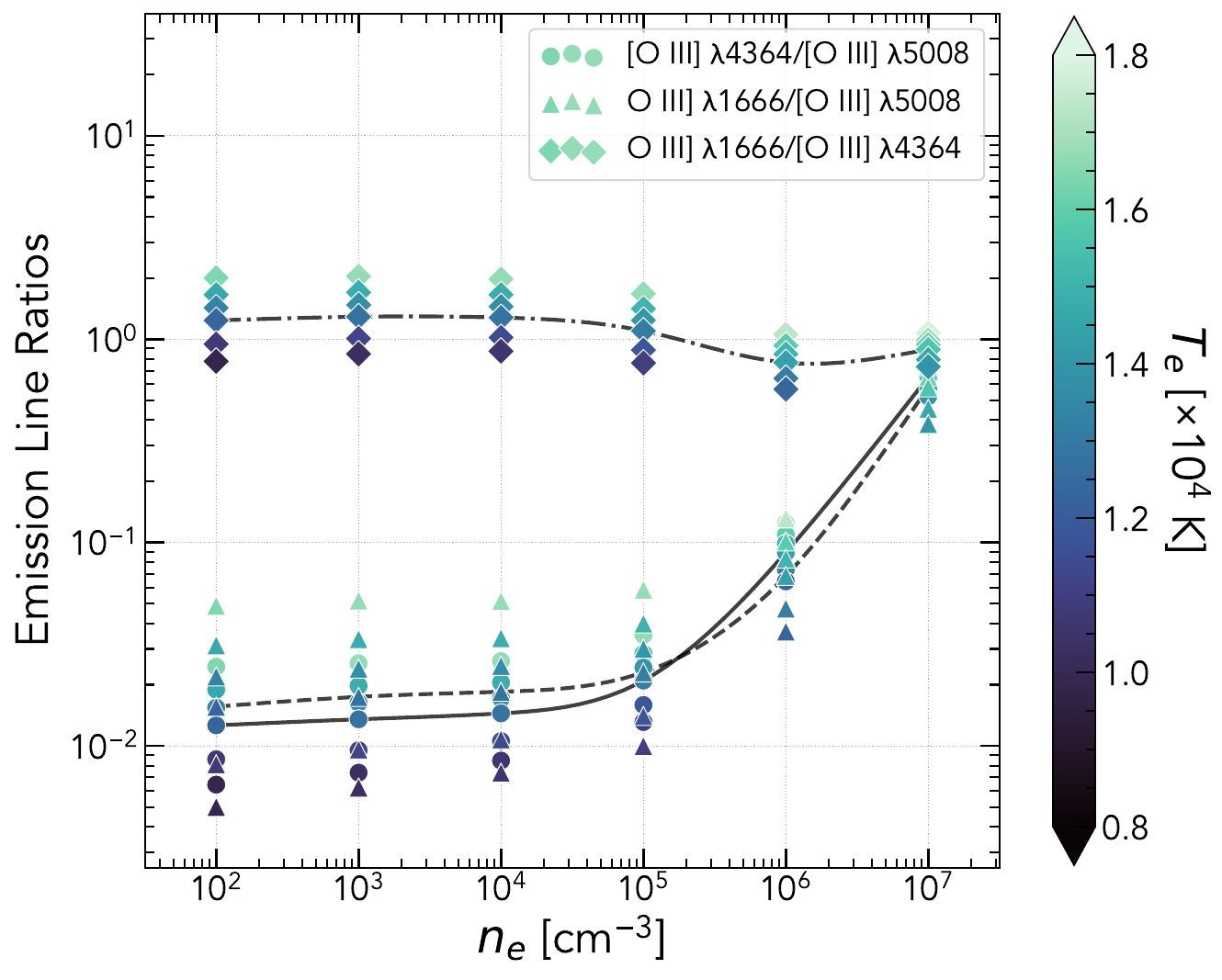}
    \caption{\texttt{Cloudy} models of \Te-sensitive [\OIII~emission line ratios versus a range of \den~$= 10^2 - 10^7$ cm$^{-3}$.
    The circle markers are the [\OIII~\W4364/[\OIII~\W5008 ratio, the triangles are \OIII~\W1666/[\OIII~\W5008, and the diamonds are \OIII~\W1666/[\OIII~\W4364.
    The color bar on the right indicates the \Te~in each model.
    The \OIII~\W1666/[\OIII~\W4364 line ratio is nearly constant over this density range due to the high critical densities of both lines.
    However, the two ratios involving [\OIII~\W5008 both change significantly above \den~$\gtrsim10^5$ cm$^{-3}$ due to its lower critical density.}
    \label{fig:o3ratios_varyden}
    \end{center}
\end{figure}

Figure \ref{fig:o3ratios_varyden} shows the effects of increasing \den~on three different $T_{e, high}$-sensitive emission line ratios ([\OIII~\W4364/[\OIII~\W5008, \OIII~\W1666/[\OIII~\W5008, and \OIII~\W1666/[\OIII~\W4364) for a set of \texttt{Cloudy} models.
The $T_{e, high}$-sensitive line ratios involving [\OIII~\W5008 begin to increase very quickly for \den~$\gtrsim 10^5$ cm$^{-3}$, near the critical \den~of [\OIII~\W5008 (see Table \ref{tab:em_characteristics} for $n_{e,crit}$).
Even the most common probe of \Te, [\OIII~\W4364/[\OIII~\W5008, which has been used in studies of \hz~galaxies with JWST, is known to be very sensitive to the \den~of the nebular gas at extreme ionized gas \den~\citep[e.g.,][]{draine11,katz23}.
Figure \ref{fig:o3ratios_varyden} shows that [\OIII~\W4364/[\OIII~\W5008 and \OIII~\W1666/[\OIII~\W5008 are robust for \den~$\lesssim 10^5$ cm$^{-3}$, but will be biased high at higher densities.
Alternatively, \OIII~\W1666/[\OIII~\W4364 is relatively constant up to den~$\lesssim 10^7$ cm$^{-3}$ making it an overall more robust tracer of the \Te~(although it has a significant dependence on reddening).

\subsection{Direct Abundances} \label{sec:abund_varyden}
Despite the importance of both N and O, significant challenges remain in quantifying robust O/H and N/O abundances across cosmic time. 
These challenges stem primarily from (1) the lack of uniform O/H and N/O abundance measurement methods in the literature, especially at high redshift, and (2) large uncertainties associated with the calculations of important values, including attenuation, electron temperature and density, and ionization correction factors.
For instance, no UV N/O abundance studies in galaxies have characterized the \Te, \den, $\log U$, {\it and} ionization correction factors in each ionization zone, which leads to samples of measurements that are simply not analogous.
We, therefore, present a uniform method for determining direct O/H and N/O abundances, including the first ionization correction factors for the UV N/O abundance.

Relative ionic abundances are determined as 
\begin{equation}
    \frac{N(X^i)}{N({\rm H}^+)} = \frac{I_{\lambda_i}}{I_{\rm H\beta}} \frac{\varepsilon_{\rm H\beta}}{\varepsilon_{\lambda_i}},
\end{equation}
where $\epsilon$ is the emissivity for ion $X^i$ at the appropriate \Te~and \den, which can be determined using the \texttt{getEmissivity} function in \texttt{PyNeb}.
For each ionic abundance, the appropriate physical conditions refers to the \Te~and \den~that trace each zone of the three-zone ionization framework.
Relative and total abundances are then determined by adding together the relevant ionic species and correcting for any unseen species using an ionization correction factor (ICF).
We describe the specifics of this process for O/H and N/O abundances below.

\begin{figure*}[ht]
    \begin{center}
    \includegraphics[width=\textwidth]{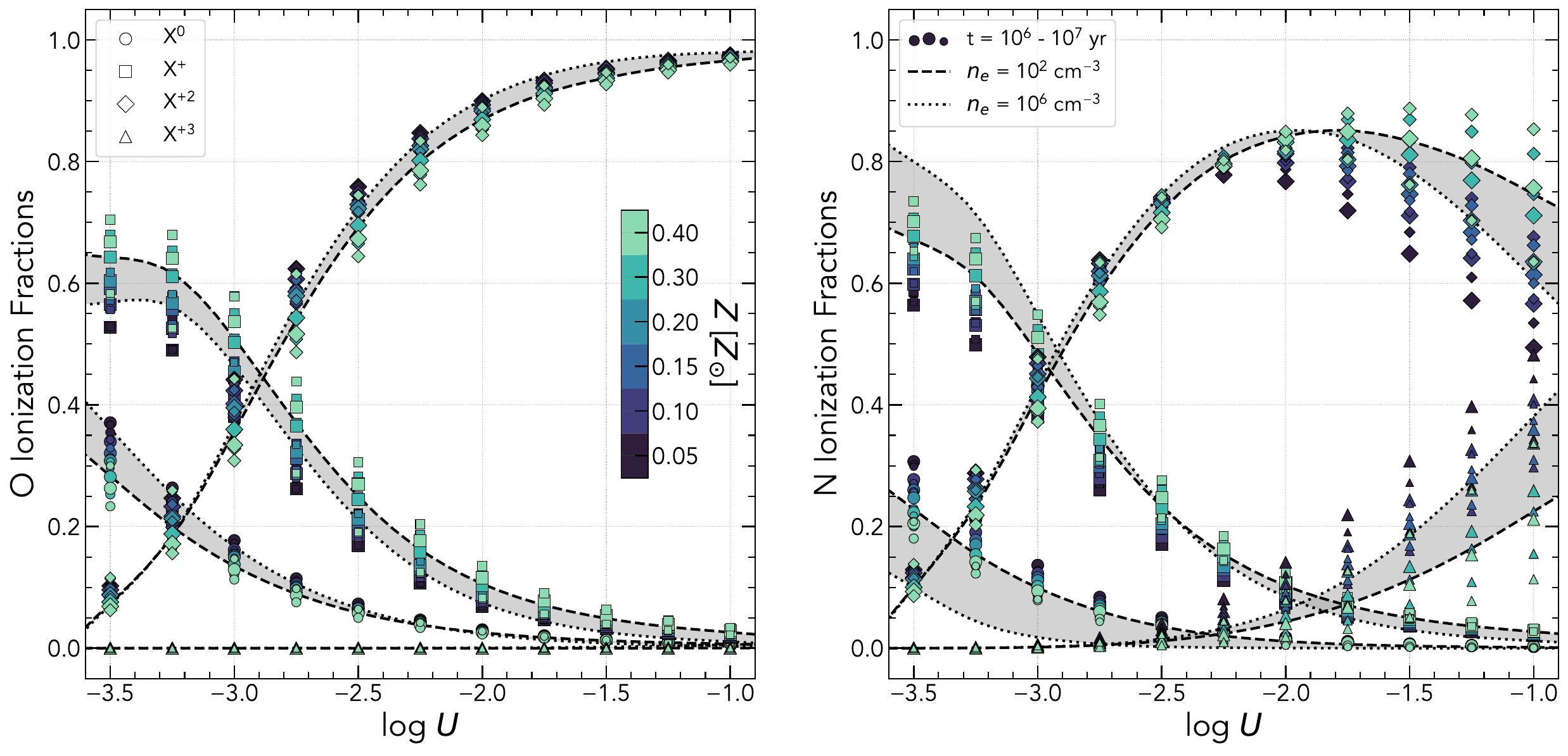}
    \caption{We show the ionization fractions as a function of $\log U$ for both O (left) and N (right).
    The different marker shapes correspond to the different ionization species, where circles are neutral, squares are singly ionized, diamonds are doubly ionized, and triangles are triply ionized species.
    The colors differentiate models of varying metallicity from 0.05 $Z_{\odot}$ - 0.40 $Z_{\odot}$.
    The marker size represents varying burst ages between 10$^{6.0}$ - 10$^{7.0}$ yr.
    Finally, the lines follow a single set of models at burst age of  10$^{6.7}$ and metallicity of 0.20 $Z_{\odot}$.
    The dashed line distinguishes the model at 10$^2$ cm$^{-3}$ from the model at 10$^6$ cm$^{-3}$, which is described by a dotted line.}
    \label{fig:ion_frac}
    \end{center}
\end{figure*}
\subsection{Oxygen Abundances} \label{sec:oxygen_varyden}
We derive O$^{+}$/H$^{+}$ and O$^{+2}$/H$^{+}$ using the appropriate \Te~and \den.
Total O/H abundances in a galaxy are determined by adding all the individual ionic species of O:
\begin{equation*}
    \rm \frac{O}{H} = \frac{O^{0}}{H^{+}} + \frac{O^{+}}{H^{+}} + \frac{O^{+2}}{H^{+}} + \frac{O^{+3}}{H^{+}}.
\end{equation*}
The relevant ionic fractions are determined as 
\begin{align*}
    \rm \frac{O^{+}}{H^{+}} &= \frac{I_{\lambda\lambda3728,3730}}{I_{\rm H\beta}} \times \frac{\epsilon_{\rm H\beta}}{\epsilon_{\lambda\lambda3728,3730}} \\
    \rm \frac{O^{+2}}{H^{+}} &= \frac{I_{\lambda5008}}{I_{\rm H\beta}} \times \frac{\epsilon_{\rm H\beta}}{\epsilon_{\lambda5008}} \\
    {\rm\ or\ } &= \frac{I_{\lambda1666}}{I_{\rm H\beta}} \times \frac{\epsilon_{\lambda1666}}{\epsilon_{\rm H\beta}},
\end{align*}
where O$^+$ is determined using $T_{e, low}$ and $n_{e, low}$ and O$^{+2}$ is determined using $T_{e, high}$ and $n_{e, high}$.

In the left panel of Figure \ref{fig:ion_frac} we plot the ionization fractions of O versus ionization parameter. 
Note that $\rm O^{+2}/H^{+}$ is the dominant species of O for most ionization parameters.
At $\log U > -3.0$, we find that the $\rm O^{0}/H^{+}$ contribution is small ($\lesssim 10$\%) and $\rm O^{+3}/H^{+}$ is negligible across all $\log U$.
This is consistent with the findings of \cite{berg21}, who directly measure [O IV] emission from the UV to conclude that $\rm O^{+3}$ only contributes about $1-2$\% to $\rm O_{tot}$ in high-ionization galaxies.
Additionally, recent work by \cite{rickardsvaught25} shows that the use of an ICF to derive a theoretical contribution of $\rm O^{+3}/H^{+}$ may be underestimating the contribution from this ion by a factor of 2, but the 2-zone and 3-zone O abundances agree within their uncertainties.
Therefore, the total O abundance is calculated from $\rm O^{+}/H^{+}$ and $\rm O^{+2}/H^{+}$.
Figure \ref{fig:ion_frac} also shows that the O ionization fractions are most sensitive to metallicity and burst age, and that each of the ionization fractions are relatively insensitive to \den, with the most change due to density at $\log U > -3.0$, in $\rm O^{+}/H^{+}$ and $\rm O^{+2}/H^{+}$.

In Section \ref{sec:tem_varyden}, we discussed the effects of high-\den~on the derived $T_{high}$, which is crucial to determining accurate total O abundances.
Since collisionally-excited line emission from O ions are the dominant cooling mechanism in the ISM at \den~$< 10^7$ cm$^{-3}$, \Te~and gas-phase metallicity are negatively correlated in this regime.
When assuming the low-\den~limit in gas with higher \den\ values, \Te~is biased high (see Figure~\ref{fig:o3ratios_varyden}), resulting in metallicity determinations that are artificially biased low. 
For example, \Te\ measurements of O$^{+}$ and S$^{+}$ have been found to be systematically higher than \Te($\rm N^{+}$) due to the much lower $n_{e,crit}$ of \OII~\W\W3727,3730 and [\ion{S}{2}] \W\W6717,6731 \citep{mendezdelgado23, rickardsvaught24}.
Therefore, it is pertinent to also exercise caution in using \Te($\rm O^{+}$) or \Te($\rm S^{+}$) as a representative \Te\ of the low-ionization zone.

\begin{figure*}[ht]
    \begin{center}
     \includegraphics[width=\textwidth]{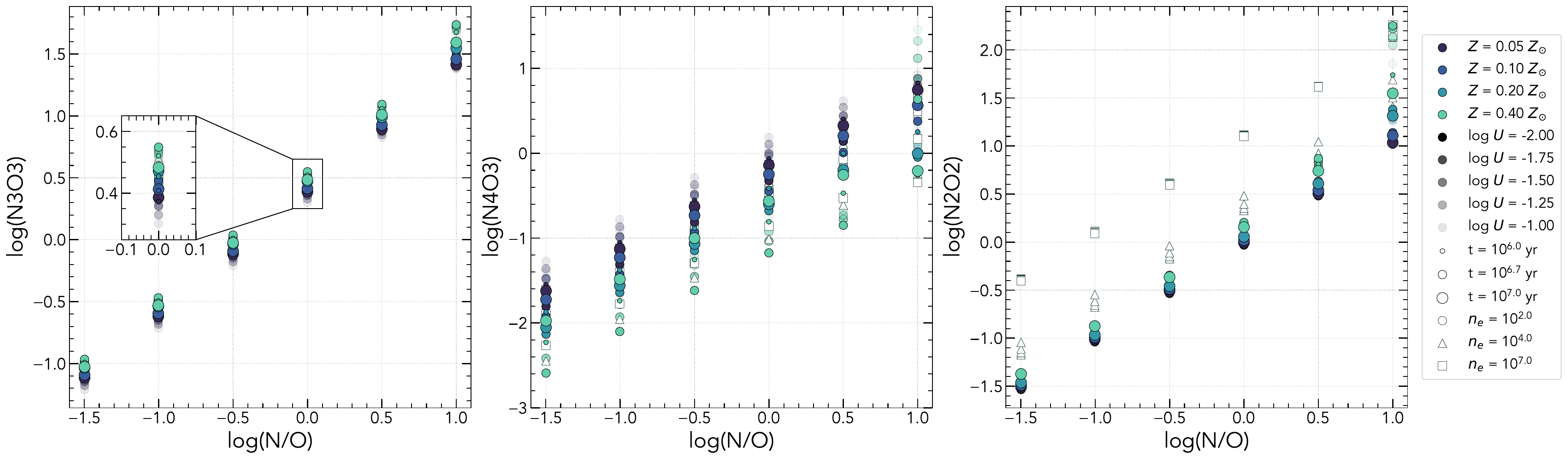}
    \caption{A grid of plots displaying different sets of \texttt{Cloudy} photoionization models at various relative N/O abundances.
    Every plot contains models that span metallicities from 0.05 Z$_{\odot}$ to 0.4 Z$_{\odot}$, which correspond to different color models going from darkest (lowest metallicity) to lightest (highest metallicity).
    Each panel has a different emission line ratio on the y-axis.
    From left to right, these are N3O3, N4O3, and N2O2.
    In each of the plots, circles correspond to models at \den~of 10$^{2.0}$, triangles are models at 10$^{4.0}$, and squares are models at 10$^{7.0}$.
    Additionally, the size of the circle markers corresponds to burst ages between 10$^{6.0}$ - 10$^{7.0}$.
    Finally, the spread in models  from $\log U$ of -2.0 to -1.0 is indicated by points of decreasing opacity.
    The N/O tracers each increase monotonically with relative N/O abundance and experience varying degrees of spread at a given N/O due to metallicity, ionization parameter, density, and burst age.}
    \label{fig:NO_cloudy_emlines}
    \end{center}
\end{figure*}
\subsection{Relative N/O Abundances} \label{sec:NO_varyden}
The N2O2, N3O3, and N4O3 emission line ratios can be used to probe the relative ionic abundance of N/O.
N/O abundances are most commonly derived from low-ionization optical emission using \NII~\W6585 and \OII~\W\W3727,3730, N2O2.
However, the increase in {\it JWST} detections of the UV \NIII~\W1750 and \NIV~\W\W1483,1487 lines has enabled N/O determinations from the N3O3 and N4O3 ratios.
We utilize \texttt{Cloudy} models to investigate the relationship between three different emission line probes of N/O (N2O2, N3O3, N4O3) and the total log(N/O) of the models.

The first panel in Figure \ref{fig:NO_cloudy_emlines} shows that there is a monotonic relationship between log(N/O) and the emission line ratio, N3O3.
However, the models show that there is a 0.24 - 0.36 dex spread in N3O3 for a given log(N/O) that is primarily driven by ionization parameter.
Other factors, such as burst age and density have little impact on the emission line ratio at a given log(N/O).

For N4O3, which is shown in the second panel of Figure \ref{fig:NO_cloudy_emlines}, there is a similar increasing relationship between the emission line ratio and log(N/O), but the 1.32 - 1.80 dex spread in models for N4O3 is much larger the spread in N3O3.
There is a bigger distinction between models of different metallicities for this emission line ratio.
At all metallicities shown, burst age, density, and ionization parameter also contribute significantly to the spread in models at varying log(N/O).

In the final panel of Figure \ref{fig:NO_cloudy_emlines}, the N2O2 for a given log(N/O) is severely impacted by high densities.
The $1.15-1.24$ dex spread in N2O2 at a given log(N/O) is driven primarily by high densities.
This variation is a result of [O II] being suppressed at $n_e \gtrsim 10^3$ \citep[e.g.,][]{rubin89, strasinska23}.
On the other hand, the N2O2 emission line ratio is nearly invariant to burst age, ionization parameter, and metallicity.

All three N/O emission line ratios are sensitive to log(N/O), however N4O3 and N2O2 are also fairly sensitive to metallicity, ionization parameter, density, and burst age.
Between these three emission line ratios, N3O3 is be the most invariant probe of log(N/O).
Given the range in sensitivity of the different N/O emission line ratios, it is important to examine the effects in observed N/O abundances.

The N/O ionic fractions are determined as
\begin{align*}
    \rm \frac{N^{+}}{O^{+}} &= \frac{I_{\lambda6585}}{I_{\lambda\lambda3727,3730}} \times \frac{\epsilon_{\lambda\lambda3727,3730}}{\epsilon_{\lambda6585}} \\
    {\rm\ or\ } &= \frac{I_{\lambda6585}}{I_{\lambda\lambda7321,7332}} \times \frac{\epsilon_{\lambda\lambda7321,7332}}{\epsilon_{\lambda6585}} \\
    \rm \frac{N^{+2}}{O^{+2}} &= \frac{I_{\lambda1750}}{I_{\lambda1666}} \times \frac{\epsilon_{\lambda1666}}{\epsilon_{\lambda1750}} \\
    \rm \frac{N^{+3}}{O^{+3}} &= \frac{I_{\lambda\lambda1483,1487}}{I_{\lambda1401}} \times \frac{\epsilon_{\lambda1401}}{\epsilon_{\lambda\lambda1483,1487}},
\end{align*}
where N$^+$ is determined using $T_{e, low}$ and $n_{e, low}$,  N$^{+2}$ is determined using $T_{e, int}$ and $n_{e, int}$, and N$^{+3}$ is determined using $T_{e, high}$ and $n_{e, high}$.
When calculating the \niiioiii~ratio for both the \lz~and \hz~samples, we consider the measured flux to be a blend of the \ion{N}{3}] \W1750 complex and therefore use the sum of all 5 emissivities in the quintuplet for our abundances calculations.

In general, the relative N/O abundance can be determined directly by adding all of the relevant species together as
\begin{equation*}
    \rm \frac{N}{O} = \frac{N^{+} + N^{+2} +  N^{+3}}{O^{+} + O^{+2} + O^{+3}}.
\end{equation*}
Since all three species of N and O are almost never observed in the same object, accurate estimates of N/O require an ICF to account for the unseen ionic species such that
\begin{equation*}
    \rm \frac{N}{O} = \frac{\sum_i N^{+i}}{\sum_j O^{+j}} \times \left[\frac{\sum_i X(N^{+i})}{\sum_j X(O^{+j})}\right]^{-1} = \frac{\sum_i N^{+i}}{\sum_j O^{+j}} \times N\;ICF.
\end{equation*}
$\rm \sum_i N^{+i}$ and $\rm \sum_j O^{+j}$ are the sum of the observed N and O ionic species, respectively, and ICF is composed of the ionization fractions of N and O, $\rm X(N^{+i})$ and $\rm X(O^{+j})$, respectively.
Therefore, we need a robust set of ICFs to enable N/O determinations for different combinations of observed N and O ionic species.


\subsubsection{N/O Ionization Correction Factors} \label{sec:ICF_varyden}
\begin{figure*}[ht]
    \begin{center}
     \includegraphics[width=0.95\textwidth]{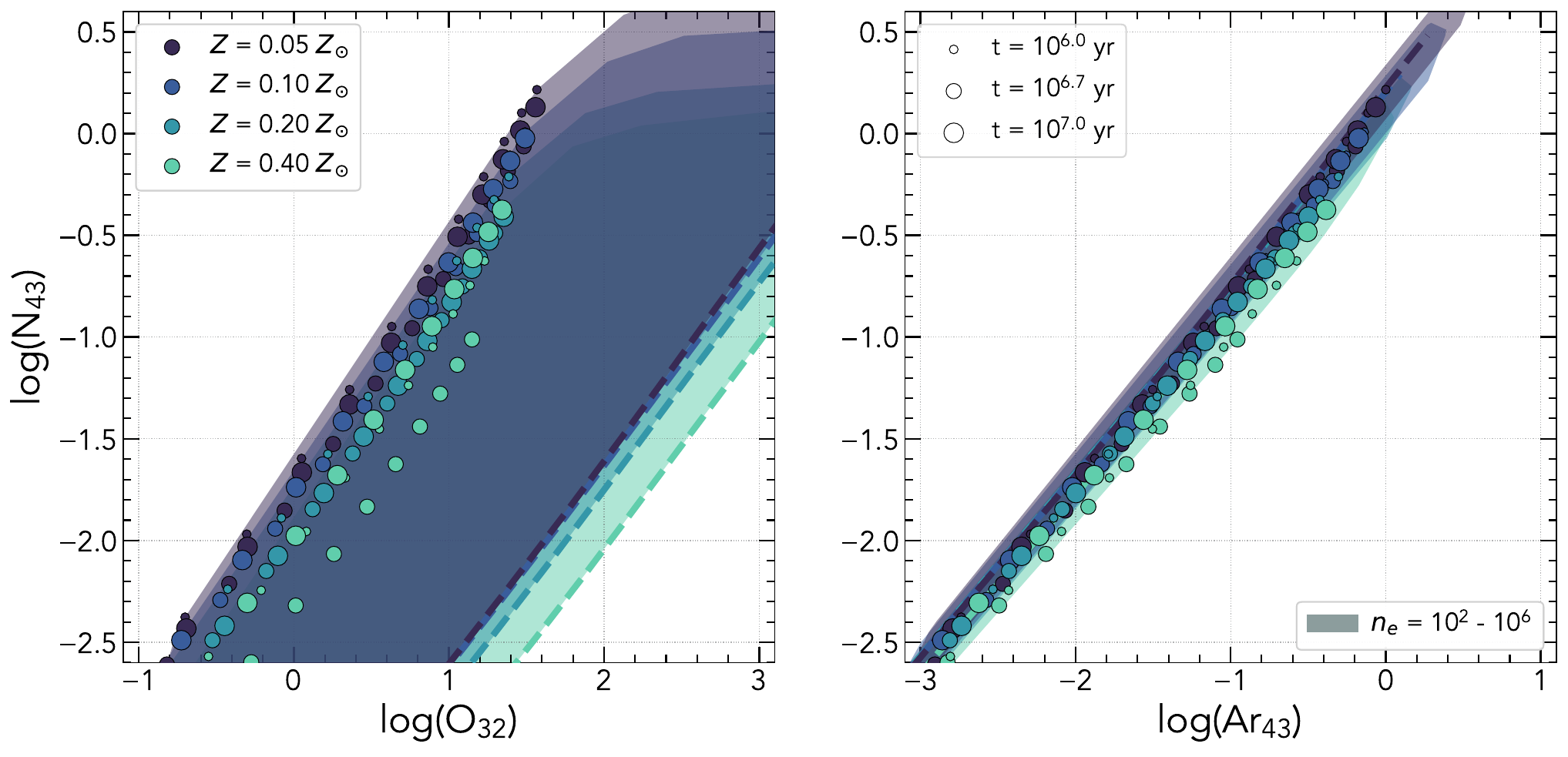}
    \caption{Comparison of different diagnostics of ionization parameter from \texttt{Cloudy} photoionization models (see Section~\ref{sec:models}). 
    The plots contains models that span metallicities from 0.05 Z$_{\odot}$ to 0.4 Z$_{\odot}$, as indicated by the color-coding, while
    the size of the circle markers corresponds to burst ages between 10$^{6.0}$ - 10$^{7.0}$.
    The spread in models by varying \den$ = 10^2 - 10^6$ cm$^{-3}$ is indicated by the shaded regions, where the dashed line represents the maximum \den$ = 10^6$ cm$^{-3}$ model.
    {\bf Left:} The UV N$_{43}$ $\log U_{high}$ diagnostic versus the classic optical O$_{32}$ $\log U_{int.}$ diagnostic.
    The large spread is due to the O$_{32}$ dependence on density, indicating it is a less reliable diagnostic of ionization parameter when the density in the high-ionization zone is unknown.
    {\bf Right:} The UV N$_{43}$ $\log U_{high}$ diagnostic versus the optical Ar$_{43}$ $\log U_{high}$ diagnostic introduced by \citet{berg21}. 
    The tight relationship between N$_{43}$ and Ar$_{43}$ is nearly density independent, suggesting N$_{43}$ can be used as a robust $\log U_{high}$ diagnostic.}
    \label{fig:IonPar_cloudy}
    \end{center}
\end{figure*}

To determine N/O abundances, we estimate the ICF as a function of the ionization parameter using our photoionization models. 
The O$_{32}$ ratio is the most commonly used ionization parameter diagnostic since O$^{+}$ and O$^{+2}$ cover the standard 3-ionization zones from $\sim$ 15 -- 55 eV.
However, \citet{berg21} showed that for galaxies with very high-ionization emission ($> 54$ eV), O$_{32}$ may not probe high enough energies to properly characterize the gas conditions.
In these cases, it is useful to examine multiple probes of the ionization parameter, where Ar$_{43}$ can be used to characterize the high-ionization zone ionization parameter ($U_{high}$), O$_{32}$ can be used to characterize the intermediate-ionization zone ionization parameter ($U_{int}$), and S$_{32}$ can be used to characterize the low-ionization zone ionization parameter ($U_{low}$).

Following \citet{berg21}, we explore ionization parameters that probe higher ionization energies than O$_{32}$, for galaxies that have very-high-ionization emission. 
However, the optical Ar$_{43}$ line ratio is rarely observed at high redshifts.
Fortunately, the N$_{43}$ line ratio offers alternative diagnostic power.
In the left panel of Figure \ref{fig:IonPar_cloudy}, we plot the optical O$_{32}$ ratio compared to the UV N$_{43}$ ratio for a constant (N/O)/(N/O)$_\odot = 0.23$ \citep[log(N/O)$\approx-1.4$, typical of metal-poor galaxies, e.g.,][]{berg19}.
The range in metallicity of the models is indicated by the color-coding, showing a small offset to higher O$_{32}$ with increasing metallicity.
Small variations are also seen with changing burst age, as shown by the increasing point size from $t = 10^6 - 10^7$ Myr. 
As metallicity increases, the variation due to burst age becomes more noticeable.
There is, however, a large spread in the models caused by varying \den, which is shown by the shaded regions extending from filled circles with \den$ = 10^2$ cm$^{-3}$ to the dashed lines marking \den$ = 10^6$ cm$^{-3}$.
As discussed in Section \ref{sec:emline_varyden}, this \den-effect on the O$_{32}$ line ratio is due to the $n_{e, crit}$ of O$^+$ at $n_e \gtrsim 10^3$ and O$^{+2}$ at $n_e \gtrsim 10^5$, suggesting it is less effective as an ionization parameter diagnostic when density is unknown.

In the right panel of Figure \ref{fig:IonPar_cloudy}, we plot the UV N$_{43}$ ratio compared to the optical Ar$_{43}$ high-ionization zone ionization parameter diagnostic.
Here, N$_{43}$ is sensitive to the high and very-high ionization zones probed by N$^{+2}$ (29.6--47.4 eV) and N$^{+3}$ (47.4--77.5 eV), respectively, similar to the Ar$^{+2}$ (27.6--40.7 eV) and Ar$^{+3}$ (40.7--59.8 eV) ionization potentials.
In general, we find a tight trend between N$_{43}$ and Ar$_{43}$ that is nearly independent of density. 
Overall, this analysis suggests that the UV N$_{43}$ ratio is consistent with optical probes of log $U_{high}$ and can, therefore, be used as an alternative diagnostic of log $U_{high}$ when these lines are available.
On the other hand, the discrepancy between $\log U_{high}$ and $\log U_{int}$ measured with N$_{43}$ and O$_{32}$, respectively, is related to \den.
This dependency renders O$_{32}$ unreliable when the high-ionization zone density is $> 10^3$.

\begin{figure*}[ht]
    \begin{center}
    \includegraphics[width=0.85\textwidth]{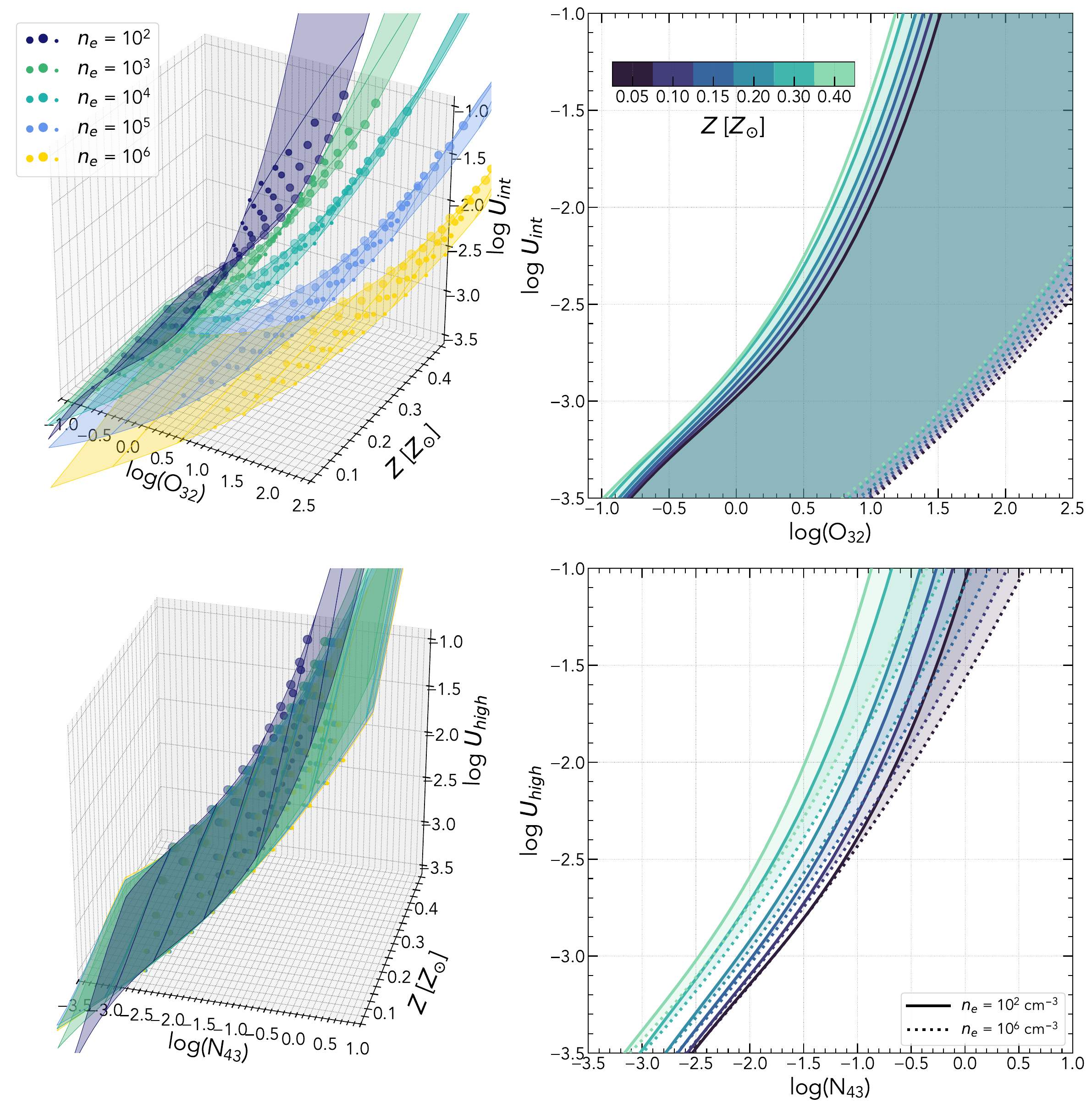}
    \caption{The \texttt{cloudy} model relationship between both O$_{32}$ (upper) and N$_{43}$ (lower) and ionization parameter, $U$.
    The left column shows 3D plots of the bicubic fits for each grid of models.
    Marker color corresponds to \den~ranging from 10$^2$ - 10$^6$ cm$^{-3}$.
    In the right column, we show a 2D projection where the range of \den~is shown by the shaded region.
    The metallicities range between 0.05 $Z_{\odot}$ and 0.40 $Z_{\odot}$ as indicated by the color bar in the upper right plot.
    Clearly, \den~has a large impact on O$_{32}$, however, N$_{43}$ is much more robust to extreme \den.
    }
    \label{fig:LU_fits}
    \end{center}
\end{figure*}

Given the strong \den-sensitivity of the O$_{32}$ ratio, we present new $\log U$ diagnostics that account for this effect.
Adopting the 4-zone ionization model presented in \citet{berg21}, we examine the O$_{32}-\log U_{int}$ and N$_{43}-\log U_{high}$ relationships from the \texttt{Cloudy} photoionization model grid (presented in Section \ref{sec:models}) in Figure~\ref{fig:LU_fits}.
The left column shows the color-coded 3-dimensional surfaces probing the dependence of $\log U$ on emission line ratio, metallicity, and density, while the right column shows a 2-dimensional projection along the metallicity axis.
The 2-dimensional plot of O$_{32}$ (upper), shows the significant sensitivity to density, while little spread is seen for N$_{43}$ (lower).

To characterize the trends in Figure \ref{fig:LU_fits}, we fit bicubic surfaces to the O$_{32}-\log U_{int}$ and N$_{43}-\log U_{high}$ relationships using the \texttt{scipy.optimize.least\_squares} function in \texttt{python}.
The resulting fit, $f(x,y$), is a bicubic surface with the form: $f(x,y) = \log U = A + Bx + Cy + Dxy + E x^2 + Fy^2 + Gxy^2 + Hyx^2 + Ix^3 + Jy^3$, where $y = Z/Z_{\odot}$ (metallicity) and $x = \log {\rm O}_{32}$ or $x = \log {\rm N}_{43}$.
The coefficients that describe these bicubic surface are reported in Table \ref{tab:LU_fits}.
The ranges of validity for these relationships are $0.05 \leq Z/Z_\odot \leq 0.4$,
$-1.0 \leq \log {\rm O_{32}} \leq 2.5$, and $-3.0 \leq \log {\rm N_{43}} \leq 0.5$.
Note that the density-insensitivity of the N$_{43}$ diagnostic suggests that it can be robustly used even in the absence of a measurement of the high-ionization zone density.
Therefore, we recommend using the N$_{43}$ diagnostic of ionization parameter over O$_{32}$ when the requisite lines are available.
For another \den-insensitive probe of $\log U$ that instead utilizes O$^{+2}$/O$^{+}$, which may be useful in studies where \NIV~and \NIII~are not present, see Appendix \ref{Appendix:C}.

\begin{deluxetable}{lDDDDD}
\setlength{\tabcolsep}{1.5pt}
\tabletypesize{\scriptsize}
\tablecaption{Ionization Parameter Fits \label{tab:LU_fits}}
\tablehead{
\CH{} & \multicolumn{10}{c}{\den~(cm$^{-3}$)} \\
\cline{2-11}
\CH{} & \twocolhead{10$^2$} & \twocolhead{10$^3$} & \twocolhead{10$^4$} & \twocolhead{10$^5$} & \twocolhead{10$^6$}
}
\decimals
\startdata
\multicolumn{11}{l}{$z = f(x,y) = \log U_{int}$; $x=$ log(O$_{32}$); $y = Z$} \\
\hline
$A$ \ldots & -3.00199	 & -3.06316	 & -3.28872	 & -3.71434	 & -4.08448	 \\
$B$ \ldots &  0.61875	 &  0.65186	 &  0.62685	 &  0.53053	 &  0.54480	 \\
$C$ \ldots &  0.41277	 &  0.45934	 &  0.32662	 &  0.07827	 &  0.17599	 \\
$D$ \ldots &  0.96910	 &  0.60495	 &  0.44956	 &  0.16911	 & -0.23221	 \\
$E$ \ldots &  0.11489	 &  0.04757	 &  0.05904	 &  0.05866	 &  0.01351	 \\
$F$ \ldots &  2.10518	 &  1.34596	 &  1.19054	 &  1.65578	 &  2.15922	 \\
$G$ \ldots & -0.88265	 & -0.61689	 & -0.61832	 & -0.62173	 & -0.47220	 \\
$H$ \ldots &  0.55524	 &  0.25114	 &  0.19987	 &  0.18713	 &  0.19535	 \\
$I$ \ldots &  0.18323	 &  0.04025	 &  0.00771	 &  0.00364	 &  0.00841	 \\
$J$ \ldots & -4.65728	 & -3.24251	 & -2.62744	 & -2.54602	 & -2.95869	 \\
\hline
$\sigma_{\rm rms}$ 	 & 0.02	 & 0.01	 & 0.01	 & 0.01	 & 0.01	 \\ 
\hline
\hline
\multicolumn{11}{l}{$z = f(x,y) = \log U_{high}$; $x=$ log(N$_{43}$); $y = Z$} \\
\hline
$A$ \ldots & -1.29754	 & -1.61212	 & -1.66787	 & -1.6783	 & -1.6839	 \\
$B$ \ldots &  1.60610	 &  1.03768	 &  0.91085	 &  0.89072	 &  0.88034	 \\
$C$ \ldots &  4.10271	 &  2.59196	 &  2.2672	 &  2.20159	 &  1.93562	 \\
$D$ \ldots &  3.82481	 &  2.04872	 &  1.63593	 &  1.56124	 &  1.55433	 \\
$E$ \ldots &  0.52310	 &  0.18279	 &  0.09245	 &  0.08156	 &  0.08954	 \\
$F$ \ldots & 12.37774	 &  9.37845	 &  8.61368	 &  8.23949	 &  8.03766	 \\
$G$ \ldots &  0.22144	 & -0.27803	 & -0.45877	 & -0.55776	 & -0.43756	 \\
$H$ \ldots &  0.74401	 &  0.32021	 &  0.20331	 &  0.17851	 &  0.22411	 \\
$I$ \ldots &  0.09083	 &  0.02670	 &  0.00723	 &  0.00484	 &  0.00874	 \\
$J$ \ldots &-17.63486  &-15.11384  &-14.59398  &-14.35798  &-13.90218  \\
\hline
$\sigma_{\rm rms}$ 	 & 0.04	 & 0.03	 & 0.03	 & 0.03	 & 0.03	 \\
\enddata
\tablecomments{The bicubic surface fit to \texttt{Cloudy} photoionization models for the ionization parameters are parameterized by the following equation: $z = f(x, y) = A + Bx + Cy + Dxy + E x^2 + Fy^2 + Gxy^2 + Hyx^2 + Ix^3 + Jy^3$.}
\end{deluxetable}

\begin{figure*}[ht]
    \begin{center}
    \includegraphics[width=\textwidth]{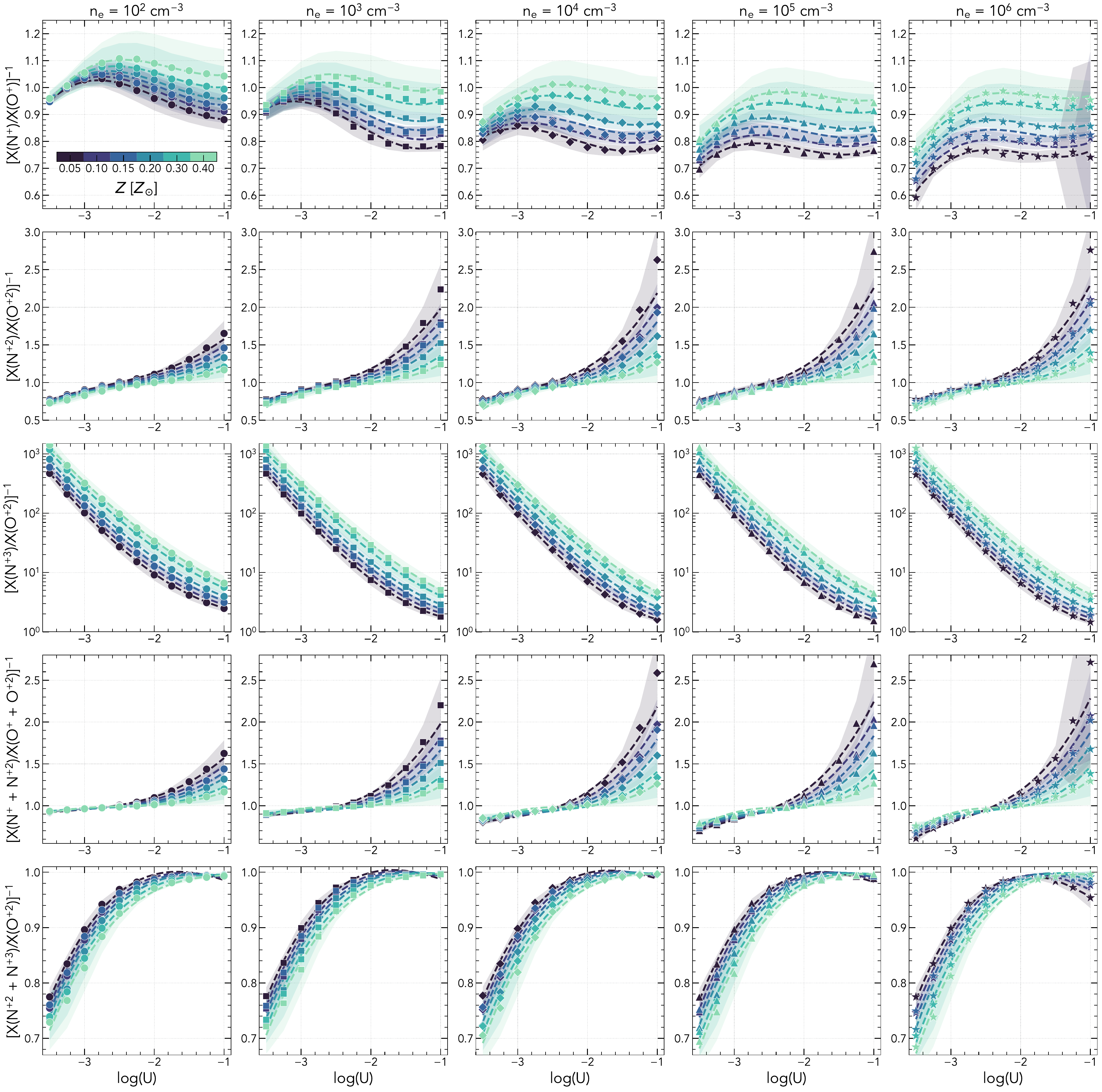}
    \caption{The five different ICFs that we fit as a function of $\log U$ are displayed in each row.
    Each is shown at a range of metallicities between 0.05 $Z_{\odot}$ and 0.40 $Z_{\odot}$ indicated by the color bar in the upper leftmost tile.
    The shaded region that corresponds to a spread in burst age ($t = 10^{6.0} - 10^{7.0}$ Myr).
    Every column corresponds to a different constant from \den~$ = 10^{2} - 10^{6}$.
    {\bf Row 1:} \niioii~is a fairly good estimate of relative N/O.
    {\bf Row 2:} \niiioiii~shows that this ionic abundance is overestimating N/O at low ionization parameter and underestimating it at high ionization parameter.
    {\bf Row 3:} \nivoiii~is underestimating N/O across the entire range of ionization parameters we consider, but most notably so at low ionization.
    {\bf Row 4:} The \noUVopt~ICF is underestimating relative N/O at all ionization parameters, but this is most severe at high ionization.
    {\bf Row 5:} Finally, the \niiinivoiii~seems to infer relative N/O very well at log(U) $>$ -2.75.}
    \label{fig:ICF_fits}
    \end{center}
\end{figure*}

We determine five ICFs for each of the following ionic abundance ratios: 
\begin{align*}
    {\rm 1.\ } {\rm \frac{N}{O}} &= {\rm \frac{N^+}{O^+}} \times {\rm ICF(N^+/O^+)} \\
    {\rm 2.\ } {\rm \frac{N}{O}} &= {\rm \frac{N^{+2}}{O^{+2}}} \times {\rm ICF(N^{+2}/O^{+2})} \\
    {\rm 3.\ } {\rm \frac{N}{O}} &= {\rm \frac{N^{+3}}{O^{+2}}} \times {\rm ICF(N^{+3}/O^{+2})} \\
    {\rm 4.\ } {\rm \frac{N}{O}} &= {\rm \frac{N^{+}+N^{+2}}{O^{+}+O^{+2}}} \times {\rm ICF((N^{+}+N^{+2})/(O^{+}+O^{+2}))} \\
    {\rm 5.\ } {\rm \frac{N}{O}} &= {\rm \frac{N^{+2}+N^{+3}}{O^{+2}}} \times {\rm ICF((N^{+2}+N^{+3})/O^{+2})},
\end{align*}
where 
O$^{+}$ is determined from \OII~\W\W3727,3729, O$^{+2}$ is determined from \OIII~\W1666, N$^{+}$ is determined from \NII~\W6585, N$^{+2}$ is determined from \NIII~\W1750, and N$^{+3}$ is determined from \NIV~\W\W1483,1487.
Note that ICF 4, \noUVopt, combines low ionization and intermediate to high ionization lines that should more accurately represent relative N/O.
Additionally, ICF 1 may also be used to in combination with the near-UV \OII~\W2470\footnote{\OII, in the near-UV, is a doublet composed of emission lines at \W2470.22 and \W2470.34, which would be unresolved in galaxy spectroscopy.} and \ion{N}{2}] \W\W2139,2143 emission lines.

Similar to the $\log U$ fits, we fit bicubic surfaces for each set of \den~grids from 10$^2 - 10^6$ cm$^{-3}$.
However, the ICFs are more sensitive to burst age than the ionization parameters, so we weight the fit by the spread in burst ages between the models at a given $\log U$.
The coefficients for these fits can be found in Table \ref{tab:ICF_fits}.
We note that the ICF fits presented in this work are only valid for $-3.5 < \log(U) < -1.0$.
Figure \ref{fig:ICF_fits} shows the five ICFs considered in this work versus ionization parameter.
Each ICF is shown for six different metallicities.

The first column in Figure \ref{fig:ICF_fits} gives each model ICF as a function of $\log U$ at \den~$ = 10^2$ cm$^{-3}$.
In this low-density limit of the \niioii~ICF, we find that a relatively small correction is needed ($<$ 10\%) across a large range of $\log U$, which is consistent with the convention that N/O $\approx$ \niioii~ \citep[e.g.][]{peimbert69,nava06,amayo21}.
The \niiioiii~ratio will overestimate N/O by $<$ 50\% at low $\log U$, and underestimate N/O by up to 200\% at high $\log U$.
While the \nivoiii~ratio underestimates N/O between 100 - 15000\% across a variety of $\log U$, the relative \noUVopt~ionic abundance will underestimate N/O ($\sim$ 150\%) at high $\log U$ and overestimate N/O ($<$ 10\%) at low $\log U$.
Finally, the \niiinivoiii~ICF indicates that this relative ionic abundance will overestimate the relative N/O abundance by as much as 30\%, however, at higher $\log U$ correction required will be much smaller.

The fifth column in Figure \ref{fig:ICF_fits} again gives each ICF as a function of $\log U$, however, the \den~$ = 10^6$ cm$^{-3}$.
The \niioii~ will overestimate  the relative N/O by nearly 40\% at the lowest metallicities and $\log U$.
For the \niiioiii~ratio, N/O is overestimated by $<$ 50\% at low $\log U$, and underestimated by up to 300\% at high $\log U$.
On the other hand, the \nivoiii~ratio underestimates relative N/O between 100 - 15000\% over all $\log U$.
The relative \noUVopt~ionic abundance will underestimate N/O ($\sim$ 250\%) at high $\log U$ and overestimate N/O ($<$ 50\%) at low $\log U$.
The \niiinivoiii~ICF reveals that this relative ionic abundance will overestimate the relative N/O abundance by $<$ 30\%; however, at $\log U > -2.0$ the correction will be closer to $>$ 5\%.

When considering the effects of \den, \niioii goes from being within 10\% of the total N/O to overestimating the N/O by up to 40\% for some $\log U$ and metallicities at the highest \den.
From low- to high-\den, \niiioiii~maintains its shape but goes from being within $\sim$50\% of the total N/O to under-estimating the N/O by up to 300\% at high $\log U$ and low metallicity.
While across the same range of \den, \nivoiii~maintains consistency in that it underestimates total N/O by as much 100 - 15000\% and continues underestimating the total N/O by roughly the same amount.
On the other hand, the \noUVopt~and \niiinivoiii~ICFs do not change significantly with \den.

\begin{deluxetable}{lDDDDD}
\setlength{\tabcolsep}{1.5pt}
\tabletypesize{\scriptsize}
\tablecaption{Ionization Correction Factor Fits \label{tab:ICF_fits}}
\tablehead{
\CH{} & \multicolumn{10}{c}{\den~($cm^{-3}$)} \\
\cline{2-11}
\CH{} & \twocolhead{10$^2$} & \twocolhead{10$^3$} & \twocolhead{10$^4$} & \twocolhead{10$^5$} & \twocolhead{10$^6$}
}
\decimals
\startdata
\multicolumn{11}{l}{$z = f(x,y) = $ \niioii~ICF; $x = log(U)$; $y = Z$} \\
\hline
$A$ \ldots &  0.99684	 &  1.11259	 &  1.04256	 &  1.05359	 &  1.05845	 \\
$B$ \ldots &  0.35805	 &  0.68396	 &  0.52349	 &  0.54522	 &  0.58814	 \\
$C$ \ldots &  0.73829	 &  0.57962	 &  0.14444	 &  0.02129	 &  0.28991	 \\
$D$ \ldots & -0.00147	 & -0.20911	 & -0.50730	 & -0.52622	 & -0.33298	 \\
$E$ \ldots &  0.25599	 &  0.37927	 &  0.27256	 &  0.28401	 &  0.31784	 \\
$F$ \ldots & -1.05333	 &  0.13139	 &  0.78050	 &  1.11581	 &  1.11112	 \\
$G$ \ldots & -0.00247	 & -0.14117	 &  0.00501	 &  0.18414	 &  0.20679	 \\
$H$ \ldots & -0.04340	 & -0.11589	 & -0.15029	 & -0.12303	 & -0.06881	 \\
$I$ \ldots &  0.04517	 &  0.05690	 &  0.04052	 &  0.04511	 &  0.05389	 \\
$J$ \ldots &  1.35162	 & -1.00523	 & -1.67611	 & -1.75450	 & -2.03291	 \\
\hline
$\sigma_{\rm rms}$ 	 & 0.01	 & 0.01	 & 0.01	 & 0.01	 & 0.02	 \\
\hline
\hline
\multicolumn{11}{l}{$z = f(x,y) = $ \niiioiii~ICF; $x = log(U)$; $y = Z$} \\
\hline
$A$ \ldots &  2.71718	 &  4.18005	 &  4.96600	 &  5.20832	 &  5.32595	 \\
$B$ \ldots &  1.40827	 &  2.73917	 &  3.48636	 &  3.71419	 &  3.82306	 \\
$C$ \ldots & -3.80233	 & -7.42258	 & -9.32333	 & -9.83354	 & -9.73777	 \\
$D$ \ldots & -1.98828	 & -4.08756	 & -5.25762	 & -5.59671	 & -5.65519	 \\
$E$ \ldots &  0.41454	 &  0.81576	 &  1.04937	 &  1.12023	 &  1.15115	 \\
$F$ \ldots &  2.86050	 &  5.91850	 &  7.28429	 &  7.46491	 &  6.66459	 \\
$G$ \ldots &  0.89889	 &  1.77628	 &  2.22999	 &  2.33000	 &  2.20890	 \\
$H$ \ldots & -0.26525	 & -0.56759	 & -0.74483	 & -0.79911	 & -0.82169	 \\
$I$ \ldots &  0.04878	 &  0.08888	 &  0.11297	 &  0.12032	 &  0.12301	 \\
$J$ \ldots &  0.13269	 & -0.43755	 & -0.42475	 & -0.22878	 &  0.41586	 \\
\hline
$\sigma_{\rm rms}$ 	 & 0.02	 & 0.03	 & 0.04	 & 0.04	 & 0.04	 \\
\hline
\hline
\multicolumn{11}{l}{$z = f(x,y) = $ log(\nivoiii~ICF)\dg; $x = log(U)$; $y = Z$} \\
\hline
$A$ \ldots &  0.33721	 &  0.32989	 &  0.27897	 &  0.27431	 &  0.24361	 \\
$B$ \ldots &  0.24224	 &  0.53349	 &  0.56120	 &  0.58497	 &  0.58882	 \\
$C$ \ldots & -0.65061	 & -0.68007	 & -0.57998	 & -0.51353	 & -0.17741	 \\
$D$ \ldots & -1.33811	 & -1.56364	 & -1.54907	 & -1.53730	 & -1.30763	 \\
$E$ \ldots &  0.27971	 &  0.45572	 &  0.48623	 &  0.49791	 &  0.50260	 \\
$F$ \ldots &  5.83889	 &  5.36918	 &  5.11709	 &  4.79966	 &  3.91905	 \\
$G$ \ldots &  0.31924	 &  0.52243	 &  0.54710	 &  0.52024	 &  0.30299	 \\
$H$ \ldots & -0.26487	 & -0.30772	 & -0.30843	 & -0.31010	 & -0.28039	 \\
$I$ \ldots &  0.00662	 &  0.03347	 &  0.03900	 &  0.04044	 &  0.04061	 \\
$J$ \ldots & -9.33063	 & -8.05623	 & -7.62178	 & -7.17482	 & -6.28983	 \\
\hline
$\sigma_{\rm rms}$ 	 & 0.03	 & 0.03	 & 0.03	 & 0.03	 & 0.03	 \\
\hline
\hline
\multicolumn{11}{l}{$z = f(x,y) = $ \noUVoptfrc~ICF;$x = log(U)$; $y = Z$} \\
\hline
$A$ \ldots &  2.80532	 &  4.28328	 &  5.00990	 &  5.23403	 &  5.38109	 \\
$B$ \ldots &  1.51853	 &  2.84543	 &  3.50850	 &  3.73697	 &  3.91911	 \\
$C$ \ldots & -4.17309	 & -7.91458	 & -9.76639	 &-10.16471	 & -9.97970	 \\
$D$ \ldots & -2.26441	 & -4.44062	 & -5.51060	 & -5.67998	 & -5.52784	 \\
$E$ \ldots &  0.42943	 &  0.81553	 &  1.01895	 &  1.11019	 &  1.19788	 \\
$F$ \ldots &  3.55046	 &  6.75686	 &  8.30719	 &  8.65627	 &  8.25633	 \\
$G$ \ldots &  1.09105	 &  2.04896	 &  2.52596	 &  2.76166	 &  2.83279	 \\
$H$ \ldots & -0.29179	 & -0.59753	 & -0.74177	 & -0.73561	 & -0.67993	 \\
$I$ \ldots &  0.04290	 &  0.08034	 &  0.10332	 &  0.11846	 &  0.13442	 \\
$J$ \ldots & -0.39350	 & -0.97013	 & -1.20540	 & -0.92040	 & -0.33026	 \\
\hline
$\sigma_{\rm rms}$ 	 & 0.01	 & 0.02	 & 0.03	 & 0.04	 & 0.04	 \\
\hline
\hline
\multicolumn{11}{l}{$z = f(x,y) = $ \niiinivoiii~ICF; $x = log(U)$; $y = Z$} \\
\hline
$A$ \ldots &  0.93486	 &  0.93562	 &  0.94769	 &  0.93356	 &  0.84213	 \\
$B$ \ldots & -0.03453	 & -0.03369	 & -0.01451	 & -0.01832	 & -0.08984	 \\
$C$ \ldots &  0.28011	 &  0.27477	 &  0.25282	 &  0.32434	 &  0.67947	 \\
$D$ \ldots &  0.24291	 &  0.22211	 &  0.16544	 &  0.17688	 &  0.32963	 \\
$E$ \ldots &  0.02786	 &  0.02993	 &  0.03802	 &  0.04019	 &  0.02188	 \\
$F$ \ldots & -0.15188	 & -0.18937	 & -0.30544	 & -0.47628	 & -1.01073	 \\
$G$ \ldots & -0.13845	 & -0.14771	 & -0.14786	 & -0.17401	 & -0.29574	 \\
$H$ \ldots &  0.02356	 &  0.01550	 & -0.00045	 & -0.00391	 &  0.00886	 \\
$I$ \ldots &  0.01429	 &  0.01478	 &  0.01579	 &  0.01642	 &  0.01486	 \\
$J$ \ldots & -0.16515	 & -0.12369	 &  0.04361	 &  0.18150	 &  0.42906	 \\
\hline
$\sigma_{\rm rms}$ 	 & 0.01	 & 0.01	 & 0.01	 & 0.01	 & 0.01	 \\
\enddata
\tablecomments{
The bicubic surface fits to \texttt{Cloudy} photoionization models for the ICFs are parameterized by the following equation: $f(x, y) = A + Bx + Cy + Dxy + E x^2 + Fy^2 + Gxy^2 + Hyx^2 + Ix^3 + Jy^3$. \\
\dg Note that the fit for the \nivoiii~ICF returns the logarithm of the ICF.
}
\end{deluxetable}

\subsection{Summary of Density Effects and Recommendations} \label{sec:den_effects}
In the previous sections, we discussed the impact of high-\den~on emission line ratios, physical conditions, ICFs, and total O/H abundances.
High values of density ($10^2 \leq n_e \leq 10^7$) will have the most effect on relative N/O abundances via biases that trickle down throughout the process of deriving N/O.
When tracing N/O with the optical tracer, N2O2, the extreme sensitivity to \den~discussed in Section \ref{sec:emline_varyden} will make this emission line ratio less representative of the relative N/O.
However, the UV tracers, N3O3 and N4O3, are relatively insensitive to \den, but dramatically dependent on the \Te~as demonstrated in Figure~\ref{fig:physcond_sensitivity}.
This becomes important when considering that deriving \Te~from [\OIII~\W4364/\W5008 in high-\den~gas will bias \Te~high.
This derived \Te~will impact the O/H and, as a result, influence the ICFs, which have a secondary dependence on metallicity.
The ICFs can also be significantly influenced by \den~effects on the O$_{32}$ ratio, which is classically used to diagnose the $\log U$ (see Sections \ref{sec:emline_varyden} and \ref{sec:ICF_varyden}).

To demonstrate the cumulative impact of density effects on derived nebular properties we explore the following three simple cases: 
(1) assuming a low density in a high density environment,
(2) a multi-phase ISM in which low-ionization gas is uniform low-density and high-ionization gas is uniform high-density, and 
(3) a multi-phase ISM in which a given ionization zone of gas is a mix of low- and high-density gas.

\subsubsection{Assuming Low Densities in a High-Density Environment}
We first determine the level of bias introduced in nebular properties if the density is underestimated.
For example, if we adopt a \den~$= 10^5$ cm$^{-3}$ nebular model, but assume a density of $10^2$ cm$^{-3}$, the \Te~determined from [\OIII~\W4364/\W5008 will be overestimated by, on average, 1800 K.
In the lowest $Z$ models the \Te~derived from [\OIII~\W4364/\W5008 is overestimated by approximately 1300 K, while this bias is closer to 2300 K in the highest $Z$ models.
This \Te~bias results in the O$^{+2}$/H$^{+}$ relative abundance and, thus, the direct gas-phase metallicity being underestimated by roughly 0.67 dex.
The O$^{+}$/H$^{+}$ is also affected by the \Te~bias, where \Te(O$^+$) is determined from the \Te(O$^{+2}$) with a \Te-\Te~relationship, but this has smaller overall effect on O/H due to the lower density of this gas and the smaller ion fraction.
These assumptions also lead to a overestimation of log(N/O) using \niiioiii of $\sim$0.16 dex.
The resulting bias from this exercise, while moderate for log(N/O), has larger ramifications on the N/O--O/H plane and our interpretations of chemical enrichment in dense, \hz~galaxies.
This underestimation is not limited to the case of making model assumptions when lacking density measurements; in practice, many standard optical density diagnostics intrinsically lose sensitivity and saturate above their critical densities, biasing observationally-inferred densities to lower values whenever dense clumps dominate the emission measure.
We will discuss the implications of these findings in Section \ref{sec:implications_sample} for a sample of UV N emitting galaxies from $0 \lesssim z \lesssim 11$.

Given the potentially significant effects of high \den, we recommend that future UV N/O abundance studies adopt the following approach.
(1) Measure multi-phase \den, when possible, to fully characterize the \den~in each ionization zone.
(2) Use the appropriate density to determine the temperature of a given ionization zone, e.g., calculate $T_{e,high}(n_{e,high})$.
In the absence of \Te-sensitive emission line ratios for the low- or intermediate-ionization zones, adopt the appropriate \Te~for each zone using a \Te-\Te~relationship \citep{garnett92, pilyugin09, croxall16, arellanocordova20, rogers21}.
(3) Calculate O$^{+}$/H$^{+}$ and O$^{+2}$/H$^{+}$ using the appropriate low- and high-ionization \Te~and \den, and determine the gas-phase metallicity directly from these total ionic abundances.
(4) Derive relative ionic N/O abundances (e.g., \niioii, \niiioiii, \nivoiii) using the correct \Te~and \den~for each ion.
(5) Determine $\log U$ using a \den-insensitive diagnostic such as N$_{43}$ or Ar$_{43}$.
If N$_{43}$ and Ar$_{43}$ are unavailable, use the O$_{32}$ diagnostic with the measured $n_{e,high}$ or use the O$^{+2}$/O$^+$ diagnostic; the latter should be used when $n_{e,high}$ is also unavailable.
(6) Finally, determine the total N/O abundance using the appropriate $\log U$, O/H abundance, and density to determine the ICFs and apply these corrections to account for unseen ionic species.
This empirically-motivated multi-phase ionization model fully accounts for the \Te~and \den~structure of the ionized gas in galaxies.

Finally, we emphasize that the recommendation outlined above is assuming the source of ionization is purely from stellar populations.
Assuming ionization from stellar populations, the extreme emission we are examining in this work can only be produced by very young stellar populations \citep{jaskot16}.
Note that these young ages are consistent with the stellar population ages from the UV continuum in the CLASSY galaxies \citep{parker25}.
However, different ionizing sources may be contributing to the high-ionization conditions and gas \den~in galaxies with extreme emission lines.
It may be that the light from objects with strong \NIV~emission is being dominated by AGN, but determining the source of ionizing radiation in our sample is topic of future work.

\subsubsection{A Multi-phase ISM with Uniform Low-Density Low-Ionization Gas and High-Density High-Ionization Gas}
In Section \ref{sec:const_den}, we discussed nebular models using a uniform \den~framework.
To better align our interpretation with observations, we explore the \texttt{Cloudy} models using a multi-phase \den~model.
There is physical evidence that gas near a SF region is higher density \citep[e.g.,][]{berg21}, often referred to as a dense core, we attempt to model this as simply as possible, where we use linear combinations of our existing \texttt{Cloudy} models as a first order approximation of dense cores.
We create linear combinations of models where low-\den~(10$^2$ cm$^{-3}$) and high-\den~(10$^5$ cm$^{-3}$) contribute varying percentages of the total emission.

\begin{figure}[ht]
    \begin{center}
    \includegraphics[width=0.45\textwidth]{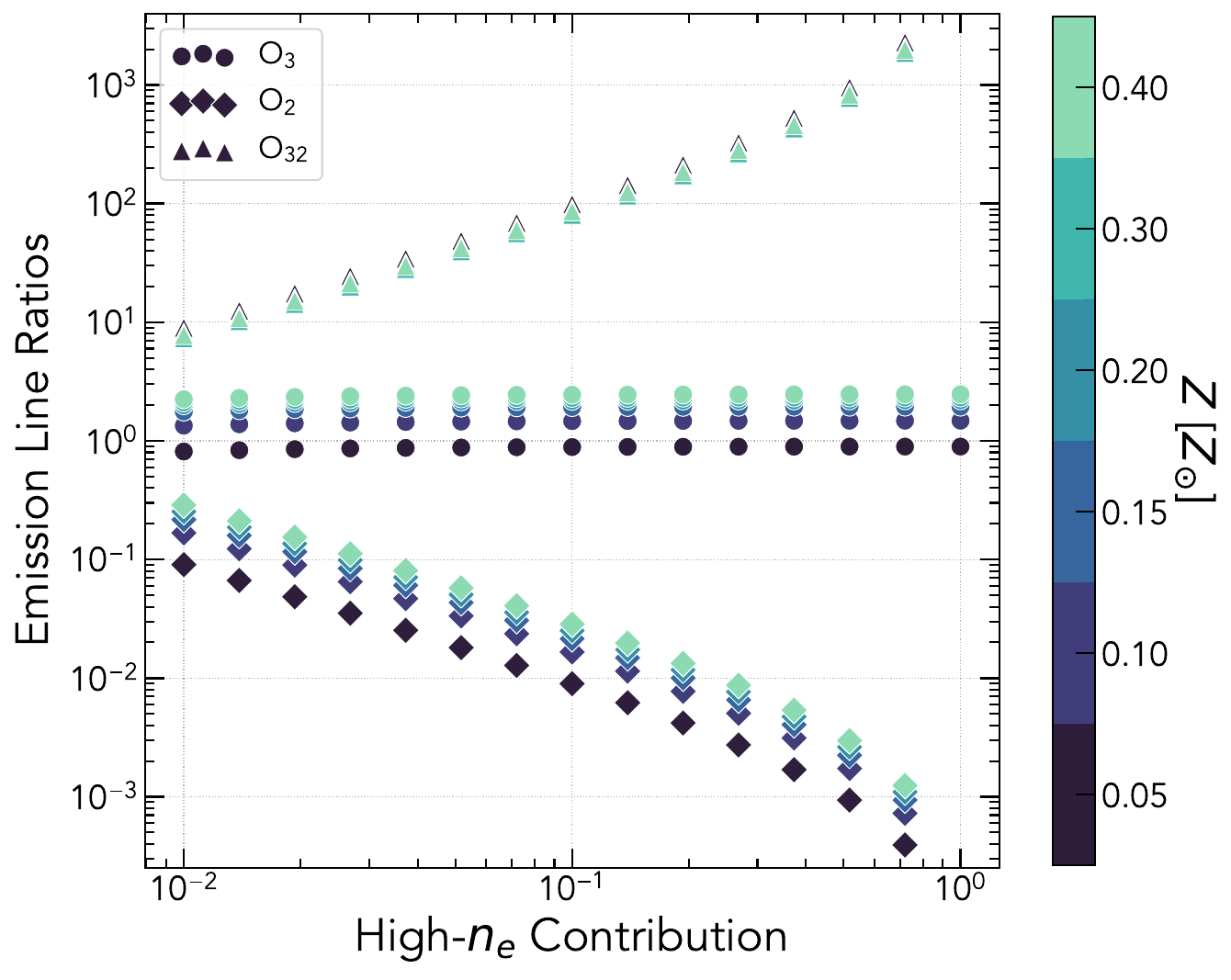}
    \caption{\texttt{Cloudy} models are able to reproduce extreme emission line ratios of O$_3$, O$_2$, and O$_{32}$ (y-axis) given varied contributions of emission from low- and high-\den~gas (x-axis).
    The circles show O$_3$, the diamonds are O$_2$, and the triangles describe O$_{32}$.
    The colors show different metallicities ranging from 0.05 $Z_{\odot}$ and 0.40 $Z_{\odot}$ given in the color bar on the right.}
    \label{fig:ratios_dencore}
    \end{center}
\end{figure}

\begin{figure*}[ht]
    \begin{center}
    \includegraphics[width=\textwidth]{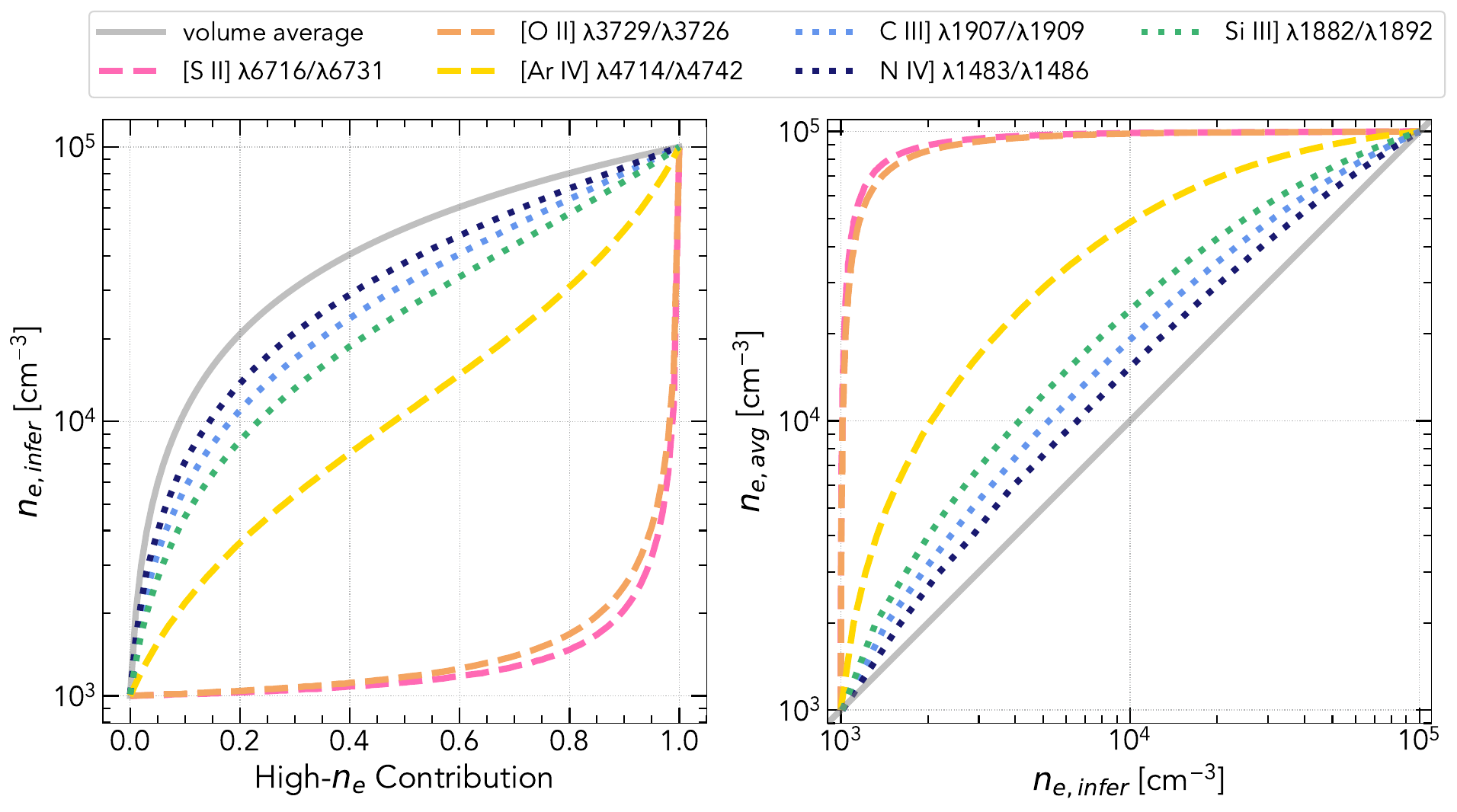}
    \end{center}
    \caption{{\bf Left:} 
    Density inferred from a given density diagnostic emission line ratio versus the volume fraction of high-density gas for a two-phase density model for a given ion, with low-density gas of $10^3$ cm$^{-3}$ and high-density gas of $10^5$ cm$^{-3}$.
    The true volumetric averaged density is plotted as the solid magenta line. 
    In comparison, nebular density diagnostics measure lower densities with the low-ionization optical diagnostics
    severely biased to the low-density value for most of the fractional range, 
    while the high-ionization UV diagnostics more closely trace the $n_{e, avg}$ due to their increased sensitivity to high-\den~gas.
    {\bf Right:} We compare the $n_{e, infer}$ for each diagnostic emission line ratio
    to the $n_{e, avg}$ to emphasize which ions serve as good tracers of the true $n_{e, avg}$ for an ionization zone with a multi-phase density.}
    \label{fig:highden_contribution_pyneb}
\end{figure*}

In Figure \ref{fig:ratios_dencore}, we plot O$_3$, O$_2$, and O$_{32}$ emission line ratios where the [\OIII~arises only from the high-\den~core and \OII~emits only from the low-\den~regions of the nebula.
We find that the O$_3$ ratio is dominated by the high-\den~gas and remains relatively constant with increasing high-\den~core contribution.
The O$_2$ begins to decrease rapidly with increasing H emission from the high-\den~region of gas.
This is because the integrated O$_2$ line ratio will only have \OII~emission from the low-\den~gas, but H emission from both the low-\den~and high-\den~gas.
Since emission will increase as $n_e^2$, the denominator of the O$_2$ fraction will quickly become dominated by H emission from the high-\den~region as the core fraction increases.
The O$_{32}$ models are able to reproduce even the most extreme observed ratios at relatively low ($\sim$ 0.01\%) contributions from high-\den~gas.
We note that this multiphase-\den~nebula model is not expected to impact the N/O emission line ratios since they mostly come from the same ionization zones.
This exercise demonstrates how easily high-\den~gas can dominate the emission in a multiphase-\den~nebula.
However, future work utilizing more complex modeling that includes an intermediate ionization zone will be useful to better understand contributions from dense cores.

\subsubsection{A Multi-Phase Ionization Zone With A Mix of Low- and High-Densities}
Previous studies have investigated the biases inherent to electron density diagnostics and showed that the interplay between an ion's ionization potential and $n_{e,crit}$ is essential to determining which gas phases each diagnostic reliably traces \citep[e.g.,][]{rubin89, mendezdelgado23}.
In this work, we have also examined the role of different densities in different ionization zones (gas of different ionization potentials).
We now shift our focus to the specific role of critical density. 
To isolate this effect, we consider a single ion (i.e., a single ionization zone) but allow the gas to have a non-uniform density distribution.
We explore how fractional contributions from low- and high-\den~gas components influence \den-diagnostics within this single-ion model by predicting emission line fluxes via \texttt{PyNeb} emissivities and combining them using a simple linear mixture weighted by the high- and low-\den~volume fractions of each component.
For this exercise, we adopt two representative density phases: a low-density component at $10^3$ cm$^{-3}$ and a high-density component at $10^5$ cm$^{-3}$.

Figure \ref{fig:highden_contribution_pyneb} illustrates the resulting effects of this multi-phase density structure of a single ion's diagnostic behavior.
The left panel shows the inferred \den~($n_{e,infer}$) from six different \den~diagnostics as a function of the fractional contribution from the high-\den~gas, along with the volumetric mean \den~($n_{e,avg}$) of the two components.
This simple two-phase model reveals that the diagnostics with the lowest critical densities (the low-ionization optical lines in this case), are strongly biased towards the low-density gas value.
This occurs because emission from the high-\den~gas is collisionally de-excited, suppressing contributions to the observed flux. 
As a result, these diagnostics trace the low-density component even when it constitutes less than $\approx15\%$ of the total volume. 
On the other hand, diagnostics with higher critical densities (e.g., the high-ionization UV lines) remain sensitive to the high-\den~gas except when it contributes $< 15$\% to the total volume.
Notably, these higher-$n_{e,crit}$ diagnostics also yield inferred densities that are in much closer agreement with $n_{e,avg}$.

The right panel of Figure \ref{fig:highden_contribution_pyneb} compares $n_{e,infer}$ and $n_{e,avg}$ for the six \den~diagnostics, where the highest ionization \den~diagnostics more closely trace $n_{e,avg}$.
Overall, this exercise highlights how collisional de-excitation in high-density gas can bias \den~diagnostics toward lower inferred values, particularly for transitions with low critical densities.
A more complex model—relaxing the simplifying assumptions adopted here—would be required for quantitative accuracy, but this approach effectively captures the underlying physical trend and diagnostic biases.

\begin{figure*}[ht]
    \begin{center}
    \includegraphics[width=\textwidth]{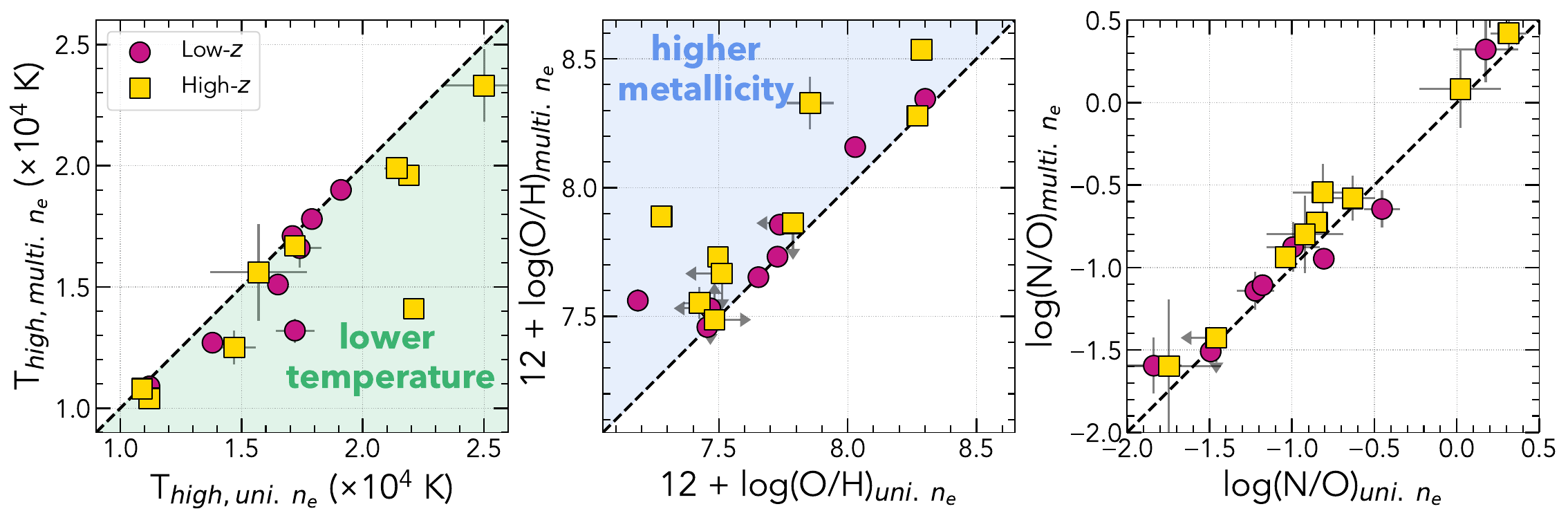}
    \end{center}
    \caption{{\bf Left:} T$_{e,high}$ when assuming a constant \den~compared to when a 3-zone \den~is assumed.
    We find that the 3-zone \den~model produces lower T$_{e,high}$ than the constant \den~model.
    {\bf Middle:} Comparing the gas-phase metallicities between these two models predictably shows that the 3-zone \den~model generates higher metallicities.
    {\bf Right:} When looking at the UV log(N/O) derived in each model, we find that the 3-zone \den~model on average drives UV log(N/O) up for both our \hz~and \lz~samples.}
    \label{fig:compare_properties}
\end{figure*}

\section{Implications for the \texorpdfstring{High$-z$ and Low$-z$}{High-z and Low-z} Samples} \label{sec:implications_sample}
In Section \ref{sec:const_den}, we showed that high densities can significantly impact our derived nebular properties, especially the temperature, ionization parameter, and oxygen abundance.
An additional implication of our results is that electron densities derived from low-$n_{e,crit}$ diagnostics should be regarded as lower limits of the true conditions of the ionized gas.
In this section, we examine two methods to determine physical conditions (i.e., $\log U$, \Te, \den) and abundances (O/H and N/O) for the Low$-z$ and High$-z$ Samples: (1) the uniform density model commonly adopted in the literature, assuming a 2-zone ionization structure for the temperature and (2) a multiphase density model adopting a \Te~and \den~for each ionization zone in the classic 3-zone ionization model.

\subsection{Multi-Phase Physical Conditions and Their Effects} \label{sec:temden}
As discussed in Section \ref{sec:tem_varyden}, accurate nebular diagnostics require correctly interpreting the electron temperature and density across the ionization structure of the nebula.
We use the \texttt{getTemDen} function in \texttt{PyNeb} to determine the temperature and density in each ionization zone.
For density determinations, we assume a \Te\ of $1.5 \times 10^4$ K, as \den~determinations are not very sensitive to \Te~\citep[e.g.,][]{osterbrock06}, and use the \SII~\W6733/\W6718 ratio for the low-ionization zone ($n_{e,{low}}$(S$^{+}$)), the \CIII~\W1909/\W1907 ratio for the intermediate-ionization zone ($n_{e,{int.}}$(C$^{+2}$)), and \NIV~\W1487/\W1483 for the high-ionization zone ($n_{e,{high}}$(N$^{+3}$)).
However, many of the objects in our \hz~sample do not have resolved low-ionization density diagnostics, so we use the $n_e-z$ relation derived in \cite{abdurrouf24} to characterize the $n_{e,{low}}$.
We note that the $n_e-z$ relation presented \cite{abdurrouf24} is consistent with other relations in the literature \citep[e.g.,][]{isobe23a, topping25b} such that this choice of relation will have little impact on the results.
The resulting low-ionization densities range from $n_{e,{low}}=(0.3 - 4.4) \times 10^2$ cm$^{-3}$ for the \lz~sample \citep[e.g.,][]{davies21, berg22} and from $n_{e,{low}}=(1.0 - 22.6) \times 10^2$ cm$^{-3}$ for the \hz~sample \citep[e.g.,][]{topping25b, stanton25}.

When comparing densities from different ionization zones it is important to consider the diagnostic range of each density diagnostic.
A conservative, but useful range to use for a given diagnostic is 
\begin{equation}
    0.1\times n_{e,crit.}^{low} \lesssim {\rm\ diagnostic\ range\ } \lesssim 10\times n_{e,crit.}^{high}, 
\end{equation}
where $n_{e,crit.}^{low}$ is the lower critical density and $n_{e,crit.}^{high}$ is the higher critical density of a given diagnostic ratio.
With this definition, we adopt the following diagnostic ranges (see Table \ref{tab:em_characteristics} for $n_{e,crit}$): 
$\sim1\times10^2-5\times10^4$ cm$^{-3}$ for \den~[\ion{O}{2}],
$\sim2\times10^2-5\times10^4$ cm$^{-3}$ for \den~[\ion{S}{2}],
$\sim9\times10^3-1\times10^{10}$ cm$^{-3}$ for \den~\ion{C}{3}], and
$\sim2\times10^4-6\times10^{10}$ cm$^{-3}$ for \den~\ion{N}{4}].
Within the diagnostic range of a given line ratio, the ratio will change sensitively as a function of density, but outside of this range the diagnostic line ratio will flatten.
The so-called ``low-density limit" defines the upper diagnostic line ratio plateau (typically $F_1/F_2 \sim 1.3 - 1.6$) at densities less then $0.1\times n_{e,crit.}^{low}$ and the ``high-density limit" defines the lower diagnostic line ratio plateau (typically $F_1/F_2 \sim 0 - 0.5$) at densities greater then $10\times n_{e,crit.}^{high}$.
This means that low densities of $\lesssim10^3$ cm$^{-3}$ will never be measured with \CIII~or \NIV~and high densities of $\gtrsim10^5$ cm$^{-3}$ will never be measured with \OII~or \SII.

Considering only measurements in the diagnostic sensitivity ranges,
the densities derived from the intermediate- and high-ionization tracers (i.e., \NIV, \CIII) are often significantly higher than those derived from low-ionization lines (i.e., \SII), especially for the high-redshift galaxies.
This implies significantly different densities in the different ionization zones, where density increases with ionization.
Note that if the low- and high-ionization zone densities were similar in reality, at least one of diagnostic line ratio measurements would be in the low- or high-density limit.
Additionally, even the measurements in the sensitivity range for $n_{e,int}$ and $n_{e,high}$ should be considered upper limits, as the $n_e^2$-dependence of emission lines means that integrated flux will be dominated by regions of higher \den~in gas with an inhomogeneous \den~distribution.
\cite{harikane25} emphasizes the presence of multiphase gas in objects that are known to have high-density gas as traced by UV and optical emission line ratios.
The authors argue that detections of [\OIII~52$\mu$m and [\OIII~88$\mu$m, which both have relatively low $n_{e, crit}$, are only possible if these systems also contain low-density gas given that [\OIII~52$\mu$m and [\OIII~88$\mu$m would be completely suppressed in an entirely high-density nebula.

We explore the possibility of using [\ion{Ar}{4}] \W\W4713,4742 to probe $n_{e,high}$ given that this \den~diagnostic is securely sensitive above 1.73 $\times 10^3$ cm$^{-3}$.
However, for the seven \lz~galaxies for which we can derive $n_{e,high}$(Ar$^{+3}$), we note that only one of them is above the sensitivity limit for this diagnostic.
In cases where we can compare these \den~to $n_{e,high}$(N$^{+3}$) and $n_{e,int}$(C$^{+2}$), we also find that the derived $n_{e,high}$(Ar$^{+3}$) values are systematically lower.
The \CIII~and \NIV~ions require higher ionization potentials, excitation energies, and originate from levels with higher $n_{e, crit}$ than [\ion{Ar}{4}] (see Table \ref{tab:em_characteristics}).
The presence of all these emission lines suggest that the UV lines may be produced in different subregions of an inhomogeneous, non-uniformly mixed high-ionization zone than the optical [\ion{Ar}{4}] lines \citep[e.g.,][]{pascale23}.
Therefore, to probe the highest \den~regimes in a given ionization zone, we opt to exclusively use $n_{e,high}$(N$^{+3}$) and $n_{e,int}$(C$^{+2}$) in this work.

In the \lz~sample, we estimate that the average high-ionization density is $\sim$20 times higher than the low-ionization density and 3.45 times higher than the intermediate-ionization density, while in the \hz~sample we find that the average high-ionization density is $\sim$260 times higher than the low-ionization density and $\sim$2 times higher than the intermediate-ionization density.
This suggests a multiphase gas with a density gradient such that the higher-ionization zones have higher densities, consistent with previous $z \sim 0$ findings \citep[e.g.,][]{berg21,mingozzi22}, but with steeper gradients at higher redshifts.
Therefore, for objects that do not have either a \CIII~or \NIV~detection, we estimate the other from using the averages above.

We derive high-ionization zone temperatures ($T_{e,{high}}$) using an [\OIII~emission line ratio with a high-ionization zone density.
We use the [\OIII~\W4364/[\OIII~\W5008 for objects with $>3\sigma$ detections.
For the other galaxies, namely GN-z11, the Sunburst arc, J1723$+$3411, SL2S J0217, and the Lynx arc, we use the \OIII~\W1666/[\OIII~\W5008 ratio.
To infer low- and intermediate-ionization temperatures, we use the \Te-\Te~relations from \cite{garnett92}.
The resulting temperatures and densities are reported in Tables \ref{tab:Lz_abunds} and \ref{tab:Hz_abunds} in Appendix~\ref{Appendix:D}. 

In the left panel of Figure \ref{fig:compare_properties}, we compare the multi-phase versus constant \den~determinations of $T_{e,{high}}$ for our \lz~and \hz~samples.
It is clear that constraining $T_{e,{high}}$ with a low-ionization \den~probe leads to higher high-ionization zone temperatures on average, which in turn lead to lower metallicities.

\subsection{Ionization Parameters} \label{sec:ionpar}
We use the bicubic relationship presented in Section \ref{sec:NO_varyden} to derive $\log U_{int}$, which describes the intermediate ionization zone, for the objects in our sample that have O$_{32}$.
For the seven \lz~objects with detections of both \ArIII~and \ArIV, we use the Ar$_{43}$ diagnostic from \cite{berg21} to determine $\log U_{high}$.
Since the \hz~sample lacks Ar$_{43}$ measurements, instead, we use the N$_{43}$ fit from Table~\ref{tab:LU_fits} for the objects that have both \NIII~and \NIV~to determine $\log U_{high}$.
Note that the N$_{43}$ ratios for our \hz~sample are higher than the range of our photoionization models, suggesting very high ionization parameters of $\log U > -1.0$. 
However, it is difficult to disentangle whether the true $\log U$ is definitely higher or if \NIV~and \NIII~are being produced by different mechanisms.
The ability to independently derive $\log U_{high}$ from another high- to intermediate-ionization emission line ratio would be helpful in distinguishing which is the case.
This leaves a total of 5 objects across the \lz~and \hz~samples for which we cannot derive a $\log U_{high}$.
For both $\log U_{int}$ and $\log U_{high}$, we use the fits from the uniform low-\den~models that are described in detail in Section \ref{sec:const_den} with the $n_{e,high}$ values reported in Tables~\ref{tab:Lz_abunds} and ~\ref{tab:Hz_abunds} in Appendix~\ref{Appendix:D}.
Ionization parameters are also reported in Tables \ref{tab:Lz_abunds} and \ref{tab:Hz_abunds} in Appendix~\ref{Appendix:D}. 

\begin{figure*}[ht]
    \begin{center}
    \includegraphics[width=0.85\textwidth]{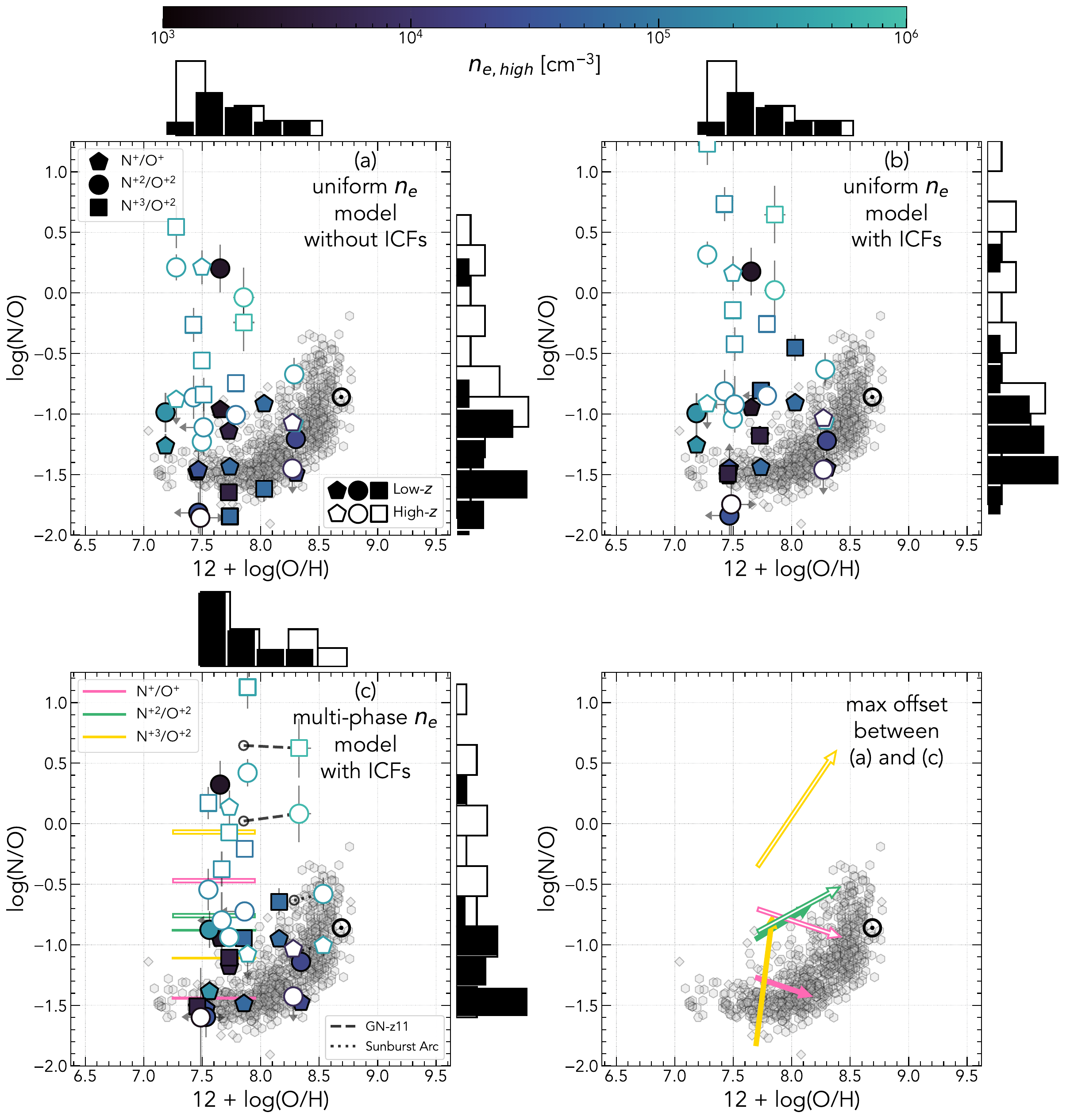}
    \caption{The gray scatter of points in both panels are the N/O-O/H scatter for local measurements of N/O in \ion{H}{2} regions \citep{pilyugin12, berg20, rogers21, rogers22}, low-mass metal-poor galaxies \citep{berg12, berg19, izotov23}, and CLASSY galaxies \citep{arellanocordova25b}.
    These measurements demonstrate clearly the expected evolution of N/O based on low-ionization optical emission.
    The various ICF corrected variations of N/O that we calculated are distinguished by marker shape.
    The pentagons represent \niioii, circles are \niiioiii, and squares show \nivoiii.
    The filled shapes are galaxies in our \lz~sample and the \hz~sample are plotted as unfilled shapes.
    The color bar gives the $n_{high}$ used in the calculations for the 3-zone \den~model.
    The histograms show the distributions of N/O and O/H values in each panel for the \lz and \hz, which are shown with black and white histograms, respectively.
    In each of these panels, the \lz~outlier at high-N/O is Mrk 996, which is being enriched by a large population of WN stars.
    {\bf Upper Left:} The uniform \den~model described in Section \ref{sec:implications_sample} without ICFs. There is a clear discrepancy between literature scatter and most of our \lz~and \hz~measurements.
    {\bf Upper Right:} We plot the uniform \den~model detailed in Section \ref{sec:implications_sample} with ICFs. We note that the discrepancy seen in the first panel is still present in the second one and is more severe for some points.
    {\bf Lower Left:} Our new physical conditions determinations, described in Section \ref{sec:implications_sample}, are used to calculate new direct gas-phase metallicities and N/O measurements, where possible, for both samples.
    Additionally, we include lines that demonstrate the median value for each type of N/O for $\rm 12 + \log(O/H) < 8.0$ in the \lz~(filled line) and \hz~(unfilled line) samples.
    These medians allow us to highlight that the \hz~sample has, on average, higher N/O than the \lz~sample suggesting a redshift evolution in the derived N/O for all tracers.
    Note, we mark with dashed and dotted lines the changes between (b) and (c) for two objects, GN-z11 and Sunburst Arc, which we discuss further in Section \ref{sec:Nevolution}.
    {\bf Lower Right:} We plot the maximum vector offsets for the each tracer of N/O for both the \lz~and \hz~sample between the uniform \den~model with no ICFs and our multiphase with ICFs \den~model.
    The \niioii~measurements on average shift down and to the right.
    The \niiioiii~measurements generally shift up and, or, to the right for our recalculated values.
    The \nivoiii~values change the most notably since these measurements require much larger ICFs across varying $\log U$.
    }
    \label{fig:NO_samp_build}
    \end{center}
\end{figure*}

\subsection{O/H and N/O Abundances} \label{sec:OH_NO_abund}
We follow the abundance methods described in Section~\ref{sec:abund_varyden}.
We use the \Te~and \den~calculated in Section \ref{sec:temden} to derive direct gas-phase abundances for our entire sample under two \den~assumptions.
For the uniform density model, we use the $n_{e,{low}}$ versions of $T_{e,{low}}$ to determine O$^{+}$, N$^{+}$, and N$^{+2}$ and $T_{e,{high}}$ to determine O$^{+2}$ and N$^{+3}$. 
For the multi-phase model, the intermediate-ionization  N$^{+2}$ ion is instead determined using the $n_{e,{int.}}$-derived $T_{e,{ int.}}$ and the high-ionization O$^{+2}$ and N$^{+3}$ ions are determined using the $n_{e,{high}}$-derived $T_{e,{high}}$.

The ionic oxygen abundances are added to determine the total oxygen abundance. 
In the absence of \OII~emission, the derived O abundance is a lower limit.
However, for weak detections of \OII~emission, we report O/H abundances as upper limits.
We determine the N/O abundance using the five ion ratios and their corresponding ICFs described in Section~\ref{sec:ICF_varyden}.
For the uniform density model, $\log U_{int.}$ is used for the ICFs.
However, for the multi-phase model, $\log U_{high}$ is used for the ICFs.
In the cases where $\log U_{high} = -1.0$ is calculated by N$_{43}$, the ICFs are near 1.0 for \niioii, between $\sim 1.1 - 1.6$ for \niiioiii, and at their least extreme for \nivoiii.
Given the high-ionization lines the true $\log U$ must be high, but if it is instead $-2.0 < \log U < -1.0$, the ICFs don't change significantly for \niioii, \niiioiii, and \nivoiii.
The total O/H and relative N/O abundances can be found in Tables \ref{tab:Lz_abunds} and \ref{tab:Hz_abunds} in Appendix~\ref{Appendix:D}.

The middle and right hand panels of Figure \ref{fig:compare_properties} compare the oxygen abundances and UV N/O abundances, respectively from the uniform density and multi-phase methods.
The uniform low density method tends to be biased towards lower oxygen abundances, which is a result of the temperatures being biased higher.
On the other hand, when comparing a uniform density to a multi-phase density framework, the derived UV N/O does not change significantly.
This is because the UV N and O emission lines have high critical densities and so are are relatively insensitive to density.
The optical N and O emission lines, on the other hand, have lower critical densities, making their ratio more sensitive to densities for $n_e > 10^3$ cm$^{-3}$ (see Figure~\ref{fig:NO_cloudy_emlines}).

\section{The Evolution of N/O} \label{sec:Nevolution}
Detailed chemical abundance studies of local \ion{H}{2} regions and galaxies have long used low-ionization optical emission to trace relative N/O abundances.
These studies have established the well-known bi-modal N/O--O/H relationship \citep[e.g.,][]{garnett90, henry00, vanzee06, nava06, pilyugin12, berg12, berg19, izotov23}.
At low metallicity ($12 + \log(\rm O/H) \lesssim$ 8.0), the N/O ratio is relatively constant due to the metallicity-independent (primary) production of both N and O. 
However, at higher metallicities ($12 + \log(\rm O/H) \gtrsim$ 8.0), the rate of N production through the metallicity-dependent (secondary) processes exceeds the rate at which O is being produced and the ratio increases \citep[e.g.,][]{edmunds78, henry00}.
There is also significant scatter in the N/O vs O/H relationship \citep[$\sigma \sim 0.5$ dex,][]{berg19} that is thought to result from various time-dependent mechanisms such as inflows of pristine gas, metal-rich outflows, and stellar yields \citep[][e.g., CCSNe, asymptotic giant branch]{perezmontero16,berg19,arellanocordova25b}.

While decades of work on low-ionization optical N/O abundances paints the familiar picture described above, how the N/O--O/H trend evolves with redshift is not well-known.
This is due, in part, to the fact that extending N/O studies to higher redshifts is difficult, as the optical lines are redshifted into the infrared.
The advent of the {\it JWST} has opened a new window onto high redshift abundance studies, with NIRSpec's sensitivity and rest-wavelength coverage ($0.6-5.3\ \mu$m) enabling \NII~\W6585 detections out to $z\sim7$.
For example, \citet{scholte25} recently presented a direct-method N/O--O/H analysis of 12 galaxies spanning $1.80 \lesssim z \lesssim 5.2$ \citep[see also][]{strom17,haydenpawson22,sanders23}, but larger studies are needed to assess the evolution statistically. 
Unfortunately, the optical \NII~line is typically weak in the metal-poor galaxies that are expected to be more common at high redshifts, and so has proven difficult to detect.
Alternatively, N/O abundances can be determined using the high-ionization rest-UV \NIV~\W\W1483,1487, \OIII~\W\W1661,1666, and \NIII~\W1750, when available.
However, recent \hz~studies find significantly elevated N/O derived from these high- and very-high-ionization UV emission lines that is in conflict with local studies.
Therefore, understanding the evolution of N/O requires robust and consistent measurements of N/O from both UV and optical lines.

The N/O--O/H trend is illustrated in Figure \ref{fig:NO_samp_build} for the uniform-density method without ICFs (upper left), uniform-density method with ICFs (upper right), mutli-phase method (lower left), and maximum vector offsets between  uniform-density method without ICF and the the multiphase method. 
The gray points are $z\sim0$ low-ionization optical N/O measurements \citep[data from][]{pilyugin12, berg12, berg19, berg20, rogers21, rogers22, izotov23, arellanocordova25b}.
In comparison, we plot the N/O abundances derived from \niioii~(pentagons), \niiioiii~(circles), and \nivoiii~(squares) for our sample, color-coded by their high-ionization zone densities. 

The \lz~sample values are fairly consistent for the uniform-density and multi-phase density panels with ICFs; however, several of the \hz~sample points change significantly owing to their higher $n_{e,high}$ values, which shift their O/H abundances to higher values in the multi-phase model.
As a result, the N/O enrichment above the $z\sim0$ trend is reduced for several of the galaxies in our sample.
For example, the Sunburst Arc's metallicity increases from a $12+\log({\rm O/H})=8.3$ to 8.5, shifting it to agree with the secondary N/O relationship. 
Further, using the mutli-phase density model for GN-z11 increases the O/H abundance by 0.47 dex and increases the N/O abundance from \niiioiii~by $\sim$0.10 dex, bringing the \niiioiii~and \nivoiii~abundances closer together and reducing the enhancement above the N/O--O/H trend to $\sim0.7$ dex.
However, we note that the lack of variation on the N/O-O/H plane does not imply a homogeneous \den~structure.
At \lz~the density inhomogeneities could be on the order of a few hundred cm$^{-3}$; this would not significantly impact the UV and optical emission lines examined here but would have a noticeable effect on infrared fine-structure lines (e.g., [\OIII~52$\mu$m, [\OIII~88$\mu$m, \NII~122$\mu$m, \NII~205$\mu$m).

While using the multi-phase density model produces more accurate N/O abundances that better agree with the expected N/O--O/H trend, significant outliers remain.
For the subset of low-metallicity galaxies in \lz~sample (12+log(O/H) $\leq 8.0$), we find that the median N/O abundance from \niioii~is $-1.44$ dex (green plateau in Figure \ref{fig:NO_samp_build}), consistent with decades of local studies examining the N/O plateau traced by low-ionization emission \citep[e.g.,][]{vanzee06, berg19}.
The UV N/O values measured from \niiioiii~and \nivoiii~for the same sample are higher, with median values of $-0.88$ and $-1.11$ dex (green and gold solid lines in Figure \ref{fig:NO_samp_build}), respectively.
This $\Delta {\rm N/O} = 0.33 - 0.56$ dex disagreement is unexpected, as different ionic abundance methods should converge on consistent elemental abundance ratios if properly calibrated. 
Thus, the unknown origin of this discrepancy raises concerns that the high-ionization UV N/O diagnostics may systematically overestimate N/O and calls into question the robustness of the high-ionization UV N/O measurements in high-redshift systems.

For the low-metallicity \hz~galaxies, the median N/O values are $-0.47, -0.76,$ and $-0.07$ dex from \niioii, \niiioiii, and \nivoiii, respectively (dashed lines in Figure \ref{fig:NO_samp_build}).
These values suggest that the UV-optical discrepancy only decreases for \niiioiii~and increases for \niiioiii~at higher-redshifts to $\Delta{\rm N/O} = -0.27$ and $0.40$ dex with higher N/O values regardless of the measurement method.
Note, however, that the \hz~\niioii~median value is only based on two measurements that are highly discrepant.
While the \hz~median \nivoiii-derived N/O value is consistent with the UV N-emitters in recent studies \citep[e.g.,][]{senchyna24, isobe23b}, there is significant scatter in the UV N/O measurements at \hz~such that the median value is close to solar, reducing its discrepancy.
Given these results, more robust N/O measurements from \niioii~are needed at high redshifts to understand the significance of any N/O evolution.

Recent work by \cite{hayes25} derives UV N/O from stacks of \hz~galaxy spectra from the {\it DAWN JWST Archive}\footnote{\url{https://dawn-cph.github.io/dja/}} (DJA).
The authors find that using a \den~$\sim 10^6$ cm$^{-3}$ in the abundance determinations for these stacks has more impact on the derived \Te, and subsequently the gas-phase metallicity, than the UV N/O.
However, this change is significant in that it brings the measured points in agreement with the high-metallicity increase in the optical N/O scatter on the N/O--O/H plane.
Additionally, this result is consistent with our findings in Figure \ref{fig:NO_samp_build}, where we determine that using the appropriate physical conditions can have a notable impact on the placement of points on the N/O--O/H plane overall.

If the N/O-enhancement of the \hz~galaxies is taken to be true, it would imply either earlier that expected star-formation such that significant AGB enrichment has occurred or a surprisingly fast enrichment of N \citep[e.g.,][]{kobayashi24, arellanocordova25a}.
Prompt enrichment of nitrogen could occur via very massive stars \citep[VMSs; $M_\star > 100 M_\odot$; e.g.,][]{nandal24a}, where rotational mixing enables hot-CNO cycling and early N production.
Then, the observed scatter in N/O at fixed O/H could be explained by the diversity of star formation modes and stellar populations in our total sample, with some galaxies enriched by early generations of VMSs and others dominated by delayed AGB enrichment.
 
\begin{figure*}[ht]
    \begin{center}
    \includegraphics[width=0.75\textwidth]{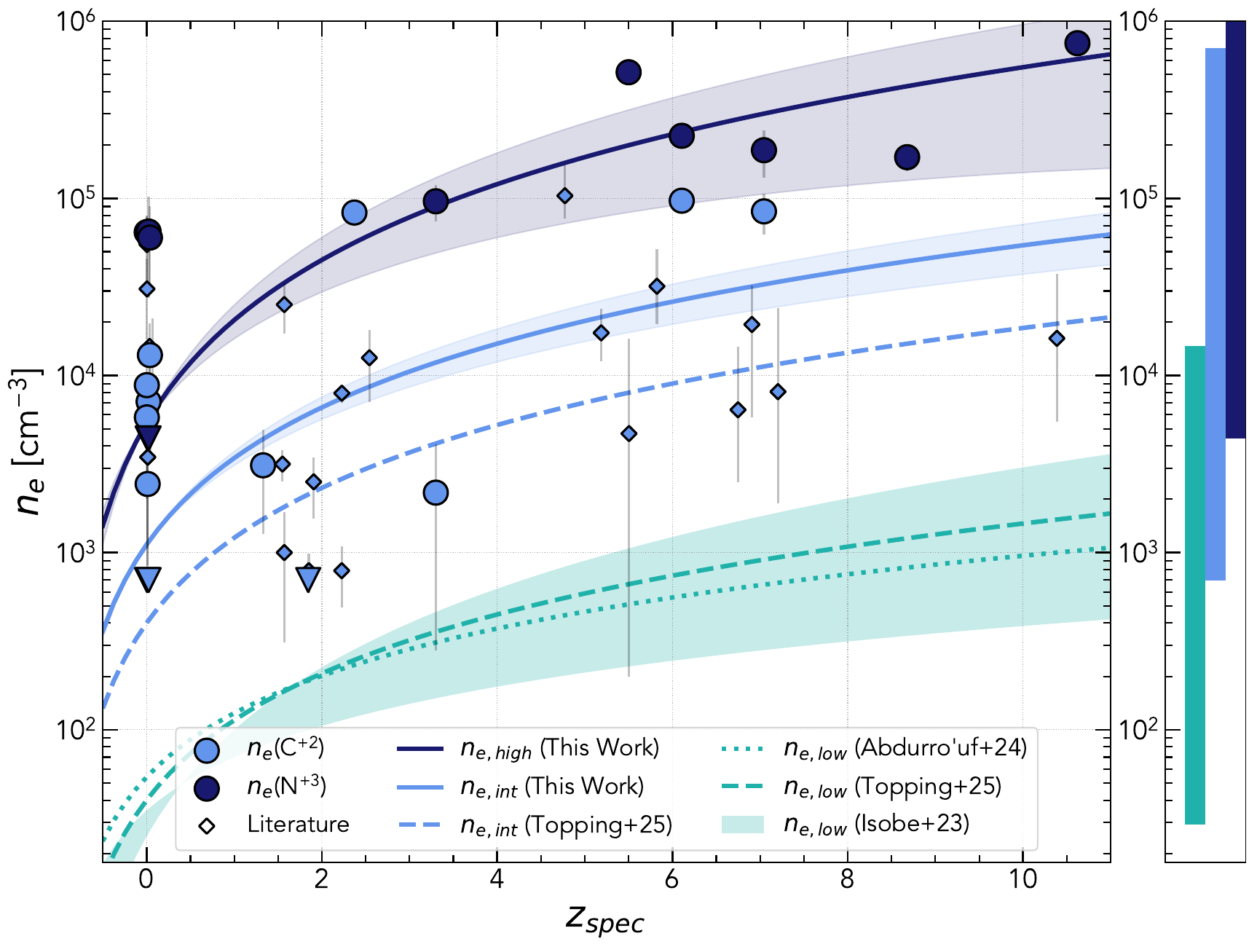}
    \caption{The redshift evolution of \den(C$^{+2}$) and \den(N$^{+3}$) for our \lz~and \hz~sample of galaxies.
    Our fits to this relationship for \den(C$^{+2}$) and \den(N$^{+3}$) are shown in solid light and dark blue lines, respectively.
    The light blue circles show \den(C$^{+2}$) values, while the dark blue circles are \den(N$^{+3}$) values.
    We include literature measurements of \den(C$^{+2}$) from \cite{maseda17}, \cite{berg19}, and \cite{topping25b} as small light blue diamonds.
    Downward pointing triangles of any color are showing where the derived \den~are upper limits.
    Additionally, we compare the $n_{e,low}-z$ trends derived by \cite{abdurrouf24} and \cite{topping25b} as dotted and dashed lines.
    In the right panel, we show the sensitivity ranges of the diagnostics used in deriving low-, intermediate-, and high-ionization \den.
    We show that the intermediate- and high-ionization \den~are higher than the low-ionization \den~across all redshifts, and the higher-ionization \den~diagnostics follow $\propto (1 + z)^{1.5 - 2.0}$, which similar to the trend for the low-ionization \den~redshift evolution.
    }
    \label{fig:ne_z_high}
    \end{center}
\end{figure*}

\subsection{Density Evolution} \label{sec:evolve_conditions}
We have shown that emission line ratios and abundances are sensitive to the derived physical conditions.
Previous works have also show that high-\den, in particular, can bias interpretations of observed galaxy properties \citep{katz23, cullen25, rusakov25}, while earlier studies of $z \sim 2 - 3$ galaxies have suggested that nebular density may evolve with redshift \citep[e.g.,][]{sanders16,kaasinen17}.
Thus, the observed high N/O abundances may also be related to the evolution of \den.

Figure \ref{fig:ne_z_high} presents the $z$ evolution of the $n_{e,low}$ (green), $n_{e, int}$ (light blue), and $n_{e, high}$ (dark blue).
From a cosmology perspective, the virial density of dark matter halos evolve according to $(1 + z)^3$.
It follows naturally that ISM conditions would scale similarly towards more extreme \den~with increasing $z$.
We fit the $z$ evolution of $n_{e, int}$ and $n_{e, high}$ using \den~$ = A(1 + z)^p$, where we fix $A$ to be the average density of the \lz~sample.
We find that the best-fit redshift evolution of the intermediate- and high-ionization densities for the combined \lz~and \hz~samples are given by
\begin{align}
    n_{e,int} &= 1.11 \times 10^3 \cdot  (1 + z)^{1.93\pm0.08}\label{eq:n_int} \\
    n_{e,high} &= 5.40 \times 10^3 \cdot (1 + z)^{1.62\pm0.12}.\label{eq:n_high}
\end{align}
In comparison, we plot the redshift-evolution trends for the low- and intermediate-ionization densities from \cite{isobe23a}, \cite{abdurrouf24}, and \cite{topping25b}.
The trends for the low-, intermediate-, and high-ionization densities have similar shapes, but offset to higher normalization with increasing ionization.
Notably, our $n_{e, int}$ trend is offset from the \cite{topping25b} relation due to our sample being generally more extreme and having higher \den.
These offset redshift evolutions of the multi-phase densities in our sample stresses the importance of accurately probing of the complete \den~structure of a galaxy especially in \hz~galaxies, where high densities are more likely to bias interpretations of galaxy properties.

While the virial density of dark matter halos increases as $(1 + z)^3$ in $\Lambda$CDM cosmology, electron density evolves as $(1 + z)^{1.93}$ for intermediate-ionization gas and $(1 + z)^{1.62}$ for high-ionization gas (Equations \ref{eq:n_int} and \ref{eq:n_high}). 
This suggests that the nebular gas density evolves more slowly than the virial density, where feedback, gas accretion, and ISM regulation can moderate how compact the star-forming gas becomes.
On the other hand, the ionized gas traced by emission lines is thought to trace a small, pressurized, ionized portion of the entire gas reservoir of a galaxy and this region may have a different scaling with redshift than the neutral gas, molecular gas, or dark matter.
Thus, measured densities may reflect more localized environmental conditions rather than global ISM properties  
and how star formation rate surface density, ionizing photon flux, and geometry evolve with redshift.
These findings imply that while cosmological density evolution sets the stage, internal galaxy physics moderate the conditions under which N enrichment proceeds.

\begin{figure}[ht]
    \begin{center}
    \includegraphics[width=0.45\textwidth]{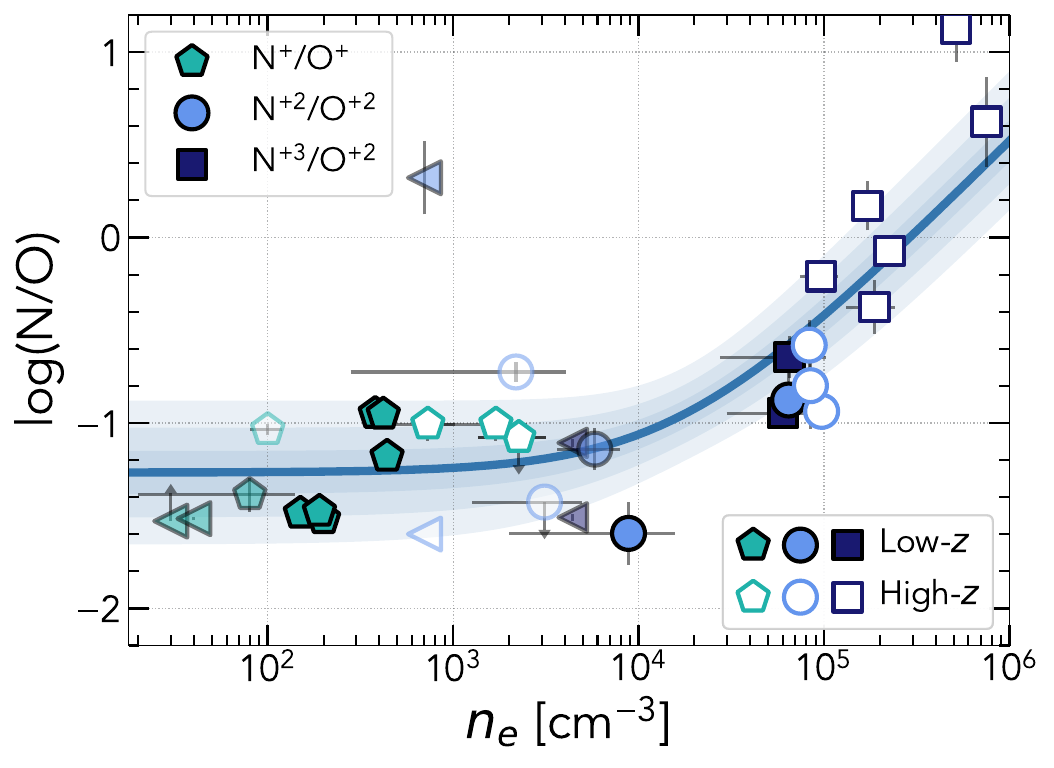}
    \caption{We explore the relationship between \den~and various tracers of N/O.
    The color and marker shape distinguish the various measured N/O, where green pentagons are the \niioii, light blue circles are \niiioiii, and dark blue squares are the high-ionization \nivoiii.
    Each N/O measurement is plotted against the \den~used to calculate the emissivity for the N ion utilized in each measurement (i.e., \niioii~vs. $n_{low}$, \niiioiii~vs. $n_{int}$, \nivoiii~vs. $n_{high}$).
    While filled points show galaxies in our \lz~sample, unfilled points are galaxies in our \hz~sample.
    As a precaution, we plot values that fall below the (conservative) sensitivity range as semi-transparent.
    Specifically, in cases where the diagnostic ratio was greater than the low-density limit, we plot the densities as triangles.
    It is therefore clear that we are robustly tracing different \den~in different ionization zones for the majority of our sample.
    We find that the overall increasing relationship between \den~and N/O in this work can be described by a logarithmic fit.
    Note the outlier in this trend is Mrk 996, which is known to have WR driven N-enhancement.
    }
    \label{fig:NO_ne}
    \end{center}
\end{figure}


\begin{figure*}[ht]
    \begin{center}
    \includegraphics[width=\textwidth]{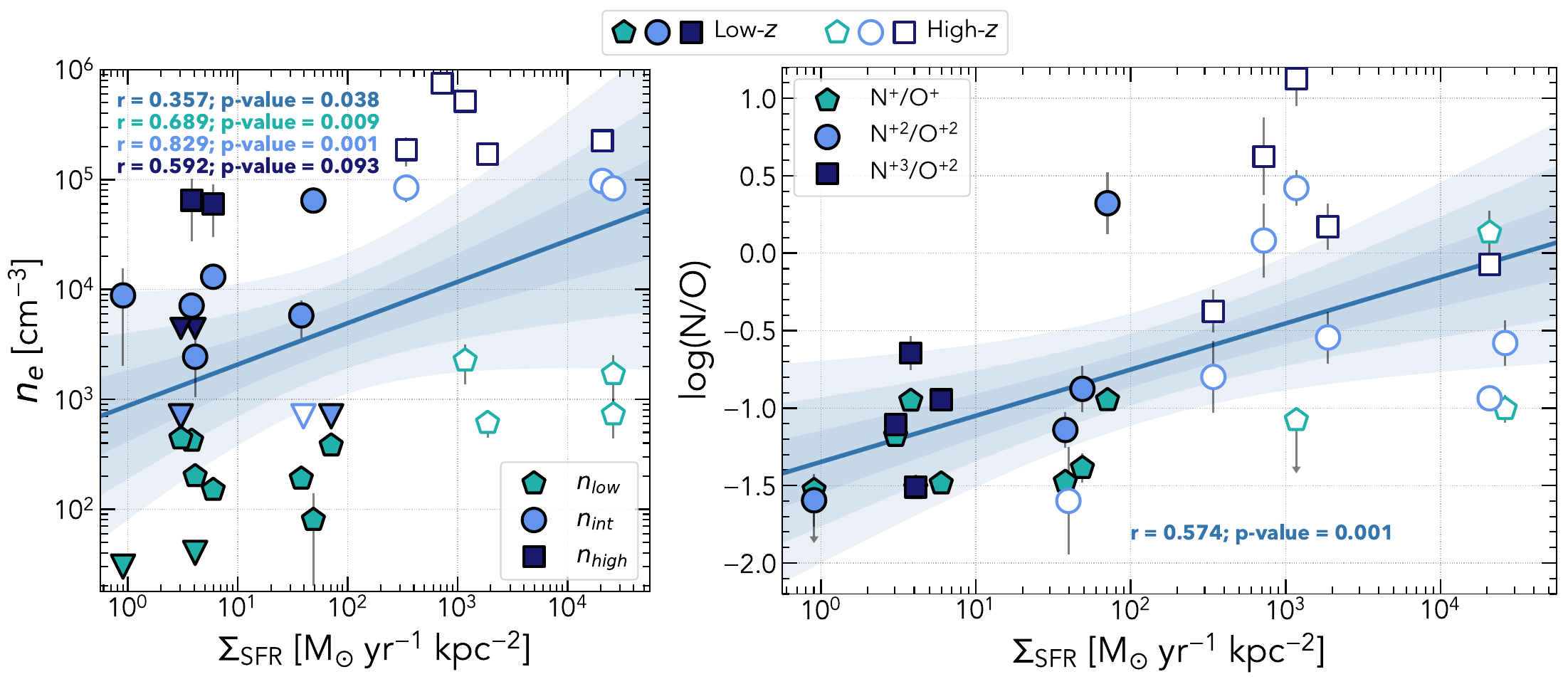}
    \caption{We demonstrate that for our sample the $\Sigma_{\rm SFR}$ correlates with both the \den~and the different tracers of N/O that we measured.
    We plot the linear best fit for the \den-$\Sigma_{\rm SFR}$ and N/O-$\Sigma_{\rm SFR}$ relationships with a solid line and show the 3$\sigma$ errors on the fit.
    Filled markers are galaxies in our \lz~sample, while unfilled markers are galaxies from our \hz~sample.
    The $\Sigma_{\rm SFR}$ clearly sorts all of our galaxies into the \lz~and \hz~samples since the \hz~sample has significantly higher $\Sigma_{\rm SFR}$ than the \lz~sample.
    {\bf Upper:} We present the \den-$\Sigma_{\rm SFR}$ for each measured \den~in our sample.
    The color and marker shape distinguish the various measured \den, where green pentagons are the low-ionization \den, light blue circles are the intermediate-ionization \den, and dark blue squares are the high-ionization \den.
    We show the Pearson's r correlation coefficient for all \den~and low-, intermediate-, and high-ionization \den~separately.
    Each of these p-values suggests a correlation between $\Sigma_{\rm SFR}$ and \den~except for the relationship between $\Sigma_{\rm SFR}$ and the intermediate-ionization \den.
    {\bf Lower:} We show the N/O-$\Sigma_{\rm SFR}$ relationship for every measure of N/O that we obtained in this work.
    In this plot, the color and marker shape distinguish the various measured N/O, where green pentagons are the \niioii, light blue circles are \niiioiii, and dark blue squares are the high-ionization \nivoiii.
    For N/O-$\Sigma_{\rm SFR}$, we only show the Pearson's r correlation coefficient between all the N/O tracers and the $\Sigma_{\rm SFR}$.
    We find that the best linear fit for N/O-$\Sigma_{\rm SFR}$ is not as steep as the relationship between \den-$\Sigma_{\rm SFR}$, but the correlation is tighter and less likely to be produced by two uncorrelated sets of data.
    Additionally, we note that the \lz~outlier at high-N/O is Mrk 996, which is known to be experiencing N-enhancement from WR stars.
    }
    \label{fig:Sig_SFR}
    \end{center}
\end{figure*}

\cite{arellanocordova25b} recently found an increasing trend between $n_{e,low}$ densities, probed by [\ion{S}{2}], and N/O values from \niioii~for both local and $z = 2 - 6$ galaxies. 
These authors suggest that this trend is driven by collisional de-excitation of [\ion{O}{2}] in the low-ionization gas.

To further explore the \den--N/O trend for all ionization zones of gas, we combine the density and N/O measurements in Figure \ref{fig:NO_ne} by plotting the N/O abundances derived from \niioii~(green pentagons), \niiioiii~(light blue circles), and \nivoiii~(dark blue squares) versus the electron density for the corresponding N ion.
Each of these \den~diagnostics have different ranges of sensitivity, and in Figure \ref{fig:NO_ne} we show which measurements fall below the lower bounds of a conservative sensitivity range for each diagnostic with semi-transparent points.
The \den~plotted as triangles are upper limits for cases where the diagnostic ratio was higher than the low-density limit, however, we emphasize that we only fit measurements that are within the conservative sensitivity range for each of the different densities.
We find that for $n_e \gtrsim 3\times10^3$ cm$^{-3}$, even after accounting for unseen ionic species and appropriate \den~structure, there is an increasing N/O trend with the \den~used to derive them.
This relationship is well fit by the same type of analytic function used to describe the scaling between O/H and N/O in \cite{groves04} and \cite{nicholls17}.
\begin{equation}\label{eq:NO_ne}
    \log({\rm N/O}) = \log(10^{-1.27\pm0.11} + 10^{\log(n_e) - (5.49\pm0.11)}),
\end{equation}
which shows that most of the extreme-N/O systems, as probed by high-ionization emission lines, also have the highest densities.
Similar trends are also seen for the low-, intermediate-, and high-ionization N/O--\den~relationships individually, suggesting that the extreme abundances in these galaxies may be driven by conditions that also drive extreme high-ionization densities.
The general trend of increasing N/O with gas density implies that denser, more pressurized environments favor efficient N enrichment, possibly through VMS formation or enhanced mixing.

Notably, the extreme \den~ and \Te~ observed in N/O enhanced galaxies at \hz~suggest a highly multi-phase ISM that may be the result of witnessing galaxies in the very early phase of an extreme burst of star formation such that thermal pressure equilibrium has not been reached.
We consider the possibility that high-ionization zones could be transiently over-dense pockets of gas, where the suppression of collisional coolants keeps the \Te~high, so these regions remain both hot ($2-2.5 \times 10^4$ K) and dense. 
These dense regions would be both dynamical and multi-phase, requiring a comprehensive picture of pressure balance to accurately model, including thermal, radiation, turbulent, ram, and magnetic pressures.
However, modeling of such a dynamic environment is outside the scope of this work.

\subsection{Relating Nebular Properties and Global \texorpdfstring{$\Sigma_{\rm SFR}$}{SSFR}} \label{sec:galprop}
Motivated by the strong correlations of N/O with density, we also examine the star formation rate surface density ($\Sigma_{\rm SFR}$).
We determine the $\Sigma_{\rm SFR}$ for our sample using the rest-UV effective radii ($R_e$) and SFRs presented in Table \ref{tab:samp_prop} along with the following equation:
\begin{equation}
    \Sigma_{\rm SFR} = \frac{\rm SFR}{2 \pi R_e^2}.
\end{equation}
The resulting $\Sigma_{\rm SFR}$ values are listed in Table \ref{tab:samp_prop}. 

\cite{topping25a} find that the \NIV~emission lines in \hz~galaxies occur only at the highest H$\beta$ equivalent widths pointing to very young stellar populations \citep{leitherer99, endsley23, matthee23} in a short-lived period during a burst of star formation. 
Additionally, \cite{schaerer24} specifically examines $\Sigma_{\rm SFR}$ in UV N-emitters and finds that these galaxies inhabit an extreme position on the $\Sigma_{\rm SFR}$--$\Sigma_{\rm M_{\star}}$ plane.
We note that, in general, spatial resolution limitations can make robust size measurements for \hz~galaxies challenging, particularly in cases where galaxies are not highly-magnified by lensing.
However, going forward, deep JWST imaging will be instrumental in reducing the scatter in the trends involving $\Sigma_{\rm SFR}$.

We examine this in our sample by looking at the \den--$\Sigma_{\rm SFR}$ relationship for the low- (green pentagons), intermediate- (light blue circles), and high-ionization (dark blue squares) zones in Figure \ref{fig:Sig_SFR}, which demonstrates that extreme densities are in fact related to the most extreme regions of SF in a galaxy.
We calculate the Pearson's r correlation coefficient along with their corresponding p-values for each set of \den~and for all of them individually.
Based on their p-values, each of the Pearson's r values suggest a significant linear correlation except the relationship between $n_{e,high}$ and $\Sigma_{\rm SFR}$.
We find a weak increasing trend between the derived \den~and the $\Sigma_{\rm SFR}$, which is consistent with the relationship found by previous works between $\Sigma_{\rm SFR}$ and $n_{e,low}$ \citep[e.g.][]{shimakawa15, davies21, topping25b}.
This is expected, as higher $\Sigma_{\rm SFR}$ corresponds to more intense and compact star formation, which increases ISM pressure and compresses gas, leading to higher local electron densities.
We fit a linear trend to the relationship between $\Sigma_{\rm SFR}$ and \den~that is best described by
\begin{equation} \label{eq:ne_Sig_SFR}
    \log(n_e) = (0.38\pm0.14) \times \log(\Sigma_{\rm SFR}) + (2.94\pm0.32).
\end{equation}
Additionally, our galaxy sample is divided such that the lower-\den~\lz~sample populates the trend at $\Sigma_{\rm SFR} < 100\ M_\odot\ {\rm yr}^{-1}\ {\rm kpc}^{-2}$ and the higher-density \hz~sample, mostly, populates the $\Sigma_{\rm SFR} > 100\ M_\odot\ {\rm yr}^{-1}\ {\rm kpc}^{-2}$ parameter space.
This is consistent with \cite{schaerer24}, who find that dense, compact, UV N-emitting \hz~galaxies can be distinguished by their extreme $\Sigma_{\rm SFR}$ and stellar mass surface densities.
These findings point to a physical link between local ISM structure and chemical enrichment and suggest that galaxies at earlier cosmic times, which exhibit higher values of $\Sigma_{SFR}$, may naturally reach temporarily elevated N/O levels through prompt enrichment channels.

In the lower panel of Figure \ref{fig:Sig_SFR}, we plot N/O abundances derived from UV and optical ion ratios relative to $\Sigma_{\rm SFR}$, finding a relatively tight positive correlation.
The best fit linear relationship between N/O and $\Sigma_{\rm SFR}$ is characterized by 
\begin{equation}\label{eq:Sig_SFR_NO}
    \log({\rm N/O}) = (0.30\pm0.08) \times \log(\Sigma_{\rm SFR}) - (1.35\pm0.20).
\end{equation}
While \cite{arellanocordova25b} finds a similarly positive, albeit weaker, trend when looking at N/O-$\Sigma_{\rm SFR}$ for the CLASSY galaxies, the authors show that the N/H-$\Sigma_{\rm SFR}$ relationship is much more significant.

Figure \ref{fig:Sig_SFR} shows that both \den~and N/O abundance increase with $\Sigma_{\rm SFR}$, indicating that compact, high-pressure star-forming regions are favorable sites for N enrichment. 
This picture suggests that N production is enhanced under the same conditions that generate intense, spatially concentrated star formation driven by strong bursts in very young stellar populations, which can be observationally distinguished by their large H$\beta$ equivalent widths. 
In such environments -- characterized by elevated densities and ionization parameters -- primary N from VMSs may be more efficiently produced and retained.

The extreme densities, star formation, and short-lived episodes of N-enhancement in these high$-z$ UV N-emitters make them tantalizing candidates for precursors to little red dots (LRDs).
Observationally, LRDs are distinguished by their v-shaped spectral energy distributions from imaging and significant broad emission components in their spectra.
Recent works find that the observational signatures that make LRDs so unique are consistent with dense gas enshrouding a massive black hole \citep{inayoshi25, naidu25, degraaff25}.
These black holes may be produced via direct collapse of VMSs or gravitational runaway in dense stellar environments \cite[e.g.,][]{greene20}, both of which could also produce the the elevated N/O in our high-$z$ sample prior to black hole formation.
Thus, we speculate that the dense gas enshrouded black holes that may be powering LRDs are remnants of stellar populations that existed in similar conditions as those that drove the elevated N/O in high-$z$ UV N emitters \citep[e.g.][]{marqueschaves24,topping25b}.

\subsection{The Source of N/O Enrichment} \label{sec:NOdata}
Although we outline a best-method approach for deriving N/O abundances from high-ionization UV N-emission in Section \ref{sec:den_effects}, these measurements are still at odds with the low-ionization optical measurements of N/O, especially at \hz.
While we acknowledge that poor or out-of-date atomic data could potentially drive this discrepancy, the atomic data would require significant changes to resolve the discrepancy between the UV and optical measurements, which seems unlikely.
Therefore, our work leads us to conclude that N/O is in fact elevated in the small fraction of \hz~systems with detected \NIII~and/or \NIV~emission lines, but the source of this N-enhancement at \hz~is still up for debate.

Since the discovery of GN-z11 \citep{bunker23}, one of the now many anomalous UV N-emitters, many works have sought to understand and characterize its unusual UV N abundance pattern.
\cite{charbonnel23} suggests that \hz~UV N-emitters are precursors to globular clusters (GCs).
Over the last several decades, studies of GCs have shown that they generally consist of two chemically-distinct populations \cite[e.g.,][]{sneden91, milone22}.
The first population (1P) shares similar light-element abundance patterns to Milky Way field stars and were born earlier, while the second population (2P) is notably enhanced in He, N, Al, and Na and depleted in O, C, and Mg compared to the field population \cite[e.g.,][]{kraft94, gratton19}.
While Mg and Si emission are sometimes observed for star forming galaxies, they vary significantly between one GC and another.
Alternatively, Al and Na could provide more secure identification of GC progenitors, but are rarely observed and so require very high S/N spectra to determine their abundances.
In the scenario presented by \cite{charbonnel23}, proto-globular clusters are polluted by super massive stars (SMSs; $M_\star > 1000\ M_\odot$) producing the abundance patterns seen in GN-z11 and other UV N-emitters.
This work also shows that GN-z11 is compatible with the stellar mass density and ionized gas mass implied by SMS enrichment, but further evidence is needed to confirm the proto-GC scenario.

Young massive stellar populations provide another possible path to N/O enrichment.
\cite{kobayashi24} used chemical evolution modeling to demonstrate that multiple bursts of star formation, interspersed with quiescent periods, allows rapid enrichment from Wolf-Rayet stars (WRs) that can reproduce the abundance pattern observed in GN-z11.
Further, \cite{senchyna24} found that the UV \NIII~profile of GN-z11 is notably similar to the $z \sim 0$ galaxy Mrk 996 that is known to host WRs based on their detailed optical spectra.
While GN-z11 is unlikely to be dominated by classical WRs, the authors argue that high-density massive star formation serves as the primary enrichment source for N-enhanced systems.
However, spectral signatures confirming WRs are often not detected or difficult to distinguish from VMSs \citep[e.g.,][]{martins23,berg24}.
Alternatively, \cite{nandal25} found that the N/O, C/O, and Ne/O abundances in GS 3037, another remarkably N-rich system, can only be explained using the theoretical yields from SMSs.
Further, \cite{nandal24b} find that, at extremely low metallicities ($Z = 10^{-5} Z_{\odot}$), the abundance patterns seen in GN-z11 can also be reproduced using a set of moderately-to-rapidly rotating MS stellar evolution models.

These previous works, taken together with our results, suggest prompt enrichment from VMSs is the most plausible explanation for the \hz~N-emitting galaxies, but further evidence is needed to confirm this theory. 
We may also be observing multiple mechanisms at play in the UV N-emitting galaxies and disentangling them with the currently available data is not yet possible.
Progress will require deep, high-resolution, multi-wavelength spectroscopy to better characterize the physical conditions, abundance patterns, kinematics, and more to constrain the N enrichment mechanisms in \hz~galaxies.

\section{Summary and Conclusions} \label{sec:conclude}
Recent JWST studies indicate that electron density (\den) evolves with redshift, especially for strong UV N-emitting galaxies. 
To assess the effects of \den~on the determination and interpretation of nebular properties and O/H and N/O abundances, we compiled a sample of \lz~($z \sim 0$) and \hz~($1 < z < 11$) UV N-emitting galaxies, probing a wide range in redshift, metallicity, and physical conditions.
By leveraging both new and archival observations from HST and JWST for this sample, we measured \den~in the low-, intermediate-, and high-ionization gas phases and found that they support a multi-phase density model, where higher-ionization gas is denser, reaching \den~of almost $10^6$ cm$^{-3}$.
Therefore, we created a custom suite of \texttt{Cloudy} photoionization models with densities spanning $n_e = 10^2 - 10^9$ cm$^{-3}$ to investigate their effects on nebular UV and optical diagnostics. 
To do so, we also created the first high-ionization $\log U_{high}$ diagnostic using the UV N$_{43}$ (= \NIV~\W\W1483,1487/\NIII~\W1750) ratio and N/O ionization correction factors (ICFs), for both UV and optical line ratios, that are calibrated up to $n_e = 10^6$ cm$^{-3}$, and so are appropriate for galaxies at all redshifts.

We showed that assuming a uniform, low-density model for a multi-phase density object results in significant biases.
The most notable effects are for nebular diagnostics using optical emission lines ratios with differing critical densities such as \OII~\W\W3727,3730/H$\beta$, [\OIII~\W5008/H$\beta$, [\OIII~\W4364/[\OIII~\W5008, and [\OIII~\W5008/\OII~\W\W3727,3730.
Consequently, assuming the uniform low-density limit ($n_e = 10^2$ cm$^{-3}$) for a high-density gas ($n_e = 10^5$ cm$^{-3}$) results in overestimated $\log U$ values by 1.13 dex, overestimated direct \Te~measurements by 1800 K, and underestimated direct 12+log(O/H) measurements by 0.67 dex.
Alternatively, diagnostics using only emission lines with high critical densities, such as UV lines, are insensitive to density over the observed \den~range, and so the diagnostics like the [\OIII~\W4363/\OIII~\W1666 \Te~diagnostic and the N$_{43}$ $\log U_{high}$ diagnostic are much more robust. 
Therefore, we emphasize the importance of adopting a mutli-phase gas model, when possible, for determining nebular gas properties and abundances, especially at \hz~where conditions may be more extreme than at \lz.
Importantly, in this work, we have assumed the ionizing source to be stellar populations, but in future work we will examine the source of ionizing radiation and it's potential role in producing the extreme objects that feature in this work.

Adopting the multi-phase abundance model for the \lz~and \hz~samples, we determined O/H and N/O abundances using appropriate electron temperatures and densities for each ionization zone and the new ICFs presented here.
For N/O, we considered five different UV and optical emission line ratio combinations that were used to determine the \niioii, \niiioiii, \nivoiii, \noUVopt, and \niiinivoiii~ion ratios.
We find that the effects of \den~are generally small for all five N/O abundance methods compared to that of O/H abundances.
While the high-ionization UV N/O diagnostics seem to be robust against \den~effects, the \niiioiii~and \nivoiii~UV tracers systematically overestimate N/O by $\sim 0.3 - 0.4$ dex compared to the low-ionization optical benchmark \niiioiii, particularly in low-metallicity systems at $z\sim0$. 
While the multi-phase density structure substantially improved the consistency among methods over the uniform density model, systematic differences remain. 
This suggests that high-ionization N/O ratios must be interpreted with caution, especially for galaxies observed only in the rest-frame UV with JWST.

Using these diagnostics, we investigated the physical drivers of N/O enrichment. 
We find that, for the \lz~sample, N/O is consistent with the well-established primary plateau at low metallicities and increasing secondary N/O trend at higher O/H. 
However, we also observe significant scatter at low O/H, including a population of galaxies with elevated N/O that deviate from the canonical plateau, and stronger deviation for the high-ionization UV-based N/O values. 
The \hz~sample shows even stronger enhancements of N/O.
While the multi-phase model did reduce this discrepancy and move some \hz~objects into agreement with the N/O--O/H trend, the significance of the offset in the remaining objects suggests a real redshift evolution in N/O for this sample.

Examining N/O as a function of \den~showed that electron density plays a critical role in shaping N/O for $n_e \gtrsim 3 \times 10^3$ cm$^{-3}$. 
Galaxies with higher gas densities exhibit higher N/O values, indicating that nitrogen production is enhanced in dense, high-pressure environments. 
While the virial density of dark matter halos increases steeply with redshift ($\propto (1 + z)^3$), we find that the electron densities of the ionized gas scale more gradually, evolving as $\propto (1 + z)^{1.9}$ for intermediate-ionization gas and $\propto (1 + z)^{1.6}$ for high-ionization gas. 
This suggests that while cosmic structure sets the overall density backdrop, baryonic feedback and gas regulation mediate how nebular gas responds to the cosmological environment.

The strongest link between local conditions and enrichment is revealed in the positive correlation between N/O and star formation rate surface density ($\Sigma_{\rm SFR}$).
This trend indicates that compact, intense star formation -- not just integrated star formation history -- drives enhanced nitrogen abundances, which is consistent with recent studies examining the properties of UV N emitters \citep[e.g.,][]{topping24, schaerer24}. 
In high-$\Sigma_{\rm SFR}$ systems, the ISM is more likely to reach the pressures and densities needed to form very massive stars, which can produce substantial primary nitrogen on short timescales through rotationally mixed winds or chemically homogeneous evolution. 
These environments may also facilitate more efficient retention of metals, enhancing localized enrichment. 
Together, the density and $\Sigma_{\rm SFR}$ correlations point to a physical enrichment channel that enables prompt N production tied to the geometry, intensity, and pressure of star formation itself.

As a whole, our findings argue for more detailed determinations physical conditions and abundances and a more nuanced view of chemical evolution in galaxies at \hz. 
N enrichment is not governed solely by metallicity or time since star formation onset but is instead shaped by the local conditions under which stars form and evolve. 
This has important implications for interpreting galaxies in the early universe, where high gas densities and elevated $\Sigma_{\rm SFR}$ are more common. 
As JWST continues to expand the frontier of UV-based abundance studies of galaxies in the reionization era, more deep, high spectral- and spatial-resolution observations are needed to better understand how density and star formation geometry influence chemical yields and interpret the origins and evolution of elemental abundances in the first billion years.

\begin{acknowledgements}
We are grateful for the public data that made this work possible from the HST programs HST-GO-15967 and HST-GO-15840 that were provided by NASA through a grant from the Space Telescope Science Institute, which is operated by the Associations of Universities for Research in Astronomy, Incorporated, under NASA contract NAS5-26555. 
This work is based, in part, on observations made with the NASA/ESA/CSA James Webb Space Telescope.
We also greatly appreciate the referee for providing thorough feedback that served to notably improve the clarity of our work.

The {\it HST} data used in this paper for Leo P can be found in MAST: 
\dataset[10.17909/kze6-rr69]{http://dx.doi.org/10.17909/kze6-rr69}.
\end{acknowledgements}

\facilities{
HST (COS),
JWST (NIRSpec)}
\software{
\texttt{Cloudy} version 23.01 \citep{chatzikos23, gunasekera23},
\texttt{astropy} \citep{astropy13, astropy18, astropy22},
\texttt{jupyter} \citep{kluyver16},
\texttt{LMFIT} version 1.2.2 \citep{newville23},
\texttt{PyNeb} version 1.1.18 \citep{luridiana15},
\texttt{BPASS} version 2.14 \citep{eldridge17},
\texttt{numpy} version 1.24.3 \citep{harris20},
\texttt{scipy} version 1.13.1 \citep{virtanen20}}

\bibliographystyle{aasjournal}
\bibliography{refs}

\appendix
\section{Rest-UV Leo P Spectrum}\label{Appendix:A}
\begin{figure*}[htp]
    \begin{center}
    \includegraphics[width=0.85\textwidth]{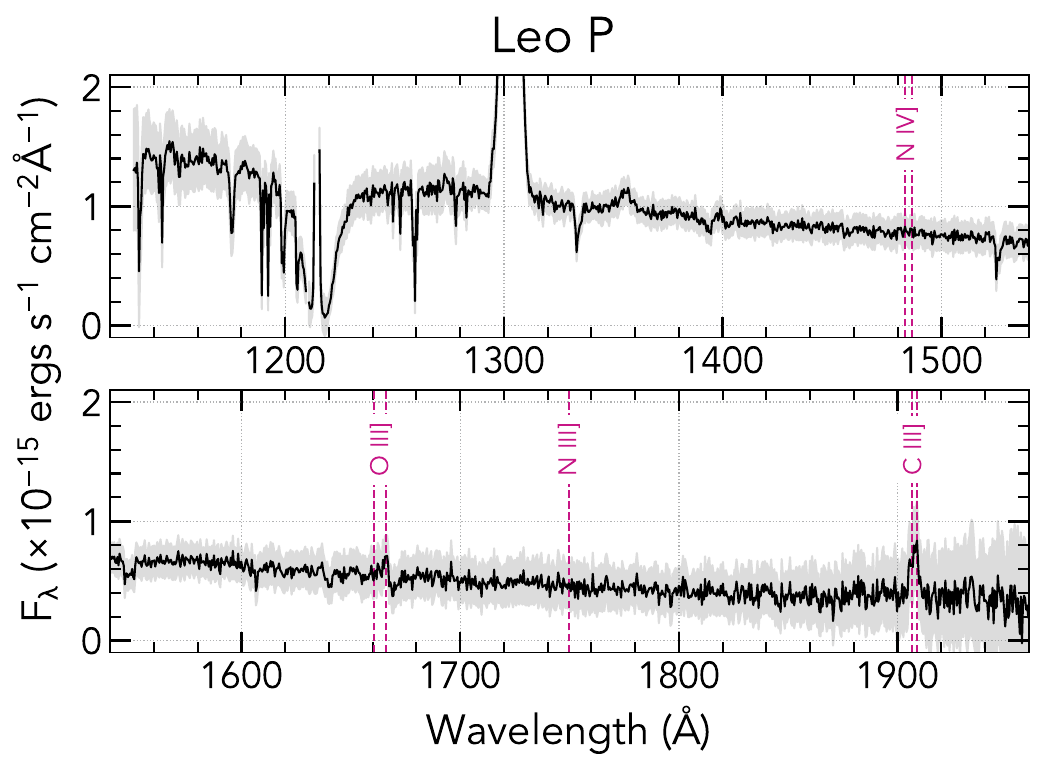}
    \caption{Final co-added rest-UV spectrum for the Leo P.
    We combined the high-resolution G130M and G160M gratings from \cite{telford23} and the G140L grating from HST-GO-17102 to obtain wavelength coverage of all key emission lines used in this study. 
    The pink dashed lines highlight the location of these key emission lines in the rest-UV spectrum of Leo P.}
    \label{fig:LeoP_spec}
    \end{center}
\end{figure*}

We present the co-added spectrum of the local extremely metal-poor galaxy Leo P \citep{skillman13}.
This process is described briefly in Section \ref{sec:lzsamp}.
We followed the steps outlined in \cite{berg22} in order to combine the high-resolution grating from \cite{telford23} and the low-resolution grating obtained in HST-GO-17102.
This final spectrum can be found in \ref{fig:LeoP_spec}, and we label the crucial emission lines in this spectrum that were used in this work.

\clearpage

\section{Emission Line Measurements}\label{Appendix:B}
We present the emission line measurements for both the \lz~and \hz~samples presented in this work.
We discuss the fitting procedure for emission lines that we measure in this work and the selection requirement for the emission lines taken from literature sources in Section \ref{sec:emlinemeasure}.
The dereddened relative intensities utilized in this work are presented in Tables \ref{tab:Lz_ELs} and \ref{tab:Hz_ELs}.
All optical emission lines are dereddened according to \cite{cardelli89}, while the UV emission lines are dereddened using \cite{reddy16}.
For the \lz~sample, we present our rest-UV intensities relative to $I$(\ion{C}{3}]) and our rest-optical intensities relative to $I(\rm H\beta)$.
However, in our \hz~sample all of the intensities that we present are given relative to $I(\rm H\beta)$.

\begin{deluxetable}{rcccccccc}[ht]
\setlength{\tabcolsep}{2.5pt}
\tabletypesize{\scriptsize}
\tablewidth{0pt}
\tablecaption{Rest UV+Optical Emission Lines for the Low$-z$ Sample\label{tab:Lz_ELs}}
\tablehead{
\CH{}                    & \CH{J0127-} & \CH{J1314+} & \CH{J1444+} & \CH{}       & \CH{J1253-} & \CH{J1044+} & \CH{J1323-} & \CH{J1545+} \\[-3.5ex]
\CH{Line}                & \CH{0619}   & \CH{3452}   & \CH{4237}   & \CH{Leo P}  & \CH{0312}   & \CH{0353}   & \CH{0132}   & \CH{0858}
}

\startdata
\multicolumn{9}{c}{$I$(\W)/$I$(\ion{C}{3}])} \\
\hline
\ion{N}{4}] \W1483.32     & \ldots       & \ldots       & \ldots       & \ldots        & 1.10\p0.35   & 1.01\p0.17   & 4.19\p0.64   & 1.25\p0.32   \\
\ion{N}{4}] \W1486.50     & \ldots       & \ldots       & \ldots       & \ldots        & 1.29\p0.36   & 0.53\p0.16   & 2.79\p0.57   & 1.42\p0.35   \\
\ion{O}{3}] \W1661.81     & 1.05\p1.06   & 4.75\p1.11   & 12.51\p2.54  & 8.36\p1.53    & 5.10\p0.66   & 17.08\p0.77  & 20.46\p2.08  & 13.08\p0.97  \\
\ion{O}{3}] \W1666.15     & 5.62\p1.19   & \ldots       & \ldots       & 15.51\p2.51   & 16.62\p0.92  & 39.73\p1.74  & 49.89\p4.88  & 30.59\p1.68  \\
\ion{N}{3}] \W1750.00\dg  & 42.92\p2.48  & 4.07\p1.06   & 2.29\p0.90   & 7.69\p2.49    & \ldots       & \ldots       & \ldots       & \ldots       \\
{[\ion{C}{3}]} \W1906.68  & 61.72\p3.02  & 57.17\p4.85  & 55.63\p15.89 & 37.15\p8.54   & 56.47\p3.43  & 59.00\p4.11  & 62.68\p9.50  & 53.60\p4.32  \\
\ion{C}{3}] \W1908.73     & 38.28\p2.51  & 42.83\p4.33  & 44.37\p14.22 & 62.85\p11.15  & 43.53\p3.13  & 41.00\p3.42  & 37.32\p7.33  & 46.40\p4.00  \\
\hline
\multicolumn{9}{c}{$I$(\W)/$I$(H$\beta$)} \\
\hline
{[\ion{O}{2}]} \W3727.10  & \ldots       & \ldots       & \ldots       & 46.51\p3.93   & 89.93\p0.89  & 9.60\p0.18  & 18.66\p0.16   & 63.90\p0.10  \\
{[\ion{O}{2}]} \W3729.88  & \ldots       & \ldots       & \ldots       & \ldots        & \ldots       & 16.77\p0.29  & \ldots       & \ldots       \\
H$\gamma$ \W4341.68       & 42.78\p0.65  & 52.29\p0.38  & 45.77\p1.27  & 45.19\p1.44   & 47.41\p0.45  & 46.55\p0.83  & 46.62\p0.11  & 47.17\p0.05  \\
{[\ion{O}{3}]} \W4364.44  & 9.15\p0.34   & 5.46\p0.10   & 6.49\p0.96   & 3.80\p0.51    & 11.15\p0.13  & 13.51\p0.25  & 20.35\p0.08  & 13.42\p0.04  \\
{[\ion{Ar}{4}]} \W4712.67 & \ldots       & 0.47\p0.05   & \ldots       & 0.94\p0.38    & 1.74\p0.05   & 164.99\p6.24 & 4.05\p0.06   & 1.42\p0.03   \\
{[\ion{Ar}{4}]} \W4741.53 & \ldots       & 0.32\p0.04   & $<$2.49      & 0.08\p0.02    & 1.35\p0.05   & 115.99\p2.34 & 3.07\p0.06   & 1.26\p0.03   \\
H$\beta$ \W4862.68        & 100.00\p1.11 & 100.00\p0.61 & 100.00\p1.66 & 100.00\p2.91  & 100.00\p0.81 & 100.00\p1.49 & 100.00\p0.16 & 100.00\p0.08 \\
{[\ion{O}{3}]} \W5008.24  & 353.51\p2.96 & 582.27\p3.32 & 241.95\p3.21 & 145.26\p4.31  & 683.70\p8.85 & 427.48\p6.22 & 720.74\p0.83 & 559.06\p0.32 \\
H$\alpha$ \W6564.61       & 340.24\p4.39 & 282.45\p1.65 & 280.59\p3.59 & 274.92\p11.06 & 277.80\p2.25 & 296.70\p5.23 & 279.63\p0.32 & 265.06\p0.15 \\
{[\ion{N}{2}]} \W6585.27  & 22.10\p0.58  & 10.28\p0.07  & 4.58\p0.65   & 2.51\p0.50    & 14.28\p0.13  & 0.81\p0.01   & 1.39\p0.01   & 2.43\p0.01   \\
{[\ion{S}{2}]} \W6718.29  & 18.72\p0.20  & 12.88\p0.09  & 13.69\p0.69  & 3.60\p0.31    & 7.91\p0.07   & 2.50\p0.05   & 1.79\p0.02   & 5.54\p0.01   \\
{[\ion{S}{2}]} \W6732.67  & 17.07\p0.18  & 10.48\p0.08  & 9.01\p0.66   & 2.70\p0.31    & 7.35\p0.08   & 2.04\p0.05   & 1.68\p0.01   & 4.37\p0.01   \\
{[\ion{Ar}{3}]} \W7137.77 & 11.02\p0.17  & 9.04\p0.07   & 2.86\p0.62   & 2.60\p0.21    & 7.91\p0.07   & 241.99\p9.35 & 3.11\p0.02   & 4.07\p0.01   \\
{[\ion{O}{2}]} \W7320.94  & 4.20\p0.17   & 2.35\p0.04   & $<$2.03      & 0.60\p0.20    & \ldots       & 0.60\p0.02   & \ldots       & \ldots       \\
{[\ion{O}{2}]} \W7331.68  & 3.54\p0.17   & 2.02\p0.03   & $<$2.10      & 0.70\p0.20    & \ldots       & 0.48\p0.07   & \ldots       & \ldots       \\
\hline
$E(B-V)$                  & 0.298        & 0.140        & 0.081        & 0.063         & 0.158        & 0.039        & 0.128        & 0.110        \\
$F_{\text{\ion{C}{3}}]}$  & 70.68\p2.24  & 21.38\p1.13  & 15.39\p2.67  & 1.75\p0.20    & 89.15\p3.37  & 27.28\p1.19  & 11.92\p1.16  & 33.44\p1.61  \\
$F_{{\rm H}\beta}$        & 423.38\p3.33 & 82.79\p0.36  & 2.47\p0.03   & 3.40\p0.07    & 189.77\p1.09 & 95.20\p1.00  & 10.38\p0.01  & 32.48\p0.0.  \\
\enddata
\tablecomments{
Reddening-corrected emission-line intensities adopted for the Low-$z$ Sample. 
The UV and optical fluxes are given relative to $F_{\text{\ion{C}{3}]}}\times100$ and $F_{{\rm H}\beta}\times100$, respectively. 
The last three rows give the dust extinction using the \citet{cardelli89} reddening law and the observed \ion{C}{3}] and H$\beta$ fluxes, 
in units of $10^{-15}$ erg s$^{-1}$ cm$^{-2}$. \\
\dg\ion{N}{3}] \W1750 is the integrated flux of the \ion{N}{3}] \W\W1746,1748,1749,1752,1754 quintuplet.\\
}
\end{deluxetable}

\begin{deluxetable}{rccccccccc}[hb]
\setlength{\tabcolsep}{2.5pt}
\tabletypesize{\scriptsize}
\tablewidth{0pt}
\tablecaption{Rest UV+Optical Emission Lines for the High$-z$ Sample\label{tab:Hz_ELs}}
\tablehead{
\CH{}                    & \CH{RXC}      & \CH{}       & \CH{GDS}    & \CH{}         & \CH{J1723+} & \CH{CEERS}  & \CH{SL2S}   & \CH{}       & \CH{A1703-} \\ [-3.5ex]
\CH{Line}                & \CH{J2248-ID} & \CH{GN-z11} & \CH{3073}   & \CH{Sunburst} & \CH{3411}   & \CH{1019}   & \CH{J0217}  & \CH{Lynx}   & \CH{zd6}
}
\startdata
\multicolumn{10}{c}{$I$(\W)/$I$(H$\beta$)} \\
\hline
\ion{N}{4}] \W1483.32    & 33.9\p 1.6  &  7.3\p 1.4   & 52.6\p 9.3  & \ldots      & \ldots      & 103.2\p19.6 & \ldots      & 17.2\p 2.8  & 17.7\p 5.9  \\
\ion{N}{4}] \W1486.50    & 81.9\p 2.4  & 47.5\p 9.2   & 245.9\p10.6 & \ldots      & \ldots      & 206.1\p26.8 & \ldots      & 24.4\p 2.9  & 37.6\p 5.1  \\
\ion{O}{3}] \W1666.15    & 66.1\p 1.3  & 16.1\p 6.2   & 13.5\p 3.2  &  5.8\p 0.6  &  3.9\p 0.8  & 89.2\p19.7  & 11.1\p 0.4  & 37.8\p 3.0  & 60.6\p 6.0  \\
\ion{N}{3}] \W1750.00\dg & 17.7\p 1.6  & 74.2\p13.3   & 99.1\p 3.9  &  6.8\p 1.0  &  $<$0.8     & 54.0\p22.9  &  0.8\p 0.2  & 17.8\p 2.8  & 21.4\p 6.8  \\
{[\ion{C}{3}]} \W1906.68 & 27.2\p 1.2  & \ldots       & \ldots      &  7.3\p 0.5  & 13.2\p 1.1  & \ldots      & 15.3\p 0.5  & 35.0\p 2.9  & 17.7\p 4.6  \\
\ion{C}{3}] \W1908.73    & 60.2\p 1.6  & 56.4\p 9.1   & 94.1\p 1.9  & 14.5\p 1.2  &  9.3\p 0.8  & 170.6\p30.8 & 10.1\p 0.3  & 24.2\p 2.9  & 35.8\p 5.7  \\
{[\ion{O}{2}]} \W3727.10 & \ldots      & \ldots       & \ldots      & 34.5\p 6.0  & 55.9\p 2.1  & 33.9\p 4.7  & \ldots      & \ldots      & \ldots      \\
{[\ion{O}{2}]} \W3729.88 &  4.3\p 1.6  & 47.8\p 7.8   & 21.8\p 0.6  & 25.7\p 6.7  & 76.9\p 2.6  & 34.6\p 4.8  & \ldots      & $<$25.0     & $<$17.7     \\
H$\gamma$ \W4341.68      & 47.3\p 0.5  & 47.3\p 9.1   & 47.3\p 0.4  & 47.3\p 0.7  & 47.3\p 1.9  & 54.7\p11.4  & 47.3\p 1.9  & 47.3\p 1.9  & 47.3\p 2.7  \\
{[\ion{O}{3}]} \W4364.44 & 42.1\p 0.9  & 14.0\p 7.8   & 15.1\p 0.6  & 11.6\p 6.4  &  2.0\p 4.3  & 20.8\p 6.3  &  6.7\p 4.1  & \ldots      & 34.6\p 3.4  \\
H$\beta$ \W4862.68$^\star$       & 100.0\p 1.1 & 100.0\p19.3  & 100.0\p 0.7 & 100.0\p 1.4 & 100.0\p 4.0 & 100.0\p14.5 & 100.0\p 4.1 & 100.0\p 3.9 & 100.0\p 5.6 \\
{[\ion{O}{3}]} \W5008.24 & 657.1\p 5.2 & 561.2\p121.3 & 377.7\p 2.1 & 697.6\p56.3 & 544.6\p17.2 & 646.7\p67.7 & 309.5\p10.4 & 750.0\p21.0 & 635.5\p26.3 \\
H$\alpha$ \W6564.61      & 254.7\p 2.2 & 277.8\p53.6  & 279.7\p 1.6 & 300.4\p25.0 & 288.4\p 9.8 & 277.8\p40.4 & 277.8\p11.5 & 277.8\p10.9 & 277.8\p15.6 \\
{[\ion{N}{2}]} \W6585.27 &  6.3\p 0.8  & \ldots       &  $<$3.2     &  9.9\p 1.1  & 17.9\p 0.6  & \ldots      & \ldots      & \ldots      & \ldots      \\
{[\ion{S}{2}]} \W6718.29 & \ldots      & \ldots       &  1.1\p 0.6  &  3.7\p 0.5  & \ldots      & \ldots      & \ldots      & \ldots      & \ldots      \\
{[\ion{S}{2}]} \W6732.67 & \ldots      & \ldots       &  1.7\p 0.7  &  4.0\p 0.5  & \ldots      & \ldots      & \ldots      & \ldots      & \ldots      \\
\hline
$E(B-V)$                 & \ldots      & \ldots       & 0.156       & 0.110       & 0.028       & 0.120       & \ldots      & \ldots      & \ldots      \\
$F_{{\rm H}\beta}$       & 25.4\p 0.2  & 18.6\p 2.5   & 37.9\p 0.2  & 100.0\p 1.0 & 142.0\p 4.0 &  2.1\p 0.2  & 78.9\p 2.3  &  3.6\p 0.1  & 32.7\p 1.3  \\
\enddata
\tablecomments{
Reddening-corrected emission-line intensities adopted for the High-$z$ Sample. 
The fluxes are given relative to $F_{{\rm H}\beta}\times100$. 
The last two rows give the dust extinction using the \citet{cardelli89} reddening law and the observed H$\beta$ fluxes, 
in units of $10^{-16}$ erg s$^{-1}$ cm$^{-2}$. \\
$^{\dagger}$\ion{N}{3}] \W1750 is the integrated flux of the \ion{N}{3}] \W\W1746,1748,1749,1752,1754 quintuplet.\\
$^\star$H$\beta$ \W4862 is not covered in the spectrum of GN-z11; instead, the H$\beta$ flux was inferred from the
measured H$\gamma$ flux.}
\end{deluxetable}
\clearpage

\section{\texorpdfstring{O$^{+2}$/O$^{+}$}{O+2/O+} Ionization Parameter Diagnostic Fit} \label{Appendix:C}
\begin{figure}[ht]
    \begin{center}
    \includegraphics[width=0.85\textwidth]{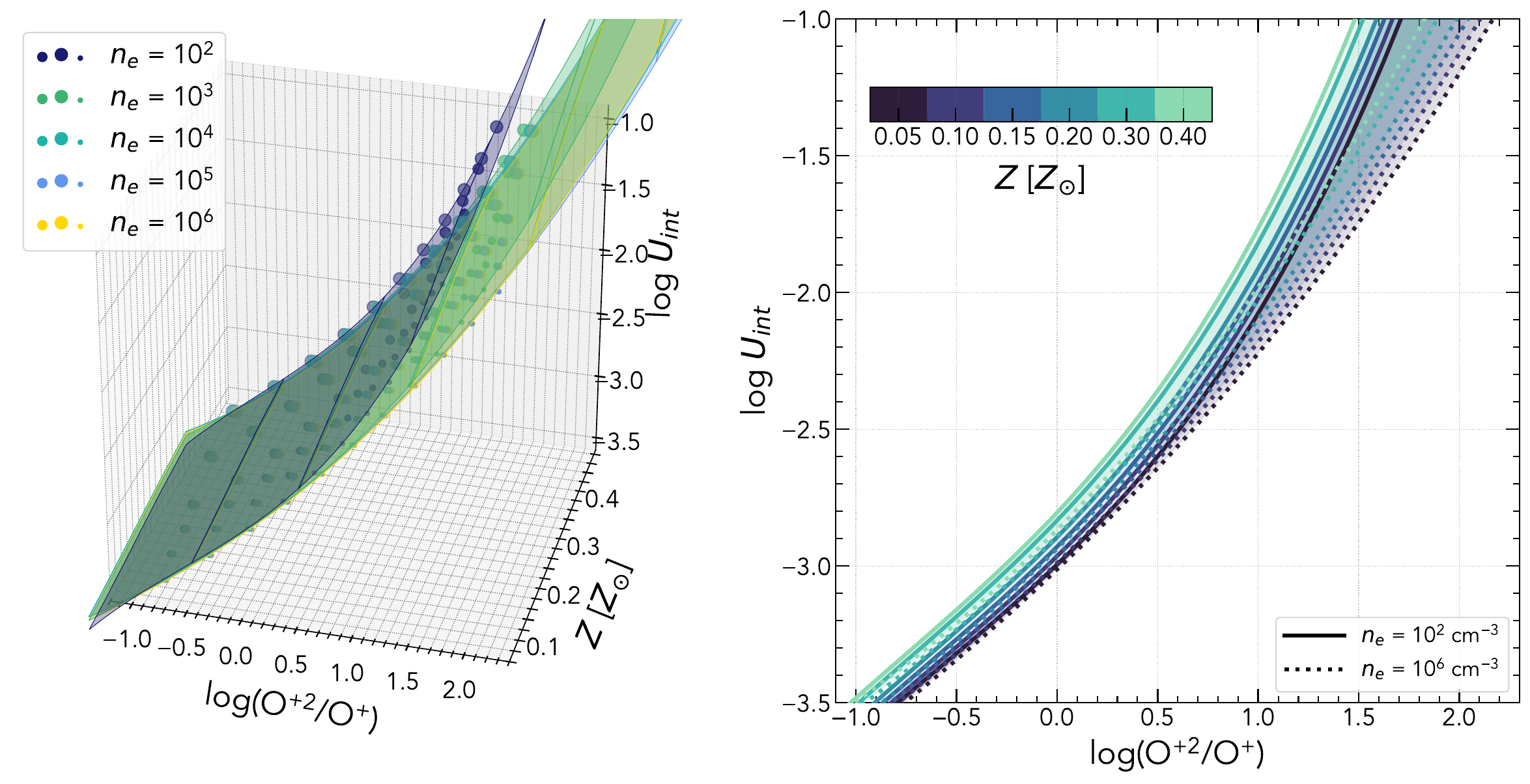}
    \caption{The \texttt{cloudy} model relationship between O$^{+2}$/O$^{+}$ and $\log U$.
    The left panel shows 3D plots of the bicubic fits for each grid of models, and the marker color denotes \den~ranging from $10^2 - 10^6$ cm$^{-3}$.
    In the right panel, we show a 2D projection where the range of \den~is shown by the shaded region.
    The metallicities range between 0.05 $Z_{\odot}$ and 0.40 $Z_{\odot}$ as indicated by the color bar in the upper right plot.
    Additionally, the solid lines shows the fits at $n_e = 10^2$ cm$^{-3}$ and the dotted lines mark where the fits are at $n_e = 10^6$ cm$^{-3}$.
    The effect of \den~on O$^{+2}$/O$^{+}$ is minimal making this probe of $\log U$ extremely robust.}
    \label{fig:LU_ION_fits}
    \end{center}
\end{figure}

We present another \den-insensitive diagnostic for $\log U$ using the relative ionic abundance, O$^{+2}$/O$^{+}$.
This probe is useful when there is not access to the lines necessary to utilize N$_{43}$ as a diagnostic that is robust to \den~or in the event it is not possible to appropriately characterize the entire \den~structure.
We used the functional form $f(x, y) = A + Bx + Cy + Dxy + E x^2 + Fy^2 + Gxy^2 + Hyx^2 + Ix^3 + Jy^3$ to fit a bicubic surface to the O$^{+2}$/O$^{+}$ -- $\log U$ relationship for \den~ranging from 10$^2$ to 10$^6$ cm$^{-3}$ using the same procedure outlined in Section \ref{sec:ICF_varyden}.
The coefficients for these fits can be found in Table \ref{tab:LU_ION_fits}.
While the left panel of Figure \ref{fig:LU_ION_fits}, we present a 3D plot of our fits along with the models color coded by \den, the right panel shows a 2D projection of the fits at $n_e = 10^2$ and 10$^6$ cm$^{-3}$, where the color bar indicates the metallicity.
Figure \ref{fig:LU_ION_fits} shows that the modeled relationship between O$^{+2}$/O$^{+}$ -- $\log U$ is very insensitive to \den, especially at $n_e \gtrsim 10^3$ cm$^{-3}$.

\begin{deluxetable}{lDDDDD}[ht]
\tabletypesize{\scriptsize}
\setlength{\tabcolsep}{2.5pt}
\tablecaption{Ionization Parameter Fits \label{tab:LU_ION_fits}}
\tablehead{
\CH{} & \multicolumn{10}{c}{\den~($cm^{-3}$)} \\
\cline{2-11}
\CH{} & \twocolhead{10$^2$} & \twocolhead{10$^3$} & \twocolhead{10$^4$} & \twocolhead{10$^5$} & \twocolhead{10$^6$}
}
\decimals
\startdata
\multicolumn{11}{l}{$z = f(x,y) = \log U_{int}$; $x=$ log(O$^{+2}$/O$^{+}$); $y = Z$} \\
\hline
$A$ \ldots & -3.01893	 & -3.02826	 & -3.03228	 & -3.03319	 & -3.03090	 \\
$B$ \ldots &  0.68142	 &  0.67769	 &  0.68678	 &  0.68770	 &  0.68578	 \\
$C$ \ldots &  0.39134	 &  0.47521	 &  0.47450	 &  0.41954	 &  0.30140	 \\
$D$ \ldots &  0.40444	 &  0.36650	 &  0.36517	 &  0.34904	 &  0.25356	 \\
$E$ \ldots &  0.13516	 &  0.08499	 &  0.08376	 &  0.08278	 &  0.07391	 \\
$F$ \ldots &  1.93922	 &  1.24708	 &  1.01559	 &  1.14266	 &  1.42633	 \\
$G$ \ldots & -0.38687	 & -0.37495	 & -0.41327	 & -0.39662	 & -0.25902	 \\
$H$ \ldots &  0.11450	 &  0.10476	 &  0.11398	 &  0.11960	 &  0.13938	 \\
$I$ \ldots &  0.07717	 &  0.02243	 &  0.00769	 &  0.00657	 &  0.01181	 \\
$J$ \ldots & -3.95339	 & -2.87921	 & -2.47978	 & -2.61133	 & -2.90696	 \\
\enddata
\tablecomments{
The bicubic surface fit to \texttt{Cloudy} photoionization models for the ionization parameters are parameterized by the following equation: $f(x, y) = A + Bx + Cy + Dxy + E x^2 + Fy^2 + Gxy^2 + Hyx^2 + Ix^3 + Jy^3$.
}
\end{deluxetable}
\clearpage

\onecolumngrid
\section{Total and Relative Ionic Abundance Measurements}\label{Appendix:D}
We present the total and relative ionic abundances for every galaxy in our \lz~and \hz~samples in Tables \ref{tab:Lz_abunds} and \ref{tab:Hz_abunds}.
Additionally, each of these tables contains the \den, \Te, and $\log U$ derived to determine each set of abundances.
We clarify the ionization zone that each measurement traces for every property that we present.
We outline the procedure for deriving all of these properties in Section \ref{sec:implications_sample}.

\movetabledown=5cm
\begin{rotatetable}
\begin{deluxetable}{lccCCCCCCCC}
\setlength{\tabcolsep}{2.5pt}
\tabletypesize{\scriptsize}
\tablewidth{0pt}
\tablecaption{Ionic and Total N Abundances for Low-$z$ Sample\label{tab:Lz_abunds}}
\tablehead{
\CH{}                   & \CH{} & \CH{Ion. Zone}            & \CH{J0127-}    & \CH{J1314+}    & \CH{J1444+}    & \CH{}          & \CH{J1253-}    & \CH{J1044+}    & \CH{J1323-}    & \CH{J1545+}    \\ [-3ex]
\CH{}                   & \CH{Model} & \CH{$T_e$ $|$ $n_e$} & \CH{0619}      & \CH{3452}      & \CH{4237}      & \CH{Leo P}     & \CH{0312}      & \CH{0353}      & \CH{0132}      & \CH{0858}
}
\startdata
\multicolumn{11}{l}{\bf Densities (cm$^{-3}$):} \\
$n_{e, low}$(S$^{+}$)($\times 10^2$)   & Both  & $|$ L     &  3.8\pm 0.1    &  1.9\pm 0.1    &  <0.3          &  0.8\pm 0.6    &  4.2\pm 0.1    &  2.0\pm 0.2    &  4.4\pm 0.1    &  1.5\pm 0.1    \\
$n_{e, int.}$(C$^{+2}$)($\times 10^3$) & Multi & $|$ I     &  <0.7          &  5.8\pm 2.2    &  8.8\pm 6.8    & 64.4\pm14.7    &  7.1\pm 2.1    &  2.4\pm 1.4    &  <0.7          & 13.0\pm 2.8    \\
$n_{e, high}$(N$^{+3}$)($\times 10^4$) & Multi & $|$ H     & {\it  0.2}     & {\it  2.0}     & {\it  3.0}     & {\it 22.2}     &  6.5\pm 3.8    &  <0.4          &  <0.4          &  6.0\pm 3.0    \\
\multicolumn{11}{l}{\bf Temperatures (K):} \\
$T_{e, low}$($\times 10^4$)$^a$        & Uni.  & L $|$ L   & 1.50\pm0.01    & 1.08\pm0.01    & 1.52\pm0.06    & 1.50\pm0.06    & 1.27\pm0.01    & 1.64\pm0.01    & 1.55\pm0.01    & 1.46\pm0.01    \\
$T_{e, high}$(O$^{+2}$)($\times 10^4$) & Uni.  & H $|$ L   & 1.71\pm0.02    & 1.12\pm0.01    & 1.74\pm0.09    & 1.72\pm0.08    & 1.38\pm0.01    & 1.91\pm0.01    & 1.79\pm0.01    & 1.65\pm0.01    \\
$T_{e, low}$($\times 10^4$)$^b$        & Multi & L $|$ H   & 1.50\pm0.01    & 1.08\pm0.01    & 1.52\pm0.06    & 1.50\pm0.03    & 1.27\pm0.01    & 1.64\pm0.01    & 1.55\pm0.01    & 1.46\pm0.01    \\
$T_{e, int.}$($\times 10^4$)$^c$       & Multi & I $|$ H   & 1.59\pm0.02    & 1.07\pm0.01    & 1.55\pm0.07    & 1.27\pm0.04    & 1.22\pm0.01    & 1.75\pm0.01    & 1.65\pm0.01    & 1.42\pm0.01    \\
$T_{e, high}$(O$^{+2}$)($\times 10^4$) & Multi & H $|$ H   & 1.71\pm0.02    & 1.09\pm0.01    & 1.66\pm0.08    & 1.32\pm0.05    & 1.27\pm0.01    & 1.90\pm0.01    & 1.78\pm0.01    & 1.51\pm0.01    \\
\multicolumn{11}{l}{\bf Ionization Parameters:} \\
log$U_{int.}$(O$_{32}$) 	           & Uni.  & LH        & -2.724         & -2.256         & -2.736         & -2.612         & -1.857         & -1.619         & -1.000         & -1.952         \\
log$U_{high}$(N$_{43}$) 	           & Multi & IH        & \ldots         & -1.738         & -1.745         & -1.979         & -1.280         & -1.563         & -1.075         & -1.619         \\
\hline
\multicolumn{11}{l}{\bf Ionic and Total O Abundances:} \\
O$^+$/H$^+$($\times 10^{-5}$) 	       & Both  & L $|$ L   & 1.677\pm0.074  & 5.408\pm0.345  & <1.064         & 0.384\pm0.061  & 1.367\pm0.042  & 0.170\pm0.004  & 0.149\pm0.003  & 0.581\pm0.011  \\
O$^{+2}$/H$^+$($\times 10^{-5}$)       & Uni.  & H $|$ L   & 2.825\pm0.069  & 14.589\pm0.417 & 1.865\pm0.164  & 1.149\pm0.121  & 9.325\pm0.226  & 2.682\pm0.046  & 5.192\pm0.067  & 4.870\pm0.074  \\
O$^{+2}$/H$^+$($\times 10^{-5}$)       & Multi & H $|$ H   & 2.816\pm0.090  & 16.087\pm0.569 & 2.133\pm0.271  & 2.878\pm0.355  & 12.641\pm0.843 & 2.703\pm0.056  & 5.249\pm0.074  & 6.454\pm0.339  \\
12+log(O/H) 	 	 	 	 	       & Uni.  & All $|$ L & 7.653\pm0.010  & 8.301\pm0.012  & <7.483         & 7.185\pm0.037  & 8.029\pm0.009  & 7.455\pm0.007  & 7.728\pm0.005  & 7.736\pm0.006  \\
12+log(O/H) 	 	 	 	 	       & Multi & All       & 7.653\pm0.011  & 8.345\pm0.014  & <7.487         & 7.561\pm0.042  & 8.157\pm0.025  & 7.459\pm0.008  & 7.732\pm0.006  & 7.856\pm0.020  \\
\hline
\multicolumn{11}{l}{\bf Ionic and Relative N Abundances:} \\
N$^+$/O$^+$ 	 	 	 	 	       & Both  & L $|$ L   & 0.108\pm0.006  & 0.032\pm0.003  & >0.035         & 0.054\pm0.013  & 0.121\pm0.004  & 0.033\pm0.001  & 0.072\pm0.002  & 0.036\pm0.001  \\
N$^{+2}$/O$^{+2}$ 	 	 	 	       & Uni.  & H $|$ L   & 1.589\pm0.913  & 0.062\pm0.018  & 0.015\pm0.007  & 0.103\pm0.039  & \ldots         & \ldots         & \ldots         & \ldots         \\
N$^{+2}$/O$^{+2}$ 	 	 	 	       & Multi & IH $|$ IH & 2.230\pm1.283  & 0.071\pm0.022  & 0.021\pm0.009  & 0.119\pm0.049  & \ldots         & \ldots         & \ldots         & \ldots         \\
N$^{+3}$/O$^{+2}$ 	 	 	 	       & Uni.  & H $|$ L   & \ldots         & \ldots         & \ldots         & \ldots         & 0.024\pm0.007  & 0.006\pm0.000  & 0.023\pm0.002  & 0.014\pm0.001  \\
N$^{+3}$/O$^{+2}$ 	 	 	 	       & Multi & H $|$ H   & \ldots         & \ldots         & \ldots         & \ldots         & 0.025\pm0.007  & 0.006\pm0.000  & 0.023\pm0.001  & 0.015\pm0.001  \\
ICF(N$^{+}$) 	 	 	 	 	       &       &           & 1.039          & 1.101          & 0.943          & 0.978          & 1.008          & 0.926          & 0.919          & 0.974          \\
ICF(N$^{+2}$) 	 	 	 	 	       &       & 	 	   & 0.942          & 1.016          & 1.209          & 1.114          & 1.120          & 1.260          & 1.385          & 1.149          \\
ICF(N$^{+3}$) 	 	 	 	 	       &       & 	 	   & 55.951         & 25.092         & 5.984          & 9.963          & 8.892          & 4.926          & 3.425          & 7.588          \\
ICF(N$^{+}$+N$^{+2}$) 	 	 	       &       & 	 	   & 0.977          & 1.029          & 1.184          & 1.095          & 1.111          & 1.236          & 1.370          & 1.131          \\
ICF(N$^{+2}$+N$^{+3}$) 	 	 	       &       & 	 	   & 0.930          & 0.972          & 1.000          & 0.994          & 0.994          & 1.000          & 0.993          & 0.996          \\
log(N/O)$_{\rm N^+}$                   & Both  & L $|$ L   & -0.948\pm0.021 & -1.476\pm0.031 & >-1.529        & -1.386\pm0.100 & -0.952\pm0.015 & -1.516\pm0.010 & -1.179\pm0.009 & -1.487\pm0.012 \\
log(N/O)$_{\rm N^{+2}}$                & Uni.  & H $|$ L   & 0.175\pm0.197  & -1.221\pm0.113 & -1.838\pm0.156 & -0.993\pm0.149 & \ldots         & \ldots         & \ldots         & \ldots         \\
log(N/O)$_{\rm N^{+2}}$                & Multi & IH $|$ IH & 0.322\pm0.197  & -1.141\pm0.115 & -1.595\pm0.157 & -0.876\pm0.140 & \ldots         & \ldots         & \ldots         & \ldots         \\
log(N/O)$_{\rm N^{+3}}$                & Uni.  & H $|$ L   & \ldots         & \ldots         & \ldots         & \ldots         & -0.454\pm0.108 & -1.493\pm0.017 & -1.178\pm0.022 & -0.807\pm0.036 \\
log(N/O)$_{\rm N^{+3}}$                & Multi & H $|$ H   & \ldots         & \ldots         & \ldots         & \ldots         & -0.645\pm0.112 & -1.509\pm0.018 & -1.107\pm0.026 & -0.947\pm0.039 \\
log(N/O)$_{\rm N^{+}\;+\;N^{+2}}$      & Multi & All       & 0.325\pm0.194  & -1.077\pm0.105 & >-1.358        & -0.852\pm0.146 & \ldots         & \ldots         & \ldots         & \ldots         \\
log(N/O)$_{\rm N^{+2}\;+\;N^{+3}}$     & Multi & IH $|$ IH & \ldots         & \ldots         & \ldots         & \ldots         & \ldots         & \ldots         & \ldots         & \ldots         \\
\enddata
\tablecomments{
N$^{+3}$ densities for galaxies without N$^{+3}$ measurements are italicized for identification and were assumed to be 3.15$\times n_e$(C$^{+2}$) for the \lz~sample (see Section \ref{sec:temden}).
Every \Te(O$^{+2}$) derived for our \lz~sample utilized the [\OIII~\W4364/[\OIII~\W5008 emission line ratio. \\
$^a$The low-ionization zone temperature for the constant-\den~model was calculated from \Te(O$^{+2}$), determined with \den(S$^{+}$) or \den(O$^{+}$), using the theoretical \cite{garnett92} $T_e - T_e$ relationship. \\
$^b$The low-ionization zone temperature for the 3-zone \den~model was calculated from \Te(O$^{+2}$), determined with \den(N$^{+3}$), using the theoretical \cite{garnett92} $T_e-T_e$ relationship. \\
$^c$The intermediate-ionization zone temperature for the 3-zone \den~model was calculated from \Te(O$^{+2}$), determined with \den(N$^{+3}$), using the theoretical \cite{garnett92} $T_e-T_e$ relationship.
}
\end{deluxetable}
\end{rotatetable}

\movetabledown=5cm
\begin{rotatetable}
\begin{deluxetable}{lccCCCCCCCCC}
\setlength{\tabcolsep}{2.5pt}
\tabletypesize{\scriptsize}
\tablewidth{0pt}
\tablecaption{Ionic and Total N Abundances for High-$z$ Sample\label{tab:Hz_abunds}}
\tablehead{
\CH{}                   & \CH{} & \CH{Ion. Zone}            & \CH{RXC}           & \CH{}           & \CH{GDS}      & \CH{}            & \CH{J1723+}    & \CH{CEERS}    & \CH{SL2S}      & \CH{}         & \CH{A1703-} \\[-3ex]
\CH{}                   & \CH{Model} & \CH{$T_e$ $|$ $n_e$} & \CH{J2248-ID$^d$}  & \CH{GN-z11$^d$} & \CH{3073$^e$} & \CH{Sunburst$^e$} & \CH{3411$^d$} & \CH{1019$^d$} & \CH{J0217$^e$} & \CH{Lynx$^d$} & \CH{zd6$^d$}
}
\startdata
\multicolumn{11}{l}{\bf Densities (cm$^{-3}$):} \\
$n_{e,{low}}$(S$^{+}$)($\times 10^2$) 	& Both  & $|$ L     & {\it  5.7}     & {\it 10.2}      & 22.6\pm 9.0   & 17.0\pm 8.1    &  1.0\pm 0.2    &  6.1\pm 1.6    & {\it  1.9}     & {\it  3.1}     & {\it  6.6}     \\
$n_{e,{int.}}$(C$^{+2}$)($\times 10^4$) & Multi & $|$ I     &  9.7\pm 0.4    & {\it 32.4}      & {\it 22.2}    &  8.3\pm 0.8    &  0.3\pm 0.2    & {\it  7.4}     &  <0.07         &  0.2\pm 0.2    &  8.4\pm 2.2    \\
$n_{e,{high}}$(N$^{+3}$)($\times 10^5$) & Multi & $|$ H     &  3.1\pm 0.1    &  7.5\pm 1.0     &  3.9\pm 0.1   & {\it  3.1}     & {\it  0.1}     &  1.7\pm 0.3    & {\it  0.0}     &  1.0\pm 0.2    &  2.0\pm 0.1    \\
\multicolumn{11}{l}{\bf Temperatures (K):} \\
$T_{e,{low}}$($\times 10^4$)$^a$        & Uni.  & L $|$ L   & 1.83\pm0.01    & 1.33\pm0.06     & 1.85\pm0.03   & 1.08\pm0.01    & 1.06\pm0.02    & 2.05\pm0.12    & 1.40\pm0.14    & 1.50\pm0.02    & 1.80\pm0.03    \\
$T_{e,{high}}$(O$^{+2}$)($\times 10^4$) & Uni.  & H $|$ L   & 2.19\pm0.01    & 1.47\pm0.09     & 2.21\pm0.04   & 1.12\pm0.02    & 1.09\pm0.03    & 2.50\pm0.17    & 1.57\pm0.20    & 1.72\pm0.03    & 2.14\pm0.05    \\
$T_{e,{low}}$($\times 10^4$)$^b$        & Multi & L $|$ H   & 1.67\pm0.01    & 1.18\pm0.05     & 1.29\pm0.01   & 1.03\pm0.01    & 1.06\pm0.02    & 1.93\pm0.10    & 1.39\pm0.14    & 1.47\pm0.01    & 1.69\pm0.03    \\
$T_{e,{int.}}$($\times 10^4$)$^c$       & Multi & I $|$ H   & 1.80\pm0.01    & 1.21\pm0.06     & 1.34\pm0.02   & 1.03\pm0.02    & 1.07\pm0.02    & 2.10\pm0.12    & 1.46\pm0.16    & 1.56\pm0.02    & 1.82\pm0.03    \\
$T_{e,{high}}$(O$^{+2}$)($\times 10^4$) & Multi & H $|$ H   & 1.96\pm0.01    & 1.25\pm0.07     & 1.41\pm0.02   & 1.04\pm0.02    & 1.08\pm0.03    & 2.33\pm0.15    & 1.56\pm0.20    & 1.67\pm0.02    & 1.99\pm0.04    \\
\multicolumn{11}{l}{\bf Ionization Parameters:} \\
log$U_{int.}$(O$_{32}$) 	            & Uni.  & LH        & -1.000         & -1.670          & -1.615        & -1.279         & -2.105         & -2.015         & -1.500         & -1.000         & -1.000         \\
log$U_{high}$(N$_{43}$) 	            & Multi & IH        & -1.000         & -1.000          & -1.000        & \ldots         & \ldots         & -1.000         & \ldots         & -1.000         & -1.000         \\
\hline
\multicolumn{11}{l}{\bf Ionic and Total O Abundances:} \\
O$^+$/H$^+$($\times 10^{-5}$) 	        & Both  & L $|$ L   & 0.022\pm0.008  & 0.678\pm0.171   & 0.136\pm0.018 & 2.026\pm0.409  & 3.777\pm0.310  & 0.265\pm0.048  & \ldots         & <0.216         & <0.096         \\
O$^{+2}$/H$^+$($\times 10^{-5}$)        & Uni.  & H $|$ L   & 3.116\pm0.035  & 6.472\pm1.636   & 1.757\pm0.053 & 17.335\pm1.683 & 14.900\pm1.137 & 2.396\pm0.363  & 3.040\pm1.029  & 5.917\pm0.266  & 3.152\pm0.188  \\
O$^{+2}$/H$^+$($\times 10^{-5}$)        & Multi & H $|$ H   & 5.360\pm0.095  & 20.263\pm5.832  & 7.320\pm0.297 & 31.814\pm3.414 & 15.238\pm1.357 & 3.262\pm0.505  & 3.072\pm1.010  & 7.051\pm0.373  & 4.525\pm0.270  \\
$12+\log(\rm O/H)$ 	 	 	 	        &  Uni. & All $|$ L & 7.497\pm0.005  & 7.854\pm0.090   & 7.277\pm0.013 & 8.287\pm0.037  & 8.271\pm0.027  & 7.425\pm0.056  & >7.483         & <7.788         & <7.512         \\
$12+\log(\rm O/H)$ 	 	 	 	        & Multi & All       & 7.731\pm0.008  & 8.328\pm0.105   & 7.888\pm0.017 & 8.535\pm0.042  & 8.279\pm0.031  & 7.553\pm0.058  & >7.487         & <7.862         & <7.666         \\
\hline
\multicolumn{11}{l}{\bf Ionic and Relative N Abundances:} \\
N$^+$/O$^+$ 	 	 	 	 	        & Both  & L $|$ L   & 1.624\pm0.627  & \ldots          & <0.131        & 0.083\pm0.020  & 0.085\pm0.009  & \ldots         & \ldots         & \ldots         & \ldots         \\
N$^{+2}$/O$^{+2}$ 	 	 	 	        & Uni.  & H $|$ L   & 0.059\pm0.003  & 0.917\pm0.721   & 1.624\pm0.450 & 0.213\pm0.075  & <0.035         & 0.137\pm0.073  & 0.014\pm0.020  & 0.098\pm0.017  & 0.078\pm0.056  \\
N$^{+2}$/O$^{+2}$ 	 	 	 	        & Multi & IH $|$ IH & 0.079\pm0.004  & 1.048\pm0.848   & 1.886\pm0.576 & 0.221\pm0.085  & <0.038         & 0.186\pm0.095  & 0.020\pm0.024  & 0.134\pm0.018  & 0.106\pm0.075  \\
N$^{+3}$/O$^{+2}$ 	 	 	 	        & Uni.  & H $|$ L   & 0.276\pm0.011  & 0.570\pm0.491   & 3.497\pm1.733 & \ldots         & \ldots         & 0.544\pm0.279  & \ldots         & 0.179\pm0.030  & 0.145\pm0.057  \\
N$^{+3}$/O$^{+2}$ 	 	 	 	        & Multi & H $|$ H   & 0.288\pm0.014  & 0.625\pm0.575   & 3.902\pm1.991 & \ldots         & \ldots         & 0.557\pm0.257  & \ldots         & 0.186\pm0.023  & 0.150\pm0.057  \\
ICF(N$^{+}$) 	 	 	 	 	        &       &           & 0.915          & 1.065           & 0.939         & 1.280          & 1.090          & 0.895          & 0.920          & 0.934          & 0.907          \\
ICF(N$^{+2}$) 	 	 	 	 	        &       & 	 	    & 1.470          & 1.152           & 1.396         & 1.191          & 0.981          & 1.534          & 1.290          & 1.409          & 1.495          \\
ICF(N$^{+3}$) 	 	 	 	 	        &       & 	 	    & 2.944          & 6.735           & 3.430         & 2.353          & 41.966         & 2.658          & 4.440          & 3.328          & 2.816          \\
ICF(N$^{+}$+N$^{+2}$) 	 	 	        &       & 	 	    & 1.460          & 1.149           & 1.386         & 1.169          & 1.006          & 1.526          & 1.267          & 1.399          & 1.487          \\
ICF(N$^{+2}$+N$^{+3}$) 	 	 	        &       & 	 	    & 0.989          & 0.993           & 0.992         & 0.956          & 0.960          & 0.987          & 0.999          & 0.991          & 0.988          \\
log(N/O)$_{\rm N^+}$                    & Both  & L $|$ L   & 0.134\pm0.142  & \ldots          & <-1.078       & -1.006\pm0.093 & -1.035\pm0.040 & \ldots         & \ldots         & \ldots         & \ldots         \\
log(N/O)$_{\rm N^{+2}}$                 & Uni.  & H $|$ L   & -1.038\pm0.021 & 0.021\pm0.245   & 0.315\pm0.105 & -0.632\pm0.128 & <-1.459        & -0.813\pm0.176 & -1.746\pm0.347 & -0.849\pm0.068 & -0.921\pm0.234 \\
log(N/O)$_{\rm N^{+2}}$                 & Multi & IH $|$ IH & -0.937\pm0.024 & 0.082\pm0.224   & 0.421\pm0.116 & -0.580\pm0.144 & <-1.426        & -0.544\pm0.185 & -1.599\pm0.326 & -0.725\pm0.060 & -0.799\pm0.234 \\
log(N/O)$_{\rm N^{+3}}$                 & Uni.  & H $|$ L   & -0.144\pm0.016 & 0.645\pm0.239   & 1.231\pm0.173 & \ldots         & \ldots         & 0.733\pm0.156  & \ldots         & -0.257\pm0.070 & -0.422\pm0.145 \\
log(N/O)$_{\rm N^{+3}}$                 & Multi & H $|$ H   & -0.072\pm0.022 & 0.624\pm0.234   & 1.127\pm0.178 & \ldots         & \ldots         & 0.170\pm0.158  & \ldots         & -0.208\pm0.056 & -0.375\pm0.146 \\
log(N/O)$_{\rm N^{+}\;+\;N^{+2}}$       & Multi & All       & -0.907\pm0.137 & \ldots          & <0.416        & -0.579\pm0.162 & <-1.236        & \ldots         & \ldots         & \ldots         & \ldots         \\
log(N/O)$_{\rm N^{+2}\;+\;N^{+3}}$      & Multi & IH $|$ IH & -0.441\pm0.018 & 0.220\pm0.176   & 0.759\pm0.132 & \ldots         & \ldots         & -0.134\pm0.132 & \ldots         & -0.499\pm0.042 & -0.597\pm0.139 \\
\enddata
\tablecomments{
N$^{+3}$ densities for galaxies without N$^{+3}$ measurements are italicized for identification and were assumed to be 2.59$\times$$n_e$(C$^{+2}$) for the \hz~sample (see Section \ref{sec:temden}). \\
$^a$The low-ionization zone temperature for the constant-\den~model was calculated from \Te(O$^{+2}$), determined with \den(S$^{+}$) or \den(O$^{+}$), using the theoretical \cite{garnett92} $T_e-T_e$ relationship. \\
$^b$The low-ionization zone temperature for the 3-zone \den~model was calculated from \Te(O$^{+2}$), determined with \den(N$^{+3}$), using the theoretical \cite{garnett92} $T_e-T_e$ relationship. \\
$^c$The intermediate-ionization zone temperature for the 3-zone \den~model was calculated from \Te(O$^{+2}$), determined with \den(N$^{+3}$), using the theoretical \cite{garnett92} $T_e-T_e$ relationship. \\
$^d$The \Te(O$^{+2}$) measured for these objects used the \OIII~\W1666/[\OIII~\W5008 emission line ratio. \\
$^e$The \Te(O$^{+2}$) measured for these objects used the [\OIII~\W4364/[\OIII~\W5008 emission line ratio.
}
\end{deluxetable}
\end{rotatetable}

\end{document}